\documentclass[acmsmall,authorversion,natbib=false]{acmart}
\usepackage[utf8]{luainputenc}

\def\DefaultCutFileName{\def\CommentCutFile{\jobname.cut}}
\DefaultCutFileName{}

\usepackage[style=ACM-Reference-Format, maxcitenames=2, bibencoding=utf8, sortcites=true, sorting=customyeartitle]{biblatex}
\DeclareSortingTemplate{customyeartitle}{
 \sort{\field{year}}
 \sort{\field{title}}
}
\DeclareFieldFormat{titlecase}{#1}
\usepackage{citeall}
\usepackage[capitalise,nameinlink,noabbrev,english]{cleveref}
\DeclareCiteCommand{\citetitle}
{\boolfalse{citetracker}%
	\boolfalse{pagetracker}%
	\usebibmacro{prenote}}
{\ifciteindex%
	{\indexfield{indextitle}}
	{}%
	\printtext[bibhyperref]{\printfield[citetitle]{labeltitle}}}
{\multicitedelim}
{\usebibmacro{postnote}}
\DeclareCiteCommand{\citeauthor}
{\boolfalse{citetracker}%
	\boolfalse{pagetracker}%
	\usebibmacro{prenote}}
{\ifciteindex%
	{\indexnames{labelname}}
	{}%
	\printtext[bibhyperref]{\printnames{labelname}}}
{\multicitedelim}
{\usebibmacro{postnote}}
\usepackage{wrapfig}
\usepackage{tabularx}
\usepackage{seqsplit}
\usepackage{tabu}
\usepackage{longtable}
\usepackage{array}
\usepackage{multirow}
\newcolumntype{M}[1]{>{\centering\arraybackslash}m{#1}}
\usepackage{arydshln}
\usepackage{ragged2e}
\usepackage{colortbl}
\usepackage{rotating}
\usepackage{tablefootnote}
\usepackage{footnote}
\makesavenoteenv{tabular}
\makesavenoteenv{table}

\usepackage{caption}
\usepackage{subcaption}
\usepackage{graphicx}
\usepackage{tikz}
\usepackage{pgfplots}
\usepackage{sparklines}
\pgfplotsset{compat=newest}
\usepgfplotslibrary{groupplots}
\usepackage{tikzscale}
\pgfplotsset{
	every tick label/.append style={scale=0.75},
	every axis/.append style={
	}
}

\usetikzlibrary{positioning,fit,shapes,decorations.shapes,decorations.markings}
\usepackage{xstring}
\usepackage{xfp}
\usepackage{expl3}
\usepackage{xparse}
\usepackage[normalem]{ulem}
\usepackage{wasysym}

\usepackage{csquotes}

\newcommand*{\priority}[1]{{\small{
\begin{tikzpicture}[scale=0.10]%
    \draw (0,0) circle (1);
    \fill[fill opacity=1,fill=black] (0,0) -- (90:1) arc (90:90-#1*3.6:1) -- cycle;
\end{tikzpicture}%
}}}

\newcommand*{\priorityss}[1]{{\scriptsize{
\begin{tikzpicture}[scale=0.10]%
    \draw (0,0) circle (1);
    \fill[fill opacity=1,fill=black] (0,0) -- (90:1) arc (90:90-#1*3.6:1) -- cycle;
\end{tikzpicture}%
}}}

\newcommand{\changed}[1]{#1}

\newcommand{\bemph}[1]{{$\RHD$\ \emph{#1}}}

\usepackage[draft]{fixme}
\fxsetup{layout=inline,theme=color}
\FXRegisterAuthor{bds}{brds}{\color{purple}bds}
\FXRegisterAuthor{ss}{ses}{\color{brown}ss}
\FXRegisterAuthor{bc}{ebc}{\color{red}bc}
\FXRegisterAuthor{pk}{pak}{\color{orange}pk}

\AtBeginDocument{%
    \providecommand\BibTeX{{%
        \normalfont{B\kern-0.5em{\scshape i\kern-0.25em b}\kern-0.8em\TeX}}%
    }%
}%

\newcommand{\totalPapers}{571}
\newcommand{\numberAspects}{113}
\newcommand{\numberDatapoints}{64523}
\newcommand{\numberProtections}{31}
\newcommand{\numberAnalyses}{28}
\newcommand{\totalSurveys}{38}
\newcommand{\totalSurveysIncluded}{25}
\newcommand{\totalObfuscation}{360}
\newcommand{\totalDeobfuscation}{102}
\newcommand{\totalAnalysis}{166}

\newcommand{\totalDiversification}{33}
\newcommand{\totalTamperproofing}{42}
\newcommand{\totalWatermarking}{12}
\newcommand{\totalMobile}{88}
\newcommand{\totalHumanExperiments}{20}
\newcommand{\totalGoodwareHumanExperiments}{19}
\newcommand{\totalMalware}{183}
\newcommand{\totalMalwareOnly}{99}
\newcommand{\MalwareOnlyOfMalwarePercentage}{54}
\newcommand{\totalGoodware}{388}
\newcommand{\totalMachinelearning}{77}
\newcommand{\totalTheory}{29}
\newcommand{\percentSurveys}{4.4}
\newcommand{\percentObfuscation}{63}
\newcommand{\percentDeobfuscation}{18}
\newcommand{\percentAnalysis}{29}
\newcommand{\percentDiversification}{5.8}
\newcommand{\percentTamperproofing}{7.4}
\newcommand{\percentWatermarking}{2.1}
\newcommand{\percentMobile}{15}
\newcommand{\percentMalware}{32}
\newcommand{\percentMalwareOnly}{17}
\newcommand{\percentHumanExperiments}{3.5}
\newcommand{\percentGoodwareHumanExperiments}{4.9}
\newcommand{\percentGoodware}{68}
\newcommand{\percentMachinelearning}{13}
\newcommand{\percentTheory}{5.1}
\newcommand{\totalObfuscationOnly}{327}
\newcommand{\totalDeobfuscationOnly}{68}
\newcommand{\totalAnalysisOnly}{121}
\newcommand{\totalAnalysisDeobfuscationOnly}{22}
\newcommand{\totalAnalysisObfuscationOnly}{21}
\newcommand{\totalDeobfuscationObfuscationOnly}{10}
\newcommand{\totalAnalysisDeobfuscationObfuscationOnly}{2}
\newcommand{\percentObfuscationOnly}{57}
\newcommand{\percentDeobfuscationOnly}{12}
\newcommand{\percentAnalysisOnly}{21}
\newcommand{\percentAnalysisDeobfuscationOnly}{3.9}
\newcommand{\percentAnalysisObfuscationOnly}{3.7}
\newcommand{\percentDeobfuscationObfuscationOnly}{1.8}
\newcommand{\percentAnalysisDeobfuscationObfuscationOnly}{0.4}

\newcommand{\totalTools}{768}
\newcommand{\totalToolsOnce}{532}
\newcommand{\percentToolsOnce}{74}
\newcommand{\totalToolsMoreThanFive}{39}
\newcommand{\percentToolsMoreThanFive}{5.7}
\newcommand{\totalToolsIdaPro}{63}
\newcommand{\totalToolsIdaProDis}{15}
\newcommand{\percentTotalToolsIdaProDis}{24}
\newcommand{\totalToolsIdaProPlugins}{20}
\newcommand{\percentTotalToolsIdaProPlugins}{32}
\newcommand{\totalToolsIdaProNotJustBindiff}{6}
\newcommand{\percentTotalToolsIdaProNotJustBindiff}{9.5}
\newcommand{\totalToolsIdaProOnlyBindiff}{14}
\newcommand{\percentTotalToolsIdaProOnlyBindiff}{22}
\newcommand{\totalToolsIdaProAfterTwentySeventeen}{19}
\newcommand{\totalToolsPinAfterTwentySeventeen}{8}
\newcommand{\totalToolsAngrAfterTwentySeventeen}{11}
\newcommand{\totalToolsTritonAfterTwentySeventeen}{7}
\newcommand{\totalToolsKleeAfterTwentySeventeen}{9}
\newcommand{\totalToolsOLLVM}{35}
\newcommand{\totalToolsProGuard}{19}
\newcommand{\totalToolsDexGuard}{3}
\newcommand{\totalToolsPlto}{4}
\newcommand{\totalToolsGhidra}{1}
\newcommand{\totalToolsBinaryninja}{2}

\newcommand{\totalToolsDava}{4}

\newcommand{\averageAnaPerAnaOrDeobfWithAna}{3.3}
\newcommand{\averageAnaPerObfWithAna}{2.0}
\newcommand{\venueAs}{57}
\newcommand{\venueA}{60}
\newcommand{\venueB}{74}
\newcommand{\venueC}{45}
\newcommand{\venueWorkshop}{81}
\newcommand{\venueUnranked}{237}
\newcommand{\venueNational}{15}
\newcommand{\venueNew}{2}
\newcommand{\percentVenueAs}{10.0}
\newcommand{\percentVenueA}{11}
\newcommand{\percentVenueB}{13}
\newcommand{\percentVenueC}{13}
\newcommand{\percentVenueWorkshop}{14}
\newcommand{\percentVenueUnranked}{42}
\newcommand{\percentVenueNational}{2.6}
\newcommand{\percentVenueNew}{0.4}

\newcommand{\stealthAllCategoriesX}{73}
\newcommand{\stealthProtectionPapersMalwareX}{3}
\newcommand{\stealthProtectionPapersNonMalwareX}{21}
\newcommand{\stealthAnaDeobfPapersMalwareX}{29}
\newcommand{\stealthAnaDeobfPapersNonMalwareX}{14}
\newcommand{\potencyAllCategoriesX}{199}
\newcommand{\potencyProtectionPapersMalwareX}{36}
\newcommand{\potencyProtectionPapersNonMalwareX}{97}
\newcommand{\potencyAnaDeobfPapersMalwareX}{63}
\newcommand{\potencyAnaDeobfPapersNonMalwareX}{13}
\newcommand{\resilienceAllCategoriesX}{122}
\newcommand{\resilienceProtectionPapersMalwareX}{3}
\newcommand{\resilienceProtectionPapersNonMalwareX}{40}
\newcommand{\resilienceAnaDeobfPapersMalwareX}{43}
\newcommand{\resilienceAnaDeobfPapersNonMalwareX}{43}
\newcommand{\costsAllCategoriesX}{245}
\newcommand{\costsProtectionPapersMalwareX}{11}
\newcommand{\costsProtectionPapersNonMalwareX}{197}
\newcommand{\costsAnaDeobfPapersMalwareX}{20}
\newcommand{\costsAnaDeobfPapersNonMalwareX}{29}
\newcommand{\malwareDetPrecAllCategoriesX}{76}
\newcommand{\malwareDetPrecProtectionPapersMalwareX}{30}
\newcommand{\malwareDetPrecProtectionPapersNonMalwareX}{0}
\newcommand{\malwareDetPrecAnaDeobfPapersMalwareX}{49}
\newcommand{\malwareDetPrecAnaDeobfPapersNonMalwareX}{0}
\newcommand{\precisionAllCategoriesX}{118}
\newcommand{\precisionProtectionPapersMalwareX}{4}
\newcommand{\precisionProtectionPapersNonMalwareX}{31}
\newcommand{\precisionAnaDeobfPapersMalwareX}{50}
\newcommand{\precisionAnaDeobfPapersNonMalwareX}{40}
\newcommand{\similarityAllCategoriesX}{39}
\newcommand{\similarityProtectionPapersMalwareX}{3}
\newcommand{\similarityProtectionPapersNonMalwareX}{16}
\newcommand{\similarityAnaDeobfPapersMalwareX}{13}
\newcommand{\similarityAnaDeobfPapersNonMalwareX}{7}
\newcommand{\entropyAllCategoriesX}{10}
\newcommand{\entropyProtectionPapersMalwareX}{1}
\newcommand{\entropyProtectionPapersNonMalwareX}{5}
\newcommand{\entropyAnaDeobfPapersMalwareX}{3}
\newcommand{\entropyAnaDeobfPapersNonMalwareX}{2}
\newcommand{\sizeAllCategoriesX}{149}
\newcommand{\sizeProtectionPapersMalwareX}{8}
\newcommand{\sizeProtectionPapersNonMalwareX}{111}
\newcommand{\sizeAnaDeobfPapersMalwareX}{17}
\newcommand{\sizeAnaDeobfPapersNonMalwareX}{18}
\newcommand{\manualAnaAllCategoriesX}{24}
\newcommand{\manualAnaProtectionPapersMalwareX}{2}
\newcommand{\manualAnaProtectionPapersNonMalwareX}{15}
\newcommand{\manualAnaDeobfPapersMalwareX}{4}
\newcommand{\manualAnaDeobfPapersNonMalwareX}{5}

\newcommand{\memoryUsageAllCategoriesX}{18}
\newcommand{\memoryUsageProtectionPapersMalwareX}{0}
\newcommand{\memoryUsageProtectionPapersNonMalwareX}{18}
\newcommand{\memoryUsageDeobfPapersMalwareX}{0}
\newcommand{\memoryUsageDeobfPapersNonMalwareX}{0}
\newcommand{\opcodeDistributionAllCategoriesX}{21}
\newcommand{\opcodeDistributionProtectionPapersMalwareX}{0}
\newcommand{\opcodeDistributionProtectionPapersNonMalwareX}{16}
\newcommand{\opcodeDistributionDeobfPapersMalwareX}{4}
\newcommand{\opcodeDistributionDeobfPapersNonMalwareX}{2}
\newcommand{\codeComplexityAllCategoriesX}{59}
\newcommand{\codeComplexityProtectionPapersMalwareX}{2}
\newcommand{\codeComplexityProtectionPapersNonMalwareX}{48}
\newcommand{\codeComplexityDeobfPapersMalwareX}{2}
\newcommand{\codeComplexityDeobfPapersNonMalwareX}{10}
\newcommand{\applicabilityAllCategoriesX}{24}
\newcommand{\applicabilityProtectionPapersMalwareX}{2}
\newcommand{\applicabilityProtectionPapersNonMalwareX}{16}
\newcommand{\applicabilityDeobfPapersMalwareX}{3}
\newcommand{\applicabilityDeobfPapersNonMalwareX}{5}
\newcommand{\powerConsumptionAllCategoriesX}{4}
\newcommand{\powerConsumptionProtectionPapersMalwareX}{0}
\newcommand{\powerConsumptionProtectionPapersNonMalwareX}{4}
\newcommand{\powerConsumptionDeobfPapersMalwareX}{0}
\newcommand{\powerConsumptionDeobfPapersNonMalwareX}{0}
\newcommand{\attackAnaOverheadTimeAllCategoriesX}{117}
\newcommand{\attackAnaOverheadTimeProtectionPapersMalwareX}{4}
\newcommand{\attackAnaOverheadTimeProtectionPapersNonMalwareX}{20}
\newcommand{\attackAnaOverheadTimeDeobfPapersMalwareX}{63}
\newcommand{\attackAnaOverheadTimeDeobfPapersNonMalwareX}{35}
\newcommand{\compilationTimeAllCategoriesX}{22}
\newcommand{\compilationTimeProtectionPapersMalwareX}{3}
\newcommand{\compilationTimeProtectionPapersNonMalwareX}{17}
\newcommand{\compilationTimeDeobfPapersMalwareX}{2}
\newcommand{\compilationTimeDeobfPapersNonMalwareX}{1}
\newcommand{\performanceOverheadTimeAllCategoriesX}{148}
\newcommand{\performanceOverheadTimeProtectionPapersMalwareX}{5}
\newcommand{\performanceOverheadTimeProtectionPapersNonMalwareX}{137}
\newcommand{\performanceOverheadTimeDeobfPapersMalwareX}{2}
\newcommand{\performanceOverheadTimeDeobfPapersNonMalwareX}{10}
\newcommand{\otherCostsAllCategoriesX}{8}
\newcommand{\otherCostsProtectionPapersMalwareX}{0}
\newcommand{\otherCostsProtectionPapersNonMalwareX}{4}
\newcommand{\otherCostsDeobfPapersMalwareX}{2}
\newcommand{\otherCostsDeobfPapersNonMalwareX}{2}
\newcommand{\totalOffensive}{151}
\newcommand{\totalDefensive}{453}

\newcommand{\offensiveVIR}{38}
\newcommand{\defensiveVIR}{59}
\newcommand{\offensiveMBA}{12}
\newcommand{\defensiveMBA}{11}
\newcommand{\totalBeforeSelection}{1491}
\newcommand{\percentTotalInScope}{38}

\newcommand{\practicalPapers}{495}
\newcommand{\practicalPapersUsingAliasing}{20}
\newcommand{\percentPracticalPapersUsingAliasing}{4.0}
\newcommand{\practicalPapersObfUsingHardware}{10}
\newcommand{\practicalPapersAnaDeobfUsingHardware}{0}
\newcommand{\practicalPapersObfUsingCodemob}{12}
\newcommand{\practicalPapersAnaDeobfUsingCodemob}{2}
\newcommand{\practicalPapersObfUsingALI}{16}
\newcommand{\practicalPapersAnaDeobfUsingALI}{5}
\newcommand{\practicalPapersObfUsingSSE}{8}
\newcommand{\practicalPapersAnaDeobfUsingSSE}{3}
\newcommand{\practicalPapersObfUsingDtC}{23}
\newcommand{\practicalPapersAnaDeobfUsingDtC}{11}
\newcommand{\practicalPapersObfUsingCBT}{32}
\newcommand{\practicalPapersAnaDeobfUsingCBT}{15}
\newcommand{\practicalPapersObfUsingRPA}{12}
\newcommand{\practicalPapersAnaDeobfUsingRPA}{24}

\newcommand{\haveSampleset}{481}
\newcommand{\haveSamplesetPercentage}{84}
\newcommand{\haveSamplesetGoodware}{308}
\newcommand{\percentHaveSamplesetGoodware}{79}
\newcommand{\haveSamplesetMalware}{175}
\newcommand{\percentHaveSamplesetMalware}{96}
\newcommand{\haveSamplesetGoodwareNoObflang}{8}
\newcommand{\percentHaveSamplesetGoodwareNoObflang}{2.6}
\newcommand{\haveSamplesetMalwareNoObflang}{44}
\newcommand{\percentHaveSamplesetMalwareNoObflang}{25}
\newcommand{\haveSamplesetGoodwareSourcenativeMeasuresomething}{194}
\newcommand{\haveSamplesetGoodwareSourcenativeMeasuresomethingDeployVirt}{63}

\newcommand{\haveSamplesetGoodwareSourcenativeMeasuresomethingDeployVirtUseTigress}{25}
\newcommand{\percentHaveSamplesetGoodwareSourcenativeMeasuresomethingDeployVirtUseTigress}{40}
\newcommand{\haveSamplesetGoodwareSourcenativeObfsn}{15}
\newcommand{\percentHaveSamplesetGoodwareSourcenativeObfsn}{7.7}
\newcommand{\haveSamplesetGoodwareSourcenativeObfsm}{11}
\newcommand{\percentHaveSamplesetGoodwareSourcenativeObfsm}{5.7}
\newcommand{\haveSamplesetGoodwareSourcenativeObfmn}{5}
\newcommand{\percentHaveSamplesetGoodwareSourcenativeObfmn}{2.6}
\newcommand{\practicalPapersGoodwareObfuscation}{248}
\newcommand{\practicalPapersGoodwareObfuscationHaveNoAnalysis}{133}
\newcommand{\percentPracticalPapersGoodwareObfuscationHaveNoAnalysis}{54}
\newcommand{\practicalPapersGoodwareObfuscationHaveOneAnalysis}{54}
\newcommand{\percentPracticalPapersGoodwareObfuscationHaveOneAnalysis}{22}
\newcommand{\practicalPapersGoodwareObfuscationHaveMoreThanOneAnalysis}{61}
\newcommand{\percentPracticalPapersGoodwareObfuscationHaveMoreThanOneAnalysis}{25}
\newcommand{\practicalPapersGoodwareObfuscationAAndAstar}{41}
\newcommand{\practicalPapersGoodwareObfuscationHaveNoAnalysisAandAstar}{11}
\newcommand{\percentPracticalPapersGoodwareObfuscationHaveNoAnalysisAandAstar}{27}
\newcommand{\practicalPapersMalwareObfuscation}{47}
\newcommand{\practicalPapersMalwareObfuscationHaveAnalysis}{38}
\newcommand{\percentPracticalPapersMalwareObfuscationHaveAnalysis}{81}
\newcommand{\practicalPapersMalwareObfuscationHaveOneAnalysis}{21}
\newcommand{\percentPracticalPapersMalwareObfuscationHaveOneAnalysis}{45}
\newcommand{\samplesetsToy}{182}
\newcommand{\samplesetsToyAfterSeventeen}{64}

\newcommand{\samplesetsToyPercentage}{38}
\newcommand{\samplesetsToyOnly}{91}

\newcommand{\samplesetsToyOnlyVirtualization}{24}
\newcommand{\samplesetsToyOnlyVirtualizationPercentage}{26}
\newcommand{\samplesetsMalware}{166}
\newcommand{\samplesetsMalwarePercentage}{35}
\newcommand{\samplesetsDrebin}{13}
\newcommand{\samplesetsDrebinPercentage}{7.8}
\newcommand{\samplesetsMalgenome}{7}
\newcommand{\samplesetsMalgenomePercentage}{4.2}
\newcommand{\samplesetsContagio}{3}
\newcommand{\samplesetsContagioPercentage}{1.8}
\newcommand{\samplesetsVirusshare}{9}
\newcommand{\samplesetsVirussharePercentage}{5.4}
\newcommand{\samplesetsAndrozoo}{5}
\newcommand{\samplesetsAndrozooPercentage}{3.0}

\newcommand{\samplesetsNomwTotal}{305}
\newcommand{\samplesetsNomwToy}{159}
\newcommand{\samplesetsNomwToyPercentage}{52}
\newcommand{\samplesetsNomwToyOnly}{90}
\newcommand{\samplesetsNomwToyOnlyPercentage}{57}
\newcommand{\samplesetsOs}{42}
\newcommand{\samplesetsOsPercentage}{8.7}
\newcommand{\samplesetsBenchmarks}{69}
\newcommand{\samplesetsBenchmarksPercentage}{14}

\newcommand{\samplesetsComplex}{129}
\newcommand{\samplesetsComplexPercentage}{27}
\newcommand{\samplesetsMobileapps}{91}
\newcommand{\samplesetsMobileappsPercentage}{19}
\newcommand{\samplesetsOther}{19}
\newcommand{\samplesetsOtherPercentage}{4.0}
\newcommand{\samplesetsUnknown}{9}
\newcommand{\samplesetsUnknownPercentage}{1.9}
\newcommand{\samplesSpec}{36}
\newcommand{\samplesSpecPercentage}{52}
\newcommand{\samplesMibench}{8}
\newcommand{\samplesMibenchPercentage}{12}
\newcommand{\samplesTum}{6}
\newcommand{\samplesTumPercentage}{8.7}
\newcommand{\samplesCoreutils}{14}
\newcommand{\samplesCoreutilsPercentage}{33}
\newcommand{\samplesWindowsosprogs}{18}
\newcommand{\samplesWindowsosprogsPercentage}{43}
\newcommand{\samplesGames}{29}
\newcommand{\samplesGamesPercentage}{22}
\newcommand{\samplesBrowsers}{6}
\newcommand{\samplesBrowsersPercentage}{4.7}
\newcommand{\samplesWebservers}{7}
\newcommand{\samplesWebserversPercentage}{5.4}
\newcommand{\samplesComprprogs}{28}
\newcommand{\samplesComprprogsPercentage}{22}
\newcommand{\samplesMediaprogs}{14}
\newcommand{\samplesMediaprogsPercentage}{11}
\newcommand{\samplesDatabaseprogs}{9}
\newcommand{\samplesDatabaseprogsPercentage}{7.0}
\newcommand{\samplesPublicrepos}{10}
\newcommand{\samplesPublicreposPercentage}{7.8}
\newcommand{\samplesMessagingprogs}{11}
\newcommand{\samplesMessagingprogsPercentage}{8.5}
\newcommand{\samplesSortingprogs}{26}
\newcommand{\samplesSortingprogsPercentage}{14}
\newcommand{\samplesEncprogs}{64}
\newcommand{\samplesEncprogsPercentage}{35}
\newcommand{\samplesMathprogs}{51}
\newcommand{\samplesMathprogsPercentage}{28}
\newcommand{\samplesTestingprogs}{71}
\newcommand{\samplesTestingprogsPercentage}{15}
\newcommand{\samplesGeneratedprogs}{15}
\newcommand{\samplesGeneratedprogsPercentage}{8.2}

\newcommand{\storeArrayPapersA}{\storedata{arrayPapersA}{{\mwc{$\text{\cite{2022_testing_detection_of_k_ary_code_obfuscated_by_metamorphic_and_polymorphic_techniques}}_{}$}}{\mwc{$\underline{\text{\cite{2022_semeo_a_semantic_equivalence_analysis_framework_for_obfuscated_android_applications}}}_{B}$}}{\gwc{$\text{\cite{2022_on_improvements_of_robustness_of_obfuscated_javascript_code_detection}}_{}$}}{\mwc{$\text{\cite{2022_obfuscation_resilient_semantic_functionality_identification_through_program_simulation}}_{}$}}{\mwc{$\overline{\text{\cite{2022_obfuscated_malware_detection_using_dilated_convolutional_network}}}_{}$}}{\mwc{$\underline{\text{\cite{2022_malware_detection_with_obfuscation_techniques_on_android_using_dynamic_analysis}}}_{}$}}{\gwc{$\overline{\text{\cite{2022_fine_grained_obfuscation_scheme_recognition_on_binary_code}}}_{}$}}{\mwc{$\overline{\text{\cite{2022_evaluating_opcodes_for_detection_of_obfuscated_android_malware}}}_{}$}}{\mwc{$\overline{\text{\cite{2022_enhancing_obfuscated_malware_detection_with_machine_learning_techniques}}}_{}$}}{\mwc{$\text{\cite{2022_construction_of_a_technological_component_to_support_isms_for_the_detection_of_obfuscation_in_computer_worm_samples}}_{}$}}{\mwc{$\underline{\text{\cite{2022_androclonium_bytecode_level_code_clone_detection_for_obfuscated_android_apps}}}_{}$}}{\mwc{$\overline{\text{\cite{2022_a_gan_based_anti_obfuscation_detection_method_for_malicious_code}}}_{}$}}{\mwc{$\text{\cite{2021_vabox_a_virtualization_based_analysis_framework_of_virtualization_obfuscated_packed_executables}}_{}$}}{\mwc{$\overline{\text{\cite{2021_utilizing_obfuscation_information_in_deep_learning_based_android_malware_detection}}}_{B}$}}{\mwc{$\overline{\text{\cite{2021_obfuscation_revealed_leveraging_electromagnetic_signals_for_obfuscated_malware_classification}}}_{}$}}{\mwc{$\overline{\text{\cite{2021_lom_lightweight_classifier_for_obfuscation_methods}}}_{w}$}}{\mwc{$\text{\cite{2021_learning_metamorphic_malware_signatures_from_samples}}_{}$}}{\gwc{$\text{\cite{2021_anti_obfuscation_binary_code_clone_detection_based_on_software_gene}}_{}$}}{\mwc{$\underline{\text{\cite{2021_androshow_a_large_scale_investigation_to_identify_the_pattern_of_obfuscated_android_malware}}}_{}$}}{\mwc{$\overline{\textbf{\text{\cite{2021_a_survey_of_binary_code_similarity}}}}_{A^*}$}}{\mwc{$\underline{\overline{\textbf{\text{\cite{2021_a_survey_of_android_malware_static_detection_technology_based_on_machine_learning}}}}}_{C}$}}{\mwc{$\text{\cite{2021_a_novel_approach_for_analysing_and_detection_of_obfuscated_malware_payloads_in_android_platform_using_dexmonitor}}_{}$}}{\mwc{$\overline{\text{\cite{2020_towards_obfuscated_malware_detection_for_low_powered_iot_devices}}}_{C}$}}{\gwc{$\text{\cite{2020_similarity_of_binaries_across_optimization_levels_and_obfuscation}}_{A}$}}{\gwc{$\overline{\text{\cite{2020_semantics_aware_obfuscation_scheme_prediction_for_binary}}}_{B}$}}{\mwc{$\underline{\overline{\text{\cite{2020_droidpdf_the_obfuscation_resilient_packer_detection_framework_for_android_apps}}}}_{}$}}{\mwc{$\overline{\text{\cite{2020_detection_of_obfuscated_mobile_malware_with_machine_learning_and_deep_learning_models}}}_{}$}}{\mwc{$\underline{\text{\cite{2020_detection_efficiency_of_static_analyzers_against_obfuscated_android_malware}}}_{}$}}{\mwc{$\text{\cite{2020_detecting_repackaged_android_applications_using_perceptual_hashing}}_{}$}}{\mwc{$\underline{\overline{\text{\cite{2020_dandroid_a_multi_view_discriminative_adversarial_network_for_obfuscated_android_malware_detection}}}}_{}$}}{\mwc{$\underline{\text{\cite{2020_code_reordering_obfuscation_technique_detection_by_means_of_weak_bisimulation}}}_{B}$}}{\mwc{$\underline{\overline{\text{\cite{2020_aomdroid_detecting_obfuscation_variants_of_android_malware_using_transfer_learning}}}}_{}$}}{\mwc{$\underline{\overline{\textbf{\text{\cite{2020_a_survey_of_android_malware_detection_with_deep_neural_models}}}}}_{A^*}$}}{\gwc{$\underline{\text{\cite{2020_a_large_scale_study_on_the_adoption_of_anti_debugging_and_anti_tampering_protections_in_android_apps}}}_{}$}}{\mwc{$\underline{\text{\cite{2020_a_large_scale_investigation_to_identify_the_pattern_of_permissions_in_obfuscated_android_malwares}}}_{}$}}{\mwc{$\underline{\text{\cite{2020_a_large_scale_investigation_to_identify_the_pattern_of_app_component_in_obfuscated_android_malwares}}}_{}$}}{\gwc{$\text{\cite{2019_understanding_the_behaviour_of_hackers_while_performing_attack_tasks_in_a_professional_setting_and_in_a_public_challenge}}_{A}$}}{\mwc{$\underline{\overline{\text{\cite{2019_seqdroid_obfuscated_android_malware_detection_using_stacked_convolutional_and_recurrent_neural_networks}}}}_{}$}}{\gwc{$\underline{\overline{\text{\cite{2019_robust_detection_of_obfuscated_strings_in_android_apps}}}}_{w}$}}{\mwc{$\underline{\overline{\text{\cite{2019_obfusifier_obfuscation_resistant_android_malware_detection_system}}}}_{}$}}{\gwc{$\underline{\text{\cite{2019_obfuscation_resilient_code_recognition_in_android_apps}}}_{B}$}}{\mwc{$\underline{\overline{\text{\cite{2019_lightweight_versus_obfuscation_resilient_malware_detection_in_android_applications}}}}_{}$}}{\gwc{$\overline{\text{\cite{2019_fine_grained_static_detection_of_obfuscation_transforms_using_ensemble_learning_and_semantic_reasoning}}}_{w}$}}{\gwc{$\text{\cite{2019_control_flow_graph_matching_for_detecting_obfuscated_programs}}_{}$}}{\mwc{$\underline{\text{\cite{2019_automatic_components_separation_of_obfuscated_android_applications_an_empirical_study_of_design_based_features}}}_{}$}}{\gwc{$\overline{\text{\cite{2019_asm2vec_boosting_static_representation_robustness_for_binary_clone_search_against_code_obfuscation_and_compiler_optimization}}}_{A^*}$}}{\gwc{$\underline{\overline{\text{\cite{2018_unmasking_android_obfuscation_tools_using_spatial_analysis}}}}_{C}$}}{\gwc{$\textbf{\text{\cite{2018_tutorial_an_overview_of_malware_detection_and_evasion_techniques}}}_{C}$}}{\mwc{$\overline{\text{\cite{2018_obfuscated_vba_macro_detection_using_machine_learning}}}_{A}$}}{\mwc{$\underline{\overline{\text{\cite{2018_lightweight_obfuscation_resilient_detection_and_family_identification_of_android_malware}}}}_{A^*}$}}{\mwc{$\overline{\text{\cite{2018_jast_fully_syntactic_detection_of_malicious_obfuscated_javascript}}}_{C}$}}{\mwc{$\underline{\overline{\text{\cite{2018_impact_of_code_obfuscation_on_android_malware_detection_based_on_static_and_dynamic_analysis}}}}_{C}$}}{\mwc{$\underline{\text{\cite{2018_evaluating_model_checking_for_cyber_threats_code_obfuscation_identification}}}_{A}$}}{\mwc{$\underline{\text{\cite{2018_effectiveness_of_android_obfuscation_on_evading_anti_malware}}}_{}$}}{\mwc{$\underline{\text{\cite{2018_dexmonitor_dynamically_analyzing_and_monitoring_obfuscated_android_applications}}}_{}$}}{\mwc{$\underline{\overline{\text{\cite{2018_detection_of_obfuscation_techniques_in_android_applications}}}}_{B}$}}{\mwc{$\overline{\text{\cite{2018_detection_of_android_malicious_obfuscation_applications_based_on_multi_class_features}}}_{}$}}{\gwc{$\text{\cite{2018_common_program_similarity_metric_method_for_anti_obfuscation}}_{}$}}{\gwc{$\underline{\overline{\text{\cite{2017_who_changed_you_obfuscator_identification_for_android}}}}_{}$}}{\mwc{$\text{\cite{2017_using_of_the_forensic_analyzing_tools_code_obfuscation}}_{}$}}{\gwc{$\text{\cite{2017_semantics_based_obfuscation_resilient_binary_code_similarity_comparison_with_applications_to_software_and_algorithm_plagiarism_detection}}_{A^*}$}}{\gwc{$\overline{\text{\cite{2017_security_assessment_of_code_obfuscation_based_on_dynamic_monitoring_in_android_things}}}_{}$}}{\mwc{$\underline{\text{\cite{2017_scalable_and_obfuscation_resilient_android_app_repackaging_detection_based_on_behavior_birthmark}}}_{B}$}}{\mwc{$\overline{\text{\cite{2017_obfuscated_malicious_javascript_detection_scheme_using_the_feature_based_on_divided_url}}}_{C}$}}{\mwc{$\text{\cite{2017_malware_secrets_de_obfuscating_in_the_cloud}}_{B}$}}{\mwc{$\underline{\text{\cite{2017_hydroid_a_hybrid_approach_for_generating_api_call_traces_from_obfuscated_android_applications_for_mobile_security}}}_{B}$}}{\gwc{$\text{\cite{2017_how_professional_hackers_understand_protected_code_while_performing_attack_tasks}}_{A}$}}{\mwc{$\underline{\text{\cite{2017_formal_methods_meet_mobile_code_obfuscation_identification_of_code_reordering_technique}}}_{C}$}}{\mwc{$\text{\cite{2017_dynodet_detecting_dynamic_obfuscation_in_malware}}_{C}$}}{\mwc{$\underline{\overline{\text{\cite{2017_droidsieve_fast_and_accurate_classification_of_obfuscated_android_malware}}}}_{}$}}{\mwc{$\text{\cite{2017_code_modification_and_obfuscation_detection_test_using_malicious_script_distributing_website_inspection_technology}}_{}$}}{\mwc{$\text{\cite{2017_binsim_trace_based_semantic_binary_diffing_via_system_call_sliced_segment_equivalence_checking}}_{A^*}$}}{\mwc{$\text{\cite{2017_a_framework_for_recognition_and_confronting_of_obfuscated_malwares_based_on_memory_dumping_and_filter_drivers}}_{C}$}}{\mwc{$\underline{\text{\cite{2016_towards_the_usage_of_invariant_based_app_behavioral_fingerprinting_for_the_detection_of_obfuscated_versions_of_known_malware}}}_{}$}}{\mwc{$\text{\cite{2016_obfuscation_code_localization_based_on_cfg_generation_of_malware}}_{}$}}{\gwc{$\overline{\text{\cite{2016_metadata_recovery_from_obfuscated_programs_using_machine_learning}}}_{w}$}}{\gwc{$\text{\cite{2016_identifying_the_applied_obfuscation_method_towards_de_obfuscation}}_{C}$}}{\gwc{$\text{\cite{2016_deviation_based_obfuscation_resilient_program_equivalence_checking_with_application_to_software_plagiarism_detection}}_{A}$}}{\mwc{$\overline{\text{\cite{2016_detection_of_obfuscation_in_java_malware}}}_{}$}}{\gwc{$\text{\cite{2016_detecting_similar_components_between_android_applications_with_obfuscation}}_{}$}}{\mwc{$\overline{\text{\cite{2016_detecting_obfuscated_suspicious_javascript_based_on_information_theoretic_measures_and_novelty_detection}}}_{}$}}{\mwc{$\overline{\text{\cite{2016_detecting_obfuscated_malware_using_reduced_opcode_set_and_optimised_runtime_trace}}}_{}$}}{\mwc{$\text{\cite{2016_binsec_se_a_dynamic_symbolic_execution_toolkit_for_binary_level_analysis}}_{A}$}}{\mwc{$\overline{\text{\cite{2016_an_efficient_method_for_detecting_obfuscated_suspicious_javascript_based_on_text_pattern_analysis}}}_{w}$}}{\mwc{$\underline{\text{\cite{2016_alterdroid_differential_fault_analysis_of_obfuscated_smartphone_malware}}}_{A^*}$}}{\mwc{$\text{\cite{2015_unveiling_metamorphism_by_abstract_interpretation_of_code_properties}}_{A}$}}{\mwc{$\text{\cite{2015_symbolic_execution_of_obfuscated_code}}_{A^*}$}}{\gwc{$\overline{\text{\cite{2015_jsobfusdetector_a_binary_pso_based_one_class_classifier_ensemble_to_detect_obfuscated_javascript_code}}}_{}$}}{\mwc{$\underline{\text{\cite{2015_enriching_reverse_engineering_through_visual_exploration_of_android_binaries}}}_{w}$}}{\mwc{$\text{\cite{2014_p_spade_gpu_accelerated_malware_packer_detection}}_{C}$}}{\mwc{$\text{\cite{2014_droidlegacy_automated_familial_classification_of_android_malware}}_{w}$}}{\mwc{$\text{\cite{2014_bit_level_taint_analysis}}_{C}$}}{\mwc{$\overline{\text{\cite{2014_back_to_static_analysis_for_kernel_level_rootkit_detection}}}_{A}$}}{\gwc{$\text{\cite{2014_a_framework_for_understanding_dynamic_anti_analysis_defenses}}_{w}$}}{\gwc{$\text{\cite{2013_towards_a_unified_software_attack_model_to_assess_software_protections}}_{A}$}}{\mwc{$\text{\cite{2013_a_static_packer_agnostic_filter_to_detect_similar_malware_samples}}_{C}$}}{\mwc{$\text{\cite{2013_a_similarity_metric_method_of_obfuscated_malware_using_function_call_graph}}_{}$}}{\gwc{$\text{\cite{2012_towards_static_analysis_of_virtualization_obfuscated_binaries}}_{}$}}{\mwc{$\text{\cite{2012_spade_signature_based_packer_detection}}_{}$}}{\gwc{$\text{\cite{2012_replacement_attacks_against_vm_protected_applications}}_{B}$}}{\mwc{$\overline{\textbf{\text{\cite{2012_a_survey_on_automated_dynamic_malware_analysis_techniques_and_tools}}}}_{A^*}$}}{\mwc{$\text{\cite{2011_the_power_of_procrastination_detection_and_mitigation_of_execution_stalling_malicious_code}}_{A^*}$}}{\mwc{$\text{\cite{2010_modelling_metamorphism_by_abstract_interpretation}}_{B}$}}{\mwc{$\text{\cite{2010_identifying_dormant_functionality_in_malware_programs}}_{A^*}$}}{\mwc{$\text{\cite{2010_context_sensitive_analysis_without_calling_context}}_{C}$}}{\mwc{$\text{\cite{2010_context_sensitive_analysis_of_obfuscated_x86_executables}}_{C}$}}{\mwc{$\text{\cite{2009_visualizing_compiled_executables_for_malware_analysis}}_{w}$}}{\mwc{$\text{\cite{2009_server_side_dynamic_code_analysis}}_{}$}}{\mwc{$\overline{\text{\cite{2009_malware_obfuscation_detection_via_maximal_patterns}}}_{}$}}{\mwc{$\text{\cite{2009_detection_of_metamorphic_and_virtualization_based_malware_using_algebraic_specification}}_{}$}}{\mwc{$\text{\cite{2009_a_heuristic_approach_for_detection_of_obfuscated_malware}}_{C}$}}{\mwc{$\text{\cite{2008_still_exploit_code_detection_via_static_taint_and_initialization_analyses}}_{A}$}}{\gwc{$\text{\cite{2007_matching_control_flow_of_program_versions}}_{A}$}}{\mwc{$\text{\cite{2007_detecting_obfuscated_viruses_using_cosine_similarity_analysis}}_{}$}}{\mwc{$\text{\cite{2007_an_automated_signature_based_approach_against_polymorphic_internet_worms}}_{A^*}$}}{\mwc{$\text{\cite{2006_using_engine_signature_to_detect_metamorphic_malware}}_{w}$}}{\mwc{$\text{\cite{2005_semantics_aware_malware_detection}}_{A^*}$}}{\gwc{$\text{\cite{2005_hybrid_static_dynamic_attacks_against_software_protection_mechanisms}}_{w}$}}{\mwc{$\text{\cite{2005_detecting_malicious_code_by_model_checking}}_{C}$}}{\gwc{$\text{\cite{2005_analyzing_memory_accesses_in_obfuscated_x86_executables}}_{C}$}}{\gwc{$\text{\cite{2004_static_disassembly_of_obfuscated_binaries}}_{A^*}$}}}}
\newcommand{\storeArrayPapersD}{\storedata{arrayPapersD}{{\gwc{$\text{\cite{2022_program_synthesis_based_simplification_of_mba_obfuscated_malware_with_restart_strategies}}_{w}$}}{\gwc{$\text{\cite{2022_overcoming_the_obfuscation_method_of_the_dynamic_name_resolution}}_{}$}}{\mwc{$\text{\cite{2022_invoke_deobfuscation_ast_based_and_semantics_preserving_deobfuscation_for_powershell_scripts}}_{A}$}}{\gwc{$\text{\cite{2022_efficient_deobfuscation_of_linear_mixed_boolean_arithmetic_expressions}}_{w}$}}{\gwc{$\underline{\text{\cite{2022_deoptfuscator_defeating_advanced_control_flow_obfuscation_using_android_runtime_art}}}_{}$}}{\gwc{$\text{\cite{2022_chosen_instruction_attack_against_commercial_code_virtualization_obfuscators}}_{A^*}$}}{\gwc{$\text{\cite{2022_cadecff_compiler_agnostic_deobfuscator_of_control_flow_flattening}}_{}$}}{\mwc{$\text{\cite{2021_the_de_obfuscation_method_in_the_static_detection_of_malicious_pdf_documents}}_{}$}}{\gwc{$\textbf{\text{\cite{2021_sok_automatic_deobfuscation_of_virtualization_protected_applications}}}_{B}$}}{\gwc{$\text{\cite{2021_rl_bin_overcoming_binary_instrumentation_challenges_in_the_presence_of_obfuscation_techniques_and_problematic_features}}_{}$}}{\mwc{$\text{\cite{2021_mba_blast_unveiling_and_simplifying_mixed_boolean_arithmetic_obfuscation}}_{A^*}$}}{\gwc{$\overline{\text{\cite{2021_input_output_example_guided_data_deobfuscation_on_binary}}}_{}$}}{\mwc{$\text{\cite{2021_building_deobfuscated_applications_from_polymorphic_binaries}}_{}$}}{\gwc{$\text{\cite{2021_a_comprehensive_solution_for_obfuscation_detection_and_removal_based_on_comparative_analysis_of_deobfuscation_tools}}_{}$}}{\gwc{$\text{\cite{2020_symsem_symbolic_execution_with_time_stamps_for_deobfuscation}}_{}$}}{\gwc{$\overline{\text{\cite{2020_similarity_features_for_the_evaluation_of_obfuscation_effectiveness}}}_{}$}}{\gwc{$\text{\cite{2020_qsynth_a_program_synthesis_based_approach_for_binary_code_deobfuscation}}_{w}$}}{\mwc{$\text{\cite{2020_optimizing_away_javascript_obfuscation}}_{C}$}}{\mwc{$\underline{\overline{\text{\cite{2020_hidden_in_plain_sight_obfuscated_strings_threatening_your_privacy}}}}_{A}$}}{\mwc{$\underline{\overline{\text{\cite{2020_an_empirical_study_of_code_deobfuscations_on_detecting_obfuscated_android_piggybacked_apps}}}}_{B}$}}{\gwc{$\text{\cite{2019_saturn_software_deobfuscation_framework_based_on_llvm}}_{w}$}}{\mwc{$\overline{\text{\cite{2019_effective_and_light_weight_deobfuscation_and_semantic_aware_attack_detection_for_powershell_scripts}}}_{A^*}$}}{\mwc{$\underline{\text{\cite{2019_deobfuscating_android_native_binary_code}}}_{A^*}$}}{\gwc{$\overline{\text{\cite{2019_defeating_opaque_predicates_statically_through_machine_learning_and_binary_analysis}}}_{w}$}}{\mwc{$\text{\cite{2019_cubismo_decloaking_server_side_malware_via_cubist_program_analysis}}_{A}$}}{\gwc{$\text{\cite{2019_analysis_of_obfuscated_code_with_program_slicing}}_{}$}}{\mwc{$\text{\cite{2018_vmhunt_a_verifiable_approach_to_partially_virtualized_binary_code_simplification}}_{A^*}$}}{\mwc{$\underline{\text{\cite{2018_tackling_runtime_based_obfuscation_in_android_with_tiro}}}_{A^*}$}}{\gwc{$\text{\cite{2018_symbolic_deobfuscation_from_virtualized_code_back_to_the_original}}_{C}$}}{\mwc{$\text{\cite{2018_psdem_a_feasible_de_obfuscation_method_for_malicious_powershell_detection}}_{B}$}}{\gwc{$\text{\cite{2018_identifying_input_dependent_jumps_from_obfuscated_execution_using_dynamic_data_flow_graphs}}_{w}$}}{\mwc{$\text{\cite{2018_dose_deobfuscation_based_on_semantic_equivalence}}_{w}$}}{\gwc{$\text{\cite{2018_deobfuscation_of_virtualization_obfuscated_code_through_symbolic_execution_and_compilation_optimization}}_{B}$}}{\mwc{$\overline{\text{\cite{2018_cluster_analysis_for_deobfuscation_of_malware_variants_during_ransomware_attacks}}}_{}$}}{\gwc{$\overline{\text{\cite{2018_bytewise_a_case_study_in_neural_network_obfuscation_identification}}}_{}$}}{\gwc{$\underline{\text{\cite{2018_building_the_de_obfuscation_platform_based_on_llvm}}}_{}$}}{\gwc{$\text{\cite{2017_vmattack_deobfuscating_virtualization_based_packed_binaries}}_{B}$}}{\gwc{$\overline{\text{\cite{2017_syntia_synthesizing_the_semantics_of_obfuscated_code}}}_{A^*}$}}{\mwc{$\text{\cite{2017_seead_a_semantic_based_approach_for_automatic_binary_code_de_obfuscation}}_{B}$}}{\gwc{$\overline{\text{\cite{2017_recovering_clear_natural_identifiers_from_obfuscated_js_names}}}_{A^*}$}}{\mwc{$\text{\cite{2017_partial_evaluation_of_string_obfuscations_for_java_malware_detection}}_{B}$}}{\gwc{$\text{\cite{2017_effectiveness_of_synthesis_in_concolic_deobfuscation}}_{B}$}}{\gwc{$\text{\cite{2017_combating_control_flow_linearization}}_{}$}}{\mwc{$\text{\cite{2017_backward_bounded_dse_targeting_infeasibility_questions_on_obfuscated_codes}}_{A^*}$}}{\gwc{$\text{\cite{2016_defeating_mba_based_obfuscation}}_{w}$}}{\gwc{$\text{\cite{2016_a_reverse_engineering_approach_of_obfuscated_array}}_{}$}}{\mwc{$\text{\cite{2015_loop_logic_oriented_opaque_predicate_detection_in_obfuscated_binary_code}}_{A^*}$}}{\mwc{$\text{\cite{2015_a_generic_approach_to_automatic_deobfuscation_of_executable_code}}_{A^*}$}}{\mwc{$\text{\cite{2014_malware_dynamic_recompilation}}_{}$}}{\gwc{$\text{\cite{2013_looking_inside_the_drop_box}}_{w}$}}{\mwc{$\text{\cite{2012_automatic_simplification_of_obfuscated_javascript_code_a_semantics_based_approach}}_{}$}}{\mwc{$\text{\cite{2011_deobfuscation_of_virtualization_obfuscated_software_a_semantics_based_approach}}_{A^*}$}}{\mwc{$\text{\cite{2010_reverse_engineering_self_modifying_code_unpacker_extraction}}_{}$}}{\mwc{$\text{\cite{2010_experiences_in_malware_binary_deobfuscation}}_{}$}}{\gwc{$\text{\cite{2009_unpacking_virtualization_obfuscators}}_{w}$}}{\mwc{$\text{\cite{2009_automatic_static_unpacking_of_malware_binaries}}_{}$}}{\mwc{$\text{\cite{2009_automatic_reverse_engineering_of_malware_emulators}}_{A^*}$}}{\mwc{$\text{\cite{2008_eureka_a_framework_for_enabling_static_malware_analysis}}_{A}$}}{\mwc{$\text{\cite{2007_renovo_a_hidden_code_extractor_for_packed_executables}}_{w}$}}{\mwc{$\text{\cite{2007_code_normalization_for_self_mutating_malware}}_{B}$}}{\gwc{$\text{\cite{2006_understanding_obfuscated_code}}_{A}$}}{\mwc{$\text{\cite{2006_polyunpack_automating_the_hidden_code_extraction_of_unpack_executing_malware}}_{A}$}}{\gwc{$\text{\cite{2006_opaque_predicates_detection_by_abstract_interpretation}}_{}$}}{\gwc{$\text{\cite{2006_on_evaluating_obfuscatory_strength_of_alias_based_transforms_using_static_analysis}}_{}$}}{\mwc{$\text{\cite{2006_normalizing_metamorphic_malware_using_term_rewriting}}_{w}$}}{\gwc{$\text{\cite{2005_deobfuscation_reverse_engineering_obfuscated_code}}_{}$}}{\mwc{$\text{\cite{2005_a_method_for_detecting_obfuscated_calls_in_malicious_binaries}}_{A^*}$}}{\gwc{$\text{\cite{2004_abstracting_stack_to_detect_obfuscated_calls_in_binaries}}_{w}$}}}}
\newcommand{\storeArrayPapersO}{\storedata{arrayPapersO}{{\mwc{$\text{\cite{2022_wobfuscator_obfuscating_javascript_malware_via_opportunistic_translation_to_webassembly}}_{A^*}$}}{\gwc{$\text{\cite{2022_source_code_obfuscation_novel_technique_and_implementation}}_{}$}}{\gwc{$\text{\cite{2022_property_driven_code_obfuscations_reinterpreting_jones_optimality_in_abstract_interpretation}}_{B}$}}{\gwc{$\text{\cite{2022_program_code_protecting_mechanism_based_on_obfuscation_tools}}_{}$}}{\gwc{$\text{\cite{2022_on_the_optimization_of_software_obfuscation_against_hardware_trojans_in_microprocessors}}_{}$}}{\gwc{$\text{\cite{2022_model_of_execution_trace_obfuscation_between_threads}}_{A}$}}{\gwc{$\ol{\text{\cite{2022_generating_effective_software_obfuscation_sequences_with_reinforcement_learning}}}_{A}$}}{\gwc{$\text{\cite{2022_flexible_software_protection}}_{B}$}}{\gwc{$\ul{\text{\cite{2022_fineobfuscator_defeating_reverse_engineering_attacks_with_context_sensitive_and_cost_efficient_obfuscation_for_android_apps}}}_{}$}}{\mwc{$\text{\cite{2022_evolution_of_macro_vba_obfuscation_techniques}}_{C}$}}{\gwc{$\text{\cite{2022_eric_an_efficient_and_practical_software_obfuscation_framework}}_{A}$}}{\gwc{$\text{\cite{2022_enhanced_obfuscation_for_software_protection_in_autonomous_vehicular_cloud_computing_platforms}}_{}$}}{\gwc{$\text{\cite{2022_desync_cc_an_automatic_disassembly_desynchronization_obfuscator}}_{A}$}}{\mwc{$\text{\cite{2022_automating_the_process_of_developing_obfuscated_variants_of_pe_through_adope_software}}_{}$}}{\mwc{$\text{\cite{2022_androobfs_time_tagged_obfuscated_android_malware_dataset_with_family_information}}_{A}$}}{\mwc{$\ul{\text{\cite{2022_android_malware_evasion_framework_for_auditing_anti_malware_resistance_against_various_obfuscation_technique_and_dynamic_code_loading}}}_{}$}}{\mwc{$\text{\cite{2022_a_systematic_approach_for_evading_antiviruses_using_malware_obfuscation}}_{}$}}{\mwc{$\ol{\text{\cite{2022_a_first_look_at_code_obfuscation_for_webassembly}}}_{}$}}{\gwc{$\text{\cite{2021_software_obfuscation_with_non_linear_mixed_boolean_arithmetic_expressions}}_{B}$}}{\mwc{$\text{\cite{2021_research_on_powershell_obfuscation_technology_based_on_abstract_syntax_tree_transformation}}_{}$}}{\gwc{$\text{\cite{2021_program_protection_through_software_based_hardware_abstraction}}_{}$}}{\gwc{$\text{\cite{2021_program_obfuscation_via_abi_debiasing}}_{}$}}{\gwc{$\text{\cite{2021_php_source_code_protection_using_layout_obfuscation_and_aes_256_encryption_algorithm}}_{w}$}}{\gwc{$\text{\cite{2021_obfus_an_obfuscation_tool_for_software_copyright_and_vulnerability_protection}}_{}$}}{\mwc{$\text{\cite{2021_novel_approach_for_concealing_penetration_testing_payloads_using_data_privacy_obfuscation_techniques}}_{}$}}{\gwc{$\ol{\text{\cite{2021_machine_learning_classification_of_obfuscation_using_image_visualization}}}_{}$}}{\gwc{$\text{\cite{2021_hiding_in_the_particles_when_return_oriented_programming_meets_program_obfuscation}}_{A}$}}{\gwc{$\text{\cite{2021_hexon_protecting_firmware_using_hardware_assisted_execution_level_obfuscation}}_{}$}}{\gwc{$\text{\cite{2021_experimental_evaluation_of_the_obfuscation_techniques_against_reverse_engineering}}_{}$}}{\gwc{$\ol{\text{\cite{2021_deepobfuscode_source_code_obfuscation_through_sequence_to_sequence_networks}}}_{}$}}{\mwc{$\ul{\text{\cite{2021_cybercrime_specialization_an_expos_of_a_malicious_android_obfuscation_as_a_service}}}_{w}$}}{\gwc{$\text{\cite{2021_code_obfuscation_based_on_inline_split_of_control_flow_graph}}_{}$}}{\mwc{$\tb{\text{\cite{2021_classification_and_update_proposal_for_modern_computer_worms_based_on_obfuscation}}}_{}$}}{\mwc{$\text{\cite{2021_binseal_linux_binary_obfuscation_against_symbolic_execution}}_{}$}}{\gwc{$\text{\cite{2021_an_efficient_control_flow_based_obfuscator_for_micropython_bytecode}}_{}$}}{\mwc{$\text{\cite{2021_an_assessment_of_obfuscated_ransomware_detection_and_prevention_methods}}_{}$}}{\mwc{$\ol{\text{\cite{2021_adversarialuscator_an_adversarial_drl_based_obfuscator_and_metamorphic_malware_swarm_generator}}}_{B}$}}{\gwc{$\ul{\text{\cite{2021_a_novel_technique_for_control_flow_obfuscation_in_jvm_applications_using_invokedynamic_with_native_bootstrapping}}}_{}$}}{\mwc{$\ol{\text{\cite{2021_a_malware_obfuscation_ai_technique_to_evade_antivirus_detection_in_counter_forensic_domain}}}_{}$}}{\gwc{$\text{\cite{2021_a_classic_multi_method_collaborative_obfuscation_strategy}}_{}$}}{\gwc{$\text{\cite{2020_vcf_virtual_code_folding_to_enhance_virtualization_obfuscation}}_{}$}}{\gwc{$\ul{\text{\cite{2020_ui_obfuscation_and_its_effects_on_automated_ui_analysis_for_android_apps}}}_{A^*}$}}{\gwc{$\text{\cite{2020_towards_formal_verification_of_program_obfuscation}}_{w}$}}{\gwc{$\text{\cite{2020_software_protection_using_dynamic_pufs}}_{A}$}}{\gwc{$\ul{\text{\cite{2020_semantic_redirection_obfuscation_a_control_flow_obfuscation_based_on_android_runtime}}}_{A}$}}{\gwc{$\text{\cite{2020_resilient_self_debugging_software_protection}}_{w}$}}{\gwc{$\text{\cite{2020_power_profiling_and_analysis_of_code_obfuscation_for_embedded_devices}}_{}$}}{\gwc{$\text{\cite{2020_on_the_automatic_analysis_of_the_practical_resistance_of_obfuscating_transformations}}_{C}$}}{\gwc{$\ol{\text{\cite{2020_obfuscation_code_technics_based_on_neural_networks_mechanism}}}_{}$}}{\gwc{$\text{\cite{2020_mechanisms_for_source_code_obfuscation_in_c_novel_techniques_and_implementation}}_{}$}}{\gwc{$\tb{\text{\cite{2020_layered_obfuscation_a_taxonomy_of_software_obfuscation_techniques_for_layered_security}}}_{}$}}{\gwc{$\text{\cite{2020_hybrid_obfuscation_technique_to_protect_source_code_from_prohibited_software_reverse_engineering}}_{}$}}{\gwc{$\text{\cite{2020_experimental_assessment_of_xor_masking_data_obfuscation_based_on_k_clique_opaque_constants}}_{A}$}}{\gwc{$\text{\cite{2020_effective_and_efficient_java_type_obfuscation}}_{B}$}}{\mwc{$\ol{\text{\cite{2020_doom_a_novel_adversarial_drl_based_op_code_level_metamorphic_malware_obfuscator_for_the_enhancement_of_ids}}}_{}$}}{\gwc{$\ul{\text{\cite{2020_dexfus_an_android_obfuscation_technique_based_on_dalvik_bytecode_translation}}}_{}$}}{\gwc{$\text{\cite{2020_creative_manual_code_obfuscation_as_a_countermeasure_against_software_reverse_engineering}}_{}$}}{\gwc{$\text{\cite{2020_code_renewability_for_native_software_protection}}_{A}$}}{\gwc{$\text{\cite{2020_code_obfuscation_technique_for_enhancing_software_protection_against_reverse_engineering}}_{}$}}{\mwc{$\ul{\ol{\text{\cite{2020_andropgan_an_opcode_gan_for_android_malware_obfuscations}}}}_{}$}}{\mwc{$\ul{\text{\cite{2020_abusing_android_runtime_for_application_obfuscation}}}_{w}$}}{\gwc{$\text{\cite{2020_a_security_model_and_implementation_of_embedded_software_based_on_code_obfuscation}}_{B}$}}{\gwc{$\text{\cite{2020_a_framework_for_evaluation_of_software_obfuscation_tools_for_embedded_devices}}_{}$}}{\gwc{$\text{\cite{2019_virtsc_combining_virtualization_obfuscation_with_self_checksumming}}_{w}$}}{\gwc{$\text{\cite{2019_the_impact_of_control_flow_obfuscation_technique_on_software_protection_against_human_attacks}}_{}$}}{\gwc{$\ul{\text{\cite{2019_resilient_user_side_android_application_repackaging_and_tampering_detection_using_cryptographically_obfuscated_logic_bombs}}}_{A}$}}{\gwc{$\text{\cite{2019_on_preventing_symbolic_execution_attacks_by_low_cost_obfuscation}}_{C}$}}{\mwc{$\ul{\text{\cite{2019_oblive_seamless_code_obfuscation_for_java_programs_and_android_apps}}}_{A}$}}{\gwc{$\text{\cite{2019_obfuscation_based_watermarking_for_mobile_service_application_copyright_protection_in_the_cloud}}_{}$}}{\gwc{$\text{\cite{2019_obfuscation_algorithms_based_on_congruence_equation_and_knapsack_problem}}_{}$}}{\gwc{$\tb{\text{\cite{2019_obfuscated_code_quality_measurement}}}_{}$}}{\mwc{$\ul{\text{\cite{2019_obfuscated_android_application_development}}}_{}$}}{\gwc{$\text{\cite{2019_low_cost_and_secure_firmware_obfuscation_method_for_protecting_electronic_systems_from_cloning}}_{}$}}{\gwc{$\text{\cite{2019_invalidating_analysis_knowledge_for_code_virtualization_protection_through_partition_diversity}}_{}$}}{\gwc{$\text{\cite{2019_how_to_kill_symbolic_deobfuscation_for_free_or_unleashing_the_potential_of_path_oriented_protections}}_{A}$}}{\mwc{$\ol{\text{\cite{2019_generation_evaluation_of_adversarial_examples_for_malware_obfuscation}}}_{C}$}}{\gwc{$\text{\cite{2019_formal_verification_of_a_program_obfuscation_based_on_mixed_boolean_arithmetic_expressions}}_{}$}}{\gwc{$\text{\cite{2019_epona_and_the_obfuscation_paradox_transparent_for_users_and_developers_a_pain_for_reversers}}_{w}$}}{\gwc{$\text{\cite{2019_enhanced_branch_obfuscation_based_on_exception_handling_and_encrypted_mapping_table}}_{}$}}{\gwc{$\text{\cite{2019_design_and_implementation_of_obfuscating_tool_for_software_code_protection}}_{}$}}{\gwc{$\text{\cite{2019_data_obfuscation_using_secret_sharing}}_{}$}}{\gwc{$\text{\cite{2019_control_flow_obfuscation_via_cps_transformation}}_{w}$}}{\gwc{$\text{\cite{2019_cfhider_control_flow_obfuscation_with_intel_sgx}}_{A^*}$}}{\gwc{$\ul{\text{\cite{2019_android_control_flow_obfuscation_based_on_dynamic_entry_points_modification}}}_{}$}}{\gwc{$\text{\cite{2019_an_adaptive_approach_to_recommending_obfuscation_rules_for_java_bytecode_obfuscators}}_{B}$}}{\gwc{$\text{\cite{2019_a_meta_model_for_software_protections_and_reverse_engineering_attacks}}_{A}$}}{\gwc{$\text{\cite{2019_a_generalized_obfuscation_method_to_protect_software_of_mobile_apps}}_{}$}}{\mwc{$\ul{\text{\cite{2019_a_deep_camouflage_evaluating_android_s_anti_malware_systems_robustness_against_hybridization_of_obfuscation_techniques_with_injection_attacks}}}_{}$}}{\gwc{$\text{\cite{2018_vodka_virtualization_obfuscation_using_dynamic_key_approach}}_{w}$}}{\gwc{$\text{\cite{2018_vmguards_a_novel_virtual_machine_based_code_protection_system_with_vm_security_as_the_first_class_design_concern}}_{}$}}{\gwc{$\text{\cite{2018_turing_obfuscation}}_{}$}}{\gwc{$\ul{\text{\cite{2018_the_performance_cost_of_software_obfuscation_for_android_applications}}}_{B}$}}{\gwc{$\text{\cite{2018_ropob_obfuscating_binary_code_via_return_oriented_programming}}_{}$}}{\gwc{$\ul{\text{\cite{2018_protecting_million_user_ios_apps_with_obfuscation_motivations_pitfalls_and_experience}}}_{A^*}$}}{\gwc{$\ul{\text{\cite{2018_progressive_control_flow_obfuscation_for_android_applications}}}_{C}$}}{\gwc{$\text{\cite{2018_programming_experience_might_not_help_in_comprehending_obfuscated_source_code_efficiently}}_{B}$}}{\gwc{$\ul{\text{\cite{2018_on_the_impact_of_code_obfuscation_to_software_energy_consumption}}}_{}$}}{\mwc{$\text{\cite{2018_obfuscation_procedure_based_on_the_insertion_of_the_dead_code_in_the_crypter_by_binary_search}}_{}$}}{\gwc{$\text{\cite{2018_obfuscating_java_programs_by_translating_selected_portions_of_bytecode_to_native_libraries}}_{w}$}}{\gwc{$\text{\cite{2018_manufacturing_resilient_bi_opaque_predicates_against_symbolic_execution}}_{A}$}}{\gwc{$\text{\cite{2018_lambda_obfuscation}}_{}$}}{\mwc{$\text{\cite{2018_implementation_of_obfuscation_technique_on_php_source_code}}_{}$}}{\gwc{$\text{\cite{2018_hybrid_obfuscation_to_protect_against_disclosure_attacks_on_embedded_microprocessors}}_{A^*}$}}{\gwc{$\text{\cite{2018_exploiting_code_diversity_to_enhance_code_virtualization_protection}}_{B}$}}{\gwc{$\text{\cite{2018_enhance_virtual_machine_based_code_obfuscation_security_through_dynamic_bytecode_scheduling}}_{B}$}}{\gwc{$\tb{\text{\cite{2018_diversification_and_obfuscation_techniques_for_software_security_a_systematic_literature_review}}}_{A}$}}{\gwc{$\text{\cite{2018_combining_obfuscation_and_optimizations_in_the_real_world}}_{}$}}{\gwc{$\text{\cite{2018_code_obfuscation_against_abstraction_refinement_attacks}}_{B}$}}{\gwc{$\text{\cite{2018_code_obfuscation_against_abstract_model_checking_attacks}}_{}$}}{\gwc{$\ul{\text{\cite{2018_code_obfuscating_a_kotlin_based_app_with_proguard}}}_{}$}}{\gwc{$\text{\cite{2018_characterizing_a_property_driven_obfuscation_strategy}}_{B}$}}{\gwc{$\text{\cite{2018_binary_obfuscation_based_reassemble}}_{}$}}{\mwc{$\text{\cite{2018_anti_emulation_trends_in_modern_packers_a_survey_on_the_evolution_of_anti_emulation_techniques_in_upa_packers}}_{}$}}{\mwc{$\text{\cite{2018_an_overview_of_obfuscation_techniques_used_by_malware_in_visual_basic_for_application_scripts}}_{}$}}{\gwc{$\text{\cite{2018_a_user_defined_code_reinforcement_technology_based_on_llvm_obfuscator}}_{}$}}{\gwc{$\ul{\text{\cite{2018_a_large_scale_investigation_of_obfuscation_use_in_google_play}}}_{A}$}}{\gwc{$\text{\cite{2018_a_comparison_of_online_javascript_obfuscators}}_{}$}}{\gwc{$\text{\cite{2017_towards_optimally_hiding_protected_assets_in_software_applications}}_{B}$}}{\gwc{$\ol{\text{\cite{2017_stochastic_optimization_of_program_obfuscation}}}_{A^*}$}}{\mwc{$\text{\cite{2017_stealth_loader_trace_free_program_loading_for_api_obfuscation}}_{}$}}{\gwc{$\ol{\text{\cite{2017_predicting_the_resilience_of_obfuscated_code_against_symbolic_execution_attacks_via_machine_learning}}}_{A^*}$}}{\gwc{$\ul{\text{\cite{2017_on_the_effectiveness_of_code_reuse_based_android_application_obfuscation}}}_{}$}}{\gwc{$\text{\cite{2017_obfuscation_maze_of_code}}_{}$}}{\mwc{$\ul{\text{\cite{2017_obfuscating_function_call_topography_to_test_structural_malware_detection_against_evasion_attacks}}}_{}$}}{\gwc{$\text{\cite{2017_obfuscating_branch_decisions_based_on_encrypted_data_using_misr_and_hash_digests}}_{}$}}{\gwc{$\text{\cite{2017_maximal_incompleteness_as_obfuscation_potency}}_{B}$}}{\gwc{$\text{\cite{2017_lightweight_dispatcher_constructions_for_control_flow_flattening}}_{w}$}}{\gwc{$\text{\cite{2017_improving_security_level_through_obfuscation_technique_for_source_code_protection_using_aes_algorithm}}_{C}$}}{\gwc{$\text{\cite{2017_hidden_path_dynamic_software_watermarking_based_on_control_flow_obfuscation}}_{C}$}}{\mwc{$\text{\cite{2017_evasion_attacks_against_statistical_code_obfuscation_detectors}}_{w}$}}{\gwc{$\text{\cite{2017_evaluating_optimal_phase_ordering_in_obfuscation_executives}}_{w}$}}{\gwc{$\ol{\text{\cite{2017_estimating_software_obfuscation_potency_with_artificial_neural_networks}}}_{w}$}}{\gwc{$\text{\cite{2017_encrypted_computing_speed_security_and_provable_obfuscation_against_insiders}}_{}$}}{\gwc{$\ul{\text{\cite{2017_control_flow_obfuscation_based_protection_method_for_android_applications}}}_{}$}}{\gwc{$\text{\cite{2017_coat_code_obfuscation_tool_to_evaluate_the_performance_of_code_plagiarism_detection_tools}}_{}$}}{\gwc{$\text{\cite{2017_cloud_security_via_virtualized_out_of_band_execution_and_obfuscation}}_{B}$}}{\gwc{$\text{\cite{2017_automatic_generation_of_opaque_constants_based_on_the_k_clique_problem_for_resilient_data_obfuscation}}_{A}$}}{\gwc{$\ul{\text{\cite{2017_an_anti_reverse_engineering_technique_using_native_code_and_obfuscator_llvm_for_android_applications}}}_{}$}}{\gwc{$\ol{\tb{\text{\cite{2017_a_tutorial_on_software_obfuscation}}}}_{}$}}{\gwc{$\text{\cite{2017_a_parameterized_flattening_control_flow_based_obfuscation_algorithm_with_opaque_predicate_for_reduplicate_obfuscation}}_{}$}}{\gwc{$\text{\cite{2017_a_code_obfuscation_technique_to_prevent_reverse_engineering}}_{}$}}{\gwc{$\text{\cite{2016_vot4cs_a_virtualization_obfuscation_tool_for_c}}_{w}$}}{\gwc{$\text{\cite{2016_unifying_the_method_descriptor_in_java_obfuscation}}_{}$}}{\gwc{$\text{\cite{2016_translingual_obfuscation}}_{}$}}{\gwc{$\ol{\text{\cite{2016_towards_better_program_obfuscation_optimization_via_language_models}}}_{w}$}}{\gwc{$\text{\cite{2016_towards_automatic_risk_analysis_and_mitigation_of_software_applications}}_{C}$}}{\gwc{$\text{\cite{2016_tightly_coupled_self_debugging_software_protection}}_{w}$}}{\gwc{$\text{\cite{2016_theoretical_foundation_for_code_obfuscation_security_a_kolmogorov_complexity_approach}}_{}$}}{\gwc{$\tb{\text{\cite{2016_the_research_and_discussion_on_effectiveness_evaluation_of_software_protection}}}_{}$}}{\mwc{$\text{\cite{2016_the_obfuscation_efficiency_measuring_schemes}}_{}$}}{\gwc{$\text{\cite{2016_server_based_code_obfuscation_scheme_for_apk_tamper_detection}}_{}$}}{\gwc{$\text{\cite{2016_search_based_clustering_for_protecting_software_with_diversified_updates}}_{B}$}}{\gwc{$\text{\cite{2016_random_table_and_hash_coding_based_binary_code_obfuscation_against_stack_trace_analysis}}_{C}$}}{\gwc{$\text{\cite{2016_probfuscation_an_obfuscation_approach_using_probabilistic_control_flows}}_{C}$}}{\mwc{$\text{\cite{2016_obfuscation_procedure_based_in_dead_code_insertion_into_crypter}}_{}$}}{\gwc{$\text{\cite{2016_obfuscating_software_puzzle_for_denial_of_service_attack_mitigation}}_{}$}}{\gwc{$\ul{\text{\cite{2016_n_version_obfuscation}}}_{w}$}}{\gwc{$\ol{\text{\cite{2016_integrated_software_fingerprinting_via_neural_network_based_control_flow_obfuscation}}}_{A}$}}{\gwc{$\text{\cite{2016_generalized_dynamic_opaque_predicates_a_new_control_flow_obfuscation_method}}_{B}$}}{\gwc{$\text{\cite{2016_formal_verification_of_control_flow_graph_flattening}}_{}$}}{\gwc{$\text{\cite{2016_exploiting_dynamic_scheduling_for_vm_based_code_obfuscation}}_{B}$}}{\gwc{$\text{\cite{2016_evaluating_obfuscation_security_a_quantitative_approach}}_{}$}}{\mwc{$\ul{\text{\cite{2016_evaluating_analysis_tools_for_android_apps_status_quo_and_robustness_against_obfuscation}}}_{}$}}{\gwc{$\ul{\text{\cite{2016_control_flow_obfuscation_for_android_applications}}}_{B}$}}{\gwc{$\text{\cite{2016_comparing_the_effectiveness_of_commercial_obfuscators_against_mate_attacks}}_{w}$}}{\gwc{$\text{\cite{2016_code_obfuscation_by_using_floating_points_and_conditional_statements}}_{}$}}{\mwc{$\text{\cite{2016_challenging_anti_virus_through_evolutionary_malware_obfuscation}}_{}$}}{\gwc{$\text{\cite{2016_binary_protection_using_dynamic_fine_grained_code_hiding_and_obfuscation}}_{C}$}}{\gwc{$\text{\cite{2016_binary_permutation_polynomial_inversion_and_application_to_obfuscation_techniques}}_{w}$}}{\gwc{$\text{\cite{2016_assessment_of_source_code_obfuscation_techniques}}_{C}$}}{\gwc{$\tb{\text{\cite{2016_a_study_review_on_code_obfuscation}}}_{}$}}{\gwc{$\text{\cite{2016_a_reference_architecture_for_software_protection}}_{A}$}}{\gwc{$\text{\cite{2016_a_new_compile_time_obfuscation_scheme_for_software_protection}}_{}$}}{\gwc{$\ul{\text{\cite{2016_a_control_flow_obfuscation_method_for_android_applications}}}_{}$}}{\gwc{$\text{\cite{2015_software_protection_with_code_mobility}}_{w}$}}{\gwc{$\tb{\text{\cite{2015_security_and_privacy_in_cloud_computing_via_obfuscation_and_diversification_a_survey}}}_{C}$}}{\gwc{$\text{\cite{2015_pinpointing_and_hiding_surprising_fragments_in_an_obfuscated_program}}_{w}$}}{\gwc{$\text{\cite{2015_ojit_a_novel_obfuscation_approach_using_standard_just_in_time_compiler_transformations}}_{w}$}}{\gwc{$\text{\cite{2015_obfuscator_llvm_software_protection_for_the_masses}}_{w}$}}{\gwc{$\text{\cite{2015_obfuscating_windows_dlls}}_{w}$}}{\gwc{$\text{\cite{2015_matryoshka_strengthening_software_protection_via_nested_virtual_machines}}_{w}$}}{\gwc{$\text{\cite{2015_impact_of_software_obfuscation_on_susceptibility_to_return_oriented_programming_attacks}}_{}$}}{\gwc{$\text{\cite{2015_from_obfuscation_to_comprehension}}_{A}$}}{\gwc{$\text{\cite{2015_diversified_remote_code_execution_using_dynamic_obfuscation_of_conditional_branches}}_{w}$}}{\gwc{$\text{\cite{2015_different_obfuscation_techniques_for_code_protection}}_{}$}}{\gwc{$\text{\cite{2015_data_tainting_and_obfuscation_improving_plausibility_of_incorrect_taint}}_{C}$}}{\gwc{$\text{\cite{2015_code_artificiality_a_metric_for_the_code_stealth_based_on_an_n_gram_model}}_{w}$}}{\gwc{$\text{\cite{2015_assessment_of_data_obfuscation_with_residue_number_coding}}_{w}$}}{\gwc{$\text{\cite{2015_a_large_study_on_the_effect_of_code_obfuscation_on_the_quality_of_java_code}}_{A}$}}{\mwc{$\ol{\text{\cite{2015_a_framework_for_empirical_evaluation_of_malware_detection_resilience_against_behavior_obfuscation}}}_{}$}}{\gwc{$\text{\cite{2014_pushing_java_type_obfuscation_to_the_limit}}_{A}$}}{\gwc{$\text{\cite{2014_obfuscation_for_object_oriented_programs_dismantling_instance_methods}}_{B}$}}{\gwc{$\text{\cite{2014_obfuscation_by_code_fragmentation_to_evade_reverse_engineering}}_{B}$}}{\gwc{$\text{\cite{2014_mixed_obfuscation_of_overlapping_instruction_and_self_modify_code_based_on_hyper_chaotic_opaque_predicates}}_{}$}}{\gwc{$\text{\cite{2014_measuring_the_robustness_of_source_program_obfuscation_studying_the_impact_of_compiler_optimizations_on_the_obfuscation_of_c_programs}}_{}$}}{\gwc{$\text{\cite{2014_function_level_control_flow_obfuscation_for_software_security}}_{C}$}}{\gwc{$\text{\cite{2014_covert_computation_hiding_code_in_code_through_compile_time_obfuscation}}_{B}$}}{\gwc{$\text{\cite{2014_cognitive_evaluation_of_intermediate_level_obfuscator}}_{}$}}{\gwc{$\text{\cite{2014_arbitrary_control_flow_embedding_into_multiple_threads_for_obfuscation_a_preliminary_complexity_and_performance_analysis}}_{w}$}}{\gwc{$\text{\cite{2014_an_other_exercise_in_measuring_the_strength_of_source_code_obfuscation}}_{w}$}}{\gwc{$\text{\cite{2014_aes_sec_improving_software_obfuscation_through_hardware_assistance}}_{B}$}}{\gwc{$\text{\cite{2014_a_proxy_like_obfuscator_for_web_application_protection}}_{}$}}{\gwc{$\text{\cite{2014_a_new_code_obfuscation_scheme_for_software_protection}}_{}$}}{\gwc{$\text{\cite{2014_a_family_of_experiments_to_assess_the_effectiveness_and_efficiency_of_source_code_obfuscation_techniques}}_{A}$}}{\gwc{$\text{\cite{2014_a_data_obfuscation_based_on_state_transition_graph_of_mealy_automata}}_{}$}}{\gwc{$\text{\cite{2014_a_code_obfuscation_framework_using_code_clones}}_{A}$}}{\gwc{$\text{\cite{2013_software_protection_with_obfuscation_and_encryption}}_{B}$}}{\gwc{$\text{\cite{2013_software_protection_for_dynamically_generated_code}}_{w}$}}{\gwc{$\text{\cite{2013_shadobf_a_c_source_obfuscator_based_on_multi_objective_optimisation_algorithms}}_{w}$}}{\gwc{$\text{\cite{2013_protecting_your_software_updates}}_{B}$}}{\gwc{$\text{\cite{2013_potent_and_stealthy_control_flow_obfuscation_by_stack_based_self_modifying_code}}_{A}$}}{\gwc{$\text{\cite{2013_nislvmp_improved_virtual_machine_based_software_protection}}_{}$}}{\gwc{$\text{\cite{2013_jshadobf_a_javascript_obfuscator_based_on_multi_objective_optimization_algorithms}}_{B}$}}{\gwc{$\text{\cite{2013_feedback_driven_binary_code_diversification}}_{B}$}}{\gwc{$\ul{\text{\cite{2013_efficient_code_obfuscation_for_android}}}_{}$}}{\mwc{$\ul{\text{\cite{2013_droidchameleon_evaluating_android_anti_malware_against_transformation_attacks}}}_{A^*}$}}{\gwc{$\text{\cite{2013_aspects_of_intermediate_level_obfuscation}}_{}$}}{\gwc{$\text{\cite{2013_a_novel_obfuscation_class_hierarchy_flattening}}_{}$}}{\gwc{$\text{\cite{2013_a_formal_framework_for_property_driven_obfuscation_strategies}}_{B}$}}{\gwc{$\text{\cite{2012_twisting_additivity_in_program_obfuscation}}_{}$}}{\gwc{$\text{\cite{2012_tinyobf_code_obfuscation_framework_for_wireless_sensor_networks}}_{}$}}{\gwc{$\text{\cite{2012_technique_of_source_code_obfuscation_based_on_data_flow_and_control_flow_tansformations}}_{}$}}{\gwc{$\text{\cite{2012_software_security_by_obscurity_a_programming_language_perspective}}_{}$}}{\gwc{$\text{\cite{2012_research_on_java_bytecode_parse_and_obfuscate_tool}}_{}$}}{\gwc{$\text{\cite{2012_program_incomprehensibility_evaluation_for_obfuscation_methods_with_queue_based_mental_simulation_model}}_{C}$}}{\gwc{$\text{\cite{2012_obfuscation_by_partial_evaluation_of_distorted_interpreters}}_{C}$}}{\gwc{$\text{\cite{2012_making_abstract_interpretation_incomplete_modeling_the_potency_of_obfuscation}}_{B}$}}{\gwc{$\text{\cite{2012_instruction_embedding_for_improved_obfuscation}}_{}$}}{\gwc{$\text{\cite{2012_code_defactoring_evaluating_the_effectiveness_of_java_obfuscations}}_{}$}}{\gwc{$\tb{\text{\cite{2012_cloud_protection_by_obfuscation_techniques_and_metrics}}}_{}$}}{\mwc{$\text{\cite{2012_c_source_code_obfuscator}}_{}$}}{\gwc{$\text{\cite{2012_branch_obfuscation_using_code_mobility_and_signal}}_{w}$}}{\gwc{$\text{\cite{2012_aucsmith_like_obfuscation_of_java_bytecode}}_{C}$}}{\gwc{$\text{\cite{2012_a_method_and_implementation_of_control_flow_obfuscation_using_seh}}_{}$}}{\gwc{$\text{\cite{2012_a_framework_for_quantitative_evaluation_of_parallel_control_flow_obfuscation}}_{B}$}}{\gwc{$\text{\cite{2011_suggesting_potency_measures_for_obfuscated_arrays_and_usage_of_source_code_obfuscators_for_intellectual_property_protection_of_java_products}}_{}$}}{\gwc{$\text{\cite{2011_software_code_obfuscation_by_hiding_control_flow_information_in_stack}}_{w}$}}{\gwc{$\text{\cite{2011_program_obfuscation_with_leaky_hardware}}_{A}$}}{\gwc{$\text{\cite{2011_multi_stage_binary_code_obfuscation_using_improved_virtual_machine}}_{B}$}}{\gwc{$\text{\cite{2011_exploiting_code_mobility_for_dynamic_binary_obfuscation}}_{}$}}{\gwc{$\text{\cite{2011_embedded_software_security_through_key_based_control_flow_obfuscation}}_{}$}}{\gwc{$\text{\cite{2011_codebender_remote_software_protection_using_orthogonal_replacement}}_{B}$}}{\gwc{$\text{\cite{2011_code_obfuscation_against_static_and_dynamic_reverse_engineering}}_{w}$}}{\gwc{$\tb{\text{\cite{2011_a_taxonomy_of_self_modifying_code_for_obfuscation}}}_{B}$}}{\gwc{$\ul{\tb{\text{\cite{2010_theory_and_practice_of_program_obfuscation}}}}_{}$}}{\gwc{$\text{\cite{2010_the_code_obfuscation_technology_based_on_class_combination}}_{}$}}{\gwc{$\text{\cite{2010_stealthy_code_obfuscation_technique_for_software_security}}_{}$}}{\gwc{$\text{\cite{2010_program_obfuscation_by_strong_cryptography}}_{B}$}}{\gwc{$\text{\cite{2010_obfuscation_techniques_for_mobile_agent_code_confidentiality}}_{}$}}{\gwc{$\text{\cite{2010_obfuscation_methods_with_controlled_calculation_amounts_and_table_function}}_{}$}}{\gwc{$\text{\cite{2010_mimimorphism_a_new_approach_to_binary_code_obfuscation}}_{A^*}$}}{\mwc{$\tb{\text{\cite{2010_malware_obfuscation_techniques_a_brief_survey}}}_{}$}}{\gwc{$\text{\cite{2010_assure_high_quality_code_using_refactoring_and_obfuscation_techniques}}_{}$}}{\gwc{$\text{\cite{2010_a_secure_and_robust_approach_to_software_tamper_resistance}}_{w}$}}{\gwc{$\text{\cite{2010_a_general_model_for_hiding_control_flow}}_{w}$}}{\gwc{$\text{\cite{2009_trading_off_security_and_performance_in_barrier_slicing_for_remote_software_entrusting}}_{B}$}}{\gwc{$\text{\cite{2009_the_effectiveness_of_source_code_obfuscation_an_experimental_assessment}}_{A}$}}{\gwc{$\text{\cite{2009_runtime_protection_via_dataflow_flattening}}_{}$}}{\gwc{$\text{\cite{2009_research_on_java_software_protection_with_the_obfuscation_in_identifier_renaming}}_{}$}}{\gwc{$\text{\cite{2009_remote_software_protection_by_orthogonal_client_replacement}}_{B}$}}{\gwc{$\ul{\text{\cite{2009_obfuscation_mechanism_in_conjunction_with_tamper_proof_module}}}_{}$}}{\gwc{$\text{\cite{2009_obfuscating_c_programs_via_control_flow_flattening}}_{}$}}{\gwc{$\text{\cite{2009_methods_and_software_for_the_program_obfuscation}}_{}$}}{\gwc{$\text{\cite{2009_jdatatrans_for_array_obfuscation_in_java_source_codes_to_defeat_reverse_engineering_from_decompiled_codes}}_{}$}}{\gwc{$\text{\cite{2009_instruction_set_limitation_in_support_of_software_diversity}}_{}$}}{\mwc{$\text{\cite{2009_english_shellcode}}_{A^*}$}}{\gwc{$\text{\cite{2009_a_new_obfuscation_scheme_in_constructing_fuzzy_predicates}}_{}$}}{\gwc{$\text{\cite{2009_a_graph_approach_to_quantitative_analysis_of_control_flow_obfuscating_transformations}}_{A}$}}{\gwc{$\text{\cite{2008_using_exception_handling_to_build_opaque_predicates_in_intermediate_code_obfuscation_techniques}}_{B}$}}{\gwc{$\text{\cite{2008_towards_experimental_evaluation_of_code_obfuscation_techniques}}_{w}$}}{\gwc{$\text{\cite{2008_tamper_resistant_software_through_intent_protection}}_{}$}}{\gwc{$\text{\cite{2008_introducing_dynamic_name_resolution_mechanism_for_obfuscating_system_defined_names_in_programs}}_{}$}}{\gwc{$\text{\cite{2008_implementation_of_an_obfuscation_tool_for_c_c_source_code_protection_on_the_xscale_architecture}}_{w}$}}{\mwc{$\text{\cite{2008_impeding_malware_analysis_using_conditional_code_obfuscation}}_{A^*}$}}{\gwc{$\text{\cite{2008_hiding_information_in_completeness_holes_new_perspectives_in_code_obfuscation_and_watermarking}}_{B}$}}{\gwc{$\text{\cite{2008_an_inter_classes_obfuscation_method_for_java_program}}_{}$}}{\gwc{$\text{\cite{2008_a_compiler_based_infrastructure_for_software_protection}}_{w}$}}{\gwc{$\text{\cite{2007_specifying_imperative_data_obfuscations}}_{B}$}}{\gwc{$\text{\cite{2007_slicing_obfuscations_design_correctness_and_evaluation}}_{w}$}}{\gwc{$\text{\cite{2007_slicing_aided_design_of_obfuscating_transforms}}_{C}$}}{\gwc{$\text{\cite{2007_security_strength_measurement_for_dongle_protected_software}}_{B}$}}{\gwc{$\text{\cite{2007_obfuscating_java_the_most_pain_for_the_least_gain}}_{B}$}}{\gwc{$\text{\cite{2007_metrics_for_measuring_the_effectiveness_of_decompilers_and_obfuscators}}_{A}$}}{\gwc{$\text{\cite{2007_metrics_based_evaluation_of_slicing_obfuscations}}_{C}$}}{\mwc{$\text{\cite{2007_limits_of_static_analysis_for_malware_detection}}_{A}$}}{\gwc{$\text{\cite{2007_information_hiding_in_software_with_mixed_boolean_arithmetic_transforms}}_{}$}}{\gwc{$\text{\cite{2007_generalising_the_array_split_obfuscation}}_{A}$}}{\gwc{$\text{\cite{2007_binary_obfuscation_using_signals}}_{A^*}$}}{\gwc{$\text{\cite{2007_barrier_slicing_for_remote_software_trusting}}_{w}$}}{\gwc{$\text{\cite{2007_array_data_transformation_for_source_code_obfuscation}}_{}$}}{\gwc{$\text{\cite{2007_applications_for_provably_secure_intent_protection_with_bounded_input_size_programs}}_{B}$}}{\gwc{$\text{\cite{2006_three_control_flow_obfuscation_methods_for_java_software}}_{}$}}{\gwc{$\text{\cite{2006_software_protection_through_dynamic_code_mutation}}_{w}$}}{\gwc{$\text{\cite{2006_software_obfuscation_from_crackers_viewpoint}}_{}$}}{\gwc{$\text{\cite{2006_self_encrypting_code_to_protect_against_analysis_and_tampering}}_{w}$}}{\gwc{$\text{\cite{2006_proteus_virtualization_for_diversified_tamper_resistance}}_{w}$}}{\gwc{$\text{\cite{2006_preventing_reverse_engineering_threat_in_java_using_byte_code_obfuscation_techniques}}_{}$}}{\gwc{$\text{\cite{2006_on_the_effectiveness_of_source_code_transformations_for_binary_obfuscation}}_{}$}}{\gwc{$\text{\cite{2006_obfuscate_arrays_by_homomorphic_functions}}_{}$}}{\gwc{$\text{\cite{2006_manufacturing_opaque_predicates_in_distributed_systems_for_code_obfuscation}}_{}$}}{\gwc{$\text{\cite{2006_an_obfuscation_scheme_using_affine_transformation_and_its_implementation}}_{}$}}{\gwc{$\text{\cite{2006_an_obfuscation_for_binary_trees}}_{C}$}}{\gwc{$\tb{\text{\cite{2006_a_survey_of_control_flow_obfuscations}}}_{}$}}{\gwc{$\text{\cite{2006_a_qualitative_analysis_of_java_obfuscation}}_{}$}}{\gwc{$\text{\cite{2005_securing_mobile_agents_control_flow_using_opaque_predicates}}_{B}$}}{\gwc{$\text{\cite{2005_novel_obfuscation_algorithms_for_software_security}}_{}$}}{\gwc{$\text{\cite{2005_java_obfuscation_approaches_to_construct_tamper_resistant_object_oriented_programs}}_{}$}}{\gwc{$\text{\cite{2005_control_flow_based_obfuscation}}_{w}$}}{\gwc{$\text{\cite{2005_a_comparative_study_of_java_obfuscators}}_{}$}}{\gwc{$\text{\cite{2004_the_obfuscation_executive}}_{B}$}}{\gwc{$\text{\cite{2004_jhide_a_tool_kit_for_code_obfuscation}}_{}$}}{\gwc{$\text{\cite{2004_hardware_assisted_control_flow_obfuscation_for_embedded_processors}}_{B}$}}{\gwc{$\text{\cite{2004_advanced_obfuscation_techniques_for_java_bytecode}}_{A}$}}{\gwc{$\text{\cite{2004_a_framework_for_obfuscated_interpretation}}_{w}$}}{\gwc{$\text{\cite{2003_software_obfuscation_on_a_theoretical_basis_and_its_implementation}}_{}$}}{\gwc{$\text{\cite{2003_sandmark_a_tool_for_software_protection_research}}_{B}$}}{\gwc{$\text{\cite{2003_obfuscation_of_executable_code_to_improve_resistance_to_static_disassembly}}_{A^*}$}}{\gwc{$\text{\cite{2003_obfuscation_of_design_intent_in_object_oriented_applications}}_{w}$}}{\gwc{$\text{\cite{2003_java_obfuscation_with_a_theoretical_basis_for_building_secure_mobile_agents}}_{}$}}{\gwc{$\text{\cite{2003_exploiting_self_modification_mechanism_for_program_protection}}_{}$}}{\gwc{$\tb{\text{\cite{2002_watermarking_tamper_proofing_and_obfuscation_tools_for_software_protection}}}_{A^*}$}}{\gwc{$\text{\cite{2001_protection_of_software_based_survivability_mechanisms}}_{A}$}}{\gwc{$\text{\cite{2001_an_approach_to_the_obfuscation_of_control_flow_of_sequential_computer_programs}}_{B}$}}{\gwc{$\text{\cite{2000_experience_with_software_watermarking}}_{A}$}}{\gwc{$\text{\cite{1998_manufacturing_cheap_resilient_and_stealthy_opaque_constructs}}_{A^*}$}}{\gwc{$\text{\cite{1998_breaking_abstractions_and_unstructuring_data_structures}}_{}$}}{\gwc{$\text{\cite{1993_operating_system_protection_through_program_evolution}}_{B}$}}}}
\newcommand{\storeArrayPapersAD}{\storedata{arrayPapersAD}{{\gwc{$\underline{\text{\cite{2022_apkdiff_matching_android_app_versions_based_on_class_structure}}}_{w}$}}{\mwc{$\text{\cite{2021_obfuscation_resilient_executable_payload_extraction_from_packed_malware}}_{A^*}$}}{\mwc{$\underline{\overline{\text{\cite{2021_impact_of_code_deobfuscation_and_feature_interaction_in_android_malware_detection}}}}_{}$}}{\gwc{$\text{\cite{2020_binrec_dynamic_binary_lifting_and_recompilation}}_{}$}}{\gwc{$\underline{\text{\cite{2019_libid_reliable_identification_of_obfuscated_third_party_android_libraries}}}_{A}$}}{\gwc{$\underline{\text{\cite{2018_orlis_obfuscation_resilient_library_detection_for_android}}}_{}$}}{\gwc{$\underline{\overline{\text{\cite{2018_obfuscation_resilient_search_through_executable_classification}}}}_{w}$}}{\gwc{$\underline{\text{\cite{2017_ordol_obfuscation_resilient_detection_of_libraries_in_android_applications}}}_{B}$}}{\mwc{$\text{\cite{2017_cryptographic_function_detection_in_obfuscated_binaries_via_bit_precise_symbolic_loop_mapping}}_{A^*}$}}{\mwc{$\underline{\text{\cite{2017_codematch_obfuscation_won_t_conceal_your_repackaged_app}}}_{A^*}$}}{\gwc{$\underline{\text{\cite{2017_anti_proguard_towards_automated_deobfuscation_of_android_apps}}}_{w}$}}{\mwc{$\text{\cite{2017_analysis_of_exception_based_control_transfers}}_{}$}}{\mwc{$\underline{\overline{\text{\cite{2016_statistical_deobfuscation_of_android_applications}}}}_{A^*}$}}{\gwc{$\text{\cite{2016_reliable_third_party_library_detection_in_android_and_its_security_applications}}_{A^*}$}}{\mwc{$\text{\cite{2016_detecting_derivative_malware_samples_using_deobfuscation_assisted_similarity_analysis}}_{}$}}{\mwc{$\text{\cite{2013_obfuscation_resilient_binary_code_reuse_through_trace_oriented_programming}}_{A^*}$}}{\mwc{$\text{\cite{2012_aligot_cryptographic_function_identification_in_obfuscated_binary_programs}}_{A^*}$}}{\gwc{$\text{\cite{2011_equational_reasoning_on_x86_assembly_code}}_{B}$}}{\mwc{$\text{\cite{2011_automated_identification_of_cryptographic_primitives_in_binary_programs}}_{w}$}}{\gwc{$\text{\cite{2010_automatic_binary_deobfuscation}}_{}$}}{\gwc{$\text{\cite{2009_semi_automatic_binary_protection_tampering}}_{}$}}{\mwc{$\text{\cite{2007_software_transformations_to_improve_malware_detection}}_{}$}}}}
\newcommand{\storeArrayPapersAO}{\storedata{arrayPapersAO}{{\gwc{$\text{\cite{2022_argon_a_toolbase_for_evaluating_software_protection_techniques_against_symbolic_execution_attacks}}_{}$}}{\mwc{$\overline{\text{\cite{2021_statically_detecting_javascript_obfuscation_and_minification_techniques_in_the_wild}}}_{A}$}}{\gwc{$\text{\cite{2021_obfuscated_integration_of_software_protections}}_{C}$}}{\gwc{$\text{\cite{2021_dynamic_taint_analysis_versus_obfuscated_self_checking}}_{}$}}{\gwc{$\overline{\text{\cite{2020_hiding_in_plain_site_detecting_javascript_obfuscation_through_concealed_browser_api_usage}}}_{A}$}}{\gwc{$\text{\cite{2020_empirical_assessment_of_the_effort_needed_to_attack_programs_protected_with_client_server_code_splitting}}_{A}$}}{\gwc{$\overline{\text{\cite{2019_anything_to_hide_studying_minified_and_obfuscated_code_in_the_web}}}_{A^*}$}}{\mwc{$\underline{\overline{\text{\cite{2018_understanding_android_obfuscation_techniques_a_large_scale_investigation_in_the_wild}}}}_{}$}}{\gwc{$\underline{\text{\cite{2018_software_protection_on_the_go_a_large_scale_empirical_study_on_mobile_app_obfuscation}}}_{A^*}$}}{\gwc{$\text{\cite{2018_probabilistic_obfuscation_through_covert_channels}}_{}$}}{\mwc{$\underline{\text{\cite{2018_a_large_scale_empirical_study_on_the_effects_of_code_obfuscations_on_android_apps_and_anti_malware_products}}}_{A^*}$}}{\mwc{$\underline{\text{\cite{2016_testing_android_malware_detectors_against_code_obfuscation_a_systematization_of_knowledge_and_unified_methodology}}}_{}$}}{\gwc{$\textbf{\text{\cite{2016_protecting_software_through_obfuscation_can_it_keep_pace_with_progress_in_code_analysis}}}_{A^*}$}}{\gwc{$\text{\cite{2016_code_obfuscation_against_symbolic_execution_attacks}}_{A}$}}{\mwc{$\underline{\text{\cite{2015_obfuscation_techniques_against_signature_based_detection_a_case_study}}}_{w}$}}{\gwc{$\text{\cite{2015_a_framework_for_measuring_software_obfuscation_resilience_against_automated_attacks}}_{w}$}}{\gwc{$\textbf{\text{\cite{2012_reverse_code_engineering_state_of_the_art_and_countermeasures}}}_{}$}}{\mwc{$\text{\cite{2010_on_the_infeasibility_of_modeling_polymorphic_shellcode}}_{A}$}}{\mwc{$\text{\cite{2008_code_obfuscation_techniques_for_metamorphic_viruses}}_{}$}}{\gwc{$\text{\cite{2005_semantic_based_code_obfuscation_by_abstract_interpretation}}_{A}$}}{\gwc{$\text{\cite{2005_control_code_obfuscation_by_abstract_interpretation}}_{B}$}}}}
\newcommand{\storeArrayPapersDO}{\storedata{arrayPapersDO}{{\gwc{$\text{\cite{2022_loki_hardening_code_obfuscation_against_automated_attacks}}_{A^*}$}}{\gwc{$\text{\cite{2021_search_based_local_black_box_deobfuscation_understand_improve_and_mitigate}}_{A^*}$}}{\mwc{$\text{\cite{2019_powerdrive_accurate_de_obfuscation_and_analysis_of_powershell_malware}}_{C}$}}{\gwc{$\textbf{\text{\cite{2019_obfuscation_where_are_we_in_anti_dse_protections_a_first_attempt}}}_{w}$}}{\gwc{$\text{\cite{2019_dynopvm_vm_based_software_obfuscation_with_dynamic_opcode_mapping}}_{B}$}}{\mwc{$\text{\cite{2017_jsdes_an_automated_de_obfuscation_system_for_malicious_javascript}}_{B}$}}{\mwc{$\textbf{\text{\cite{2013_binary_code_obfuscations_in_prevalent_packer_tools}}}_{A^*}$}}{\gwc{$\text{\cite{2012_complexity_of_a_special_deobfuscation_problem}}_{}$}}{\gwc{$\text{\cite{2007_program_obfuscation_a_quantitative_approach}}_{w}$}}{\gwc{$\text{\cite{2007_on_the_possibility_of_practically_obfuscating_programs_towards_a_unified_perspective_of_code_protection}}_{}$}}}}
\newcommand{\storeArrayPapersADO}{\storedata{arrayPapersADO}{{\gwc{$\underline{\textbf{\text{\cite{2022_a_survey_of_obfuscation_and_deobfuscation_techniques_in_android_code_protection}}}}_{}$}}{\gwc{$\text{\cite{2006_loco_an_interactive_code_de_obfuscation_tool}}_{C}$}}}}
\newcommand{\CTtotalDENC}{138}

\newcommand{\CTtotalCDIV}{112}

\newcommand{\CTtotalOP}{116}

\newcommand{\CTtotalIREN}{126}
\newcommand{\CTtotalJCI}{117}

\newcommand{\CTtotalENC}{105}

\newcommand{\totalCFANAonlyX}{96}

\newcommand{\totalMLANAonlyX}{66}

\newcommand{\totalPATMonlyX}{96}

\newcommand{\totalSTATonlyX}{72}

\newcommand{\totalTRACEonlyX}{79}

\newcommand{\obfMLANAonlyX}{15}

\newcommand{\obfPATMonlyX}{33}
\newcommand{\obfPATMonlyXonlyGW}{9}

\newcommand{\obfSTATonlyX}{16}

\newcommand{\anaDeobfMLANAonlyX}{55}

\newcommand{\anaDeobfPATMonlyX}{72}
\newcommand{\anaDeobfPATMonlyXonlyGW}{22}

\newcommand{\anaDeobfSTATonlyX}{60}

\newcommand{\averageStrengthMeasurementsGWobfX}{1.0}

\newcommand{\zeroStrengthMeasurementsAllX}{125}
\newcommand{\zeroStrengthMeasurementsGWobfX}{92}
\newcommand{\zeroStrengthMeasurementsGWanadeobfX}{13}
\newcommand{\zeroStrengthMeasurementsMWobfX}{9}
\newcommand{\zeroStrengthMeasurementsMWanadeobfX}{16}

\newcommand{\averageStrengthMeasurementsGWobfXAAstar}{1.4}

\newcommand{\papersProtectionsX}{495}
\newcommand{\papersNonMalwareProtectionsXObfX}{248}
\newcommand{\papersNonMalwareProtectionsXAnadeobfX}{87}
\newcommand{\papersMalwareProtectionsXObfX}{47}
\newcommand{\papersMalwareProtectionsXAnadeobfX}{137}
\newcommand{\haveLessThanFourProtectionsXGoodware}{278}
\newcommand{\percentHaveLessThanFourProtectionsXGoodware}{72}
\newcommand{\haveLessThanFourProtectionsXMalware}{112}
\newcommand{\percentHaveLessThanFourProtectionsXMalware}{61}
\newcommand{\haveLessThanSixProtectionsX}{484}
\newcommand{\percentHaveLessThanSixProtectionsX}{85}
\newcommand{\goodwareCfg}{73}
\newcommand{\goodwareCfgOpa}{46}
\newcommand{\percentGoodwareCfgOpa}{63}
\newcommand{\goodwareLoo}{11}
\newcommand{\goodwareLooDat}{8}
\newcommand{\percentGoodwareLooDat}{73}
\newcommand{\goodwareMba}{19}
\newcommand{\goodwareMbaVir}{12}
\newcommand{\percentGoodwareMbaVir}{63}

\makeatletter
\NewDocumentCommand\textoverline{m}{$\overline{\hbox{#1}}$}
\makeatother

\ExplSyntaxOn{}
\NewDocumentCommand{\storedata}{mm}{%
    \bcp_store_data:nn { #1 } { #2 }%
}
\NewExpandableDocumentCommand{\getlength}{m}{%
    \seq_count:c { l_bcp_data_#1_seq }%
}
\NewExpandableDocumentCommand{\getdata}{O{1}m}{%
    \bcp_get_data:nn { #1 } { #2 }%
}
\cs_new_protected:Npn \bcp_store_data:nn #1 #2 {%
    \seq_clear_new:c { l_bcp_data_#1_seq }
    \__bcp_append_data:nn { #1 } { #2 }
}
\cs_new_protected:Npn \bcp_append_data:nn #1 #2 {%
    \seq_if_exist:cF { l_bcp_data_#1_seq }
    { \seq_new:c { l_bcp_data_#1_seq } }
    \__bcp_append_data:nn { #1 } { #2 }
}
\cs_new_protected:Npn \__bcp_append_data:nn #1 #2 {%
        \tl_map_inline:nn { #2 } {%
        \seq_put_right:cn { l_bcp_data_#1_seq } { ##1 }
    }
}
\cs_new:Npn \bcp_get_data:nn #1 #2 {%
    \seq_item:cn { l_bcp_data_#2_seq } { #1 }
}
\cs_new_protected:Nn \bcp_remove_last:Nn {%
    \seq_pop_right:cN { l_bcp_data_#2_seq } #1
}
\ExplSyntaxOff{}

\newcommand{\mwc}[1]{\textcolor{red}{#1}}
\newcommand{\gwc}[1]{\textcolor{blue}{#1}}
\newcommand{\ol}[1]{\overline{#1}}
\newcommand{\ul}[1]{\underline{#1}}
\newcommand{\tb}[1]{\textbf{#1}}

\begin{document}

\storeArrayPapersA{}
\storeArrayPapersD{}
\storeArrayPapersO{}
\storeArrayPapersAD{}
\storeArrayPapersAO{}
\storeArrayPapersDO{}
\storeArrayPapersADO{}

\newcommand{\pargap}{\vspace{2mm}}

\tikzstyle{venncirclish} = [
draw,
circle,
text width= 5cm,
]
\tikzstyle{legend} = [
align=left,
rectangle,
minimum width = 9cm,
text width = 9cm,
node distance = 0.1cm,
]
\tikzstyle{none} = [
]
\tikzstyle{pathline} = [
draw=black,
draw=none,
]
\tikzstyle{textstyle} = [
text width = 5cm,
circle,
align = center,
font=\Huge,
]
\tikzstyle{anaellipsestyle} = [
fill=none,
draw=black,
shape=ellipse,
minimum width=26cm,
minimum height=18cm,
text width=2cm,
align=center,
]
\tikzstyle{obfellipsestyle} = [
fill=none,
draw=black,
shape=ellipse,
minimum width=36cm,
minimum height=20.4cm,
text width=2cm,
align=center,
]
\tikzstyle{dobellipsestyle} = [
fill=none,
draw=black,
shape=ellipse,
minimum width=18cm,
minimum height=18cm,
text width=2cm,
]
\tikzstyle{referencenode} = [
font=\Large,
]
\title{Evaluation Methodologies in Software Protection Research}

\author{Bjorn De Sutter}
\orcid{0000-0003-0317-2089}
\authornote{B.\ De Sutter and S. Schrittwieser share dual first authorship.}
\email{bjorn.desutter@ugent.be}
\affiliation{%
    \institution{Computing Systems Lab, Ghent University}
    \streetaddress{Technologiepark Zwijnaarde 126}
    \postcode{9052}
    \city{Gent}
    \state{Belgium}
}

\author{Sebastian Schrittwieser}
\orcid{0000-0003-2115-2022}
\authornotemark[1]
\email{sebastian.schrittwieser@univie.ac.at}
\affiliation{%
    \institution{Christian Doppler Laboratory for Assurance and Transparency in Software Protection, Faculty of Computer Science, University of Vienna}
    \streetaddress{Kolingasse 14-16}
    \postcode{1090}
    \city{Wien}
    \state{Austria}
}

\author{Bart Coppens}
\orcid{0000-0002-7628-9264}
\email{bart.coppens@ugent.be}
\affiliation{%
    \institution{Computing Systems Lab, Ghent University}
    \streetaddress{Technologiepark Zwijnaarde 126}
    \postcode{9052}
    \city{Gent}
    \state{Belgium}
}

\author{Patrick Kochberger}
\orcid{0000-0002-0898-9824}
\email{patrick.kochberger@fhstp.ac.at}
\affiliation{%
    \institution{St.\ P\"olten University of Applied Sciences}
    \streetaddress{Matthias Corvinus-Straße 15}
    \postcode{3100}
    \city{St.\ P\"olten}
    \state{Austria}
}

\begin{abstract}
    \emph{Man-at-the-end} (MATE) attackers have full control over the system on which the attacked software runs, and try to break the confidentiality or integrity of assets embedded in the software. Both companies and malware authors want to prevent such attacks. This has driven an arms race between attackers and defenders, resulting in a plethora of different protection and analysis methods.
However, it remains difficult to measure the strength of protections because MATE attackers can reach their goals in many different ways and a universally accepted evaluation methodology does not exist.
This survey systematically reviews the evaluation methodologies of papers on obfuscation, a major class of protections against MATE attacks.
For \totalPapers{} papers, we collected \numberAspects{} aspects of their evaluation methodologies, ranging from sample set types and sizes, over sample treatment, to performed measurements. We provide detailed insights into how the academic state of the art evaluates both the protections and analyses thereon. In summary, there is a clear need for better evaluation methodologies. We identify nine challenges for software protection evaluations, which represent threats to the validity, reproducibility, and interpretation of research results in the context of MATE attacks and formulate a number of concrete recommendations for improving the evaluations reported in future research papers.

\end{abstract}


\authorsaddresses{%
    Bjorn De Sutter, bjorn.desutter@ugent.be, and Bart Coppens, bart.coppens@ugent.be, Computing Systems Lab, Ghent University, Technologiepark-Zwijnaarde 126, 9052, Gent, Belgium; Sebastian Schrittwieser, sebastian.schrittwieser@univie.ac.at, %
    Faculty of Computer Science, University of Vienna, Kolingasse 14-16, 1090, Wien, Austria; Patrick Kochberger, patrick.kochberger@fhstp.ac.at, St.\ P\"olten University of Applied Sciences, Campus-Platz 1, 3100, St.\ P\"olten, Austria.
}


\maketitle{}

\section{Introduction}
The desire and need to protect software from analysis, reverse engineering, and tampering have always existed.
Software vendors want to keep the exact implementation of their product and the data embedded therein, such as cryptographic keys, secret.
They want to prevent the removal of copy protections, they want to prevent cheating in multi-player games, etc.
Not only benign developers try to protect implementation details.
Malware authors also often employ \emph{software protection} (SP) mechanisms to prevent automatic tools and human analysts from detecting and analyzing the malicious nature and behavior of their code.

Over the past three decades, a vast number of obfuscation, tamperproofing, and anti-analysis techniques for both goodware and malware have been introduced in the literature~\cite{2016_protecting_software_through_obfuscation_can_it_keep_pace_with_progress_in_code_analysis,2018_diversification_and_obfuscation_techniques_for_software_security_a_systematic_literature_review,2021_measuring_software_obfuscation_quality_a_systematic_literature_review} and in industry to meet the described needs.
The deployment of such protections has in turn triggered an arms race in which many new deobfuscation, tampering, and analysis techniques have been proposed.
These are collectively called \emph{man-at-the-end} (MATE) attacks, because the \textit{human attackers} using the deobfuscation, tampering, and analysis techniques have full control over the end system on which they run, analyze, and alter the software and malware under attack.
Their tools include emulators, debuggers, disassemblers, custom operating systems, fuzzers, symbolic execution, sandboxes, and instrumentation among a wide range of static and dynamic analysis and tampering tools.
These give them white-box access to the software and malware internals, including their internal execution state.
To a considerable degree, but not completely, benign software and malware analysis techniques overlap, and hence so do the deployed protections.

In the MATE attack model, and with the current state of the art in practical SPs, complete prevention of attacks on software and malware protections is impossible~\cite{cc_book}.
With enough resources and time to exploit their white-box access, attackers will always succeed.
The goal of benign MATE SP is hence to make potential attackers' return on investment negative.
The already mentioned SPs can help to achieve this by delaying the successful identification and engineering of attacks, and by limiting the exploitability of identified attacks.
Software diversity also plays an important goal, as it can limit the attacker's use of a priori knowledge to identify an attack, and hinder its wide-scale exploitation.
The goal of malware protection is of course the inverse, namely to make the malware authors' return on investment positive by delaying analysis and detection to the point where the malware has generated enough money or caused enough damage.

These goals are much fuzzier than the goals in some other security domains such as cryptography, which builds on precisely defined mathematical foundations to protect against man-in-the-middle attacks.
While cryptographic techniques and well-defined, cryptographically-sound concepts have also been developed in the domain of obfuscation, the results (be it negative~\cite{2016_new_negative_results_on_differing_inputs_obfuscation} or positive~\cite{2004_positive_results_and_techniques_for_obfuscation}) are far from ready for widespread, general use in practice.
Because of the mentioned fuzziness, and because MATE attackers have so many alternative attack strategies and techniques at their disposal, it is difficult to measure the strength of SPs and to make founded statements about them.
As attackers can reach their goal in so many ways, there is not even a universally applicable and accepted evaluation methodology.
For example, while there is a common understanding that the relevant aspects to evaluate include potency, resilience, stealth, and cost~\cite{1997_a_taxonomy_of_obfuscating_transformations,1998_manufacturing_cheap_resilient_and_stealthy_opaque_constructs}, there is no universally agreed and applicable method to define, qualify, or quantify the former three.

This and related issues were discussed extensively at the 2019 Dagstuhl Seminar on Software Protection Decision Support and Evaluation Methodologies~\cite{2019_software_protection_decision_support_and_evaluation_methodologies_dagstuhl_seminar_19331}.
Most participants felt that there is a lack of good benchmarks for measuring the strength of SPs, and that the significance and validity of evaluations of proposed methods remains unclear.
The fact that this gut feeling and (collective) subjective concern based on anecdotal experience was not backed up with systematic research motivated us to perform this survey.
Our overall goals are to validate whether the mentioned concerns really exist, to draw a clear picture of the currently used methods to evaluate SPs, and to identify challenges and outline directions for better comparability and reproducibility of publications in the field of SP.

To that extent, we systematically analyze the content and especially the evaluation sections of publications on SP and code analysis that target software confidentiality.
In other words, we focus on publications on the topics of obfuscation and deobfuscation and techniques to perform and prevent analysis of software and malware.
Despite limiting the scope of our literature review in this way, with an initial set of \totalBeforeSelection{} publications and an in-depth analysis of \totalPapers{} papers, it is by far the largest literature study in this field to date.

This paper is organized as follows:
\Cref{sec:methodology} presents the methodology we adopted to include, exclude, and categorize papers, and for extracting data from them.
\Cref{sec:sample_sets} analyzes the sample sets used for evaluating the papers' contributions, including the sample set sizes and the nature of the used samples.
\Cref{sec:treatment} discusses the treatment that the samples underwent as discussed in the papers, i.e., how the protected (and unprotected) samples were generated.
\Cref{sec:measurements} studies how measurements were performed on the samples.
\Cref{sec:human} reports our findings about experiments involving human subjects.
\changed{\Cref{sec:recommendations} presents our recommendations for future research.}
In \cref{sec:related_work}, we list related literature surveys and reviews.
Finally, \cref{sec:conclusion} concludes this work.

\section{Methodology}\label{sec:methodology}

This section presents the methodology we used to select papers in scope of our study and to collect data from the selected papers.
We also discuss some top-level features of the selected papers, such as their perspective, the types of programming languages they target, and the quality of the venues at which the publications were published.

\subsection{Scope}\label{sec:scope}
This survey comprises \emph{peer-reviewed} literature on \emph{practical general-purpose SP} techniques that aim to protect the \emph{confidentiality} of assets embedded in distributed software (goodware or malware) by preventing or hindering software analyses and reverse engineering in the MATE (man-at-the-end) attack model.

The SPs in scope target software formats ranging from source code over bytecode/intermediate representations (IRs) to compiled software (binary code).
Such SPs include various obfuscation techniques (e.g., opaque predicates) and preventive protections (i.e., that prevent or hinder the use of reverse engineering tools and analysis tools, such as anti-debugging techniques).
Our SP scope also includes methods that use hardware to protect software (e.g., relying on physically unclonable functions).
However, schemes that protect hardware (i.e., integrated circuits) are excluded.
Furthermore, out-of-scope are SPs that prevent the exploitation of vulnerabilities (e.g., ASLR), as well as anti-tampering schemes that exclusively focus on protecting the integrity of embedded assets, such as code guards and remote attestation.
In addition, methods for obfuscating data that is processed by, but not included in the distributed software are out of scope (e.g., network traffic and location data obfuscations), as well as software-based steganography (i.e., techniques to inject external data into a program in a hidden manner).
Because our focus is on practical general-purpose SP techniques, we exclude papers focusing on special-purpose obfuscation (e.g., white-box cryptography~\cite{2020_how_to_reveal_the_secrets_of_an_obscure_white_box_implementation} and obfuscations of point-functions, pseudo-random functions, and finite automata) as well as cryptographic obfuscation (e.g., indistinguishability obfuscation, virtual black box obfuscation, fully homomorphic encryption)~\cite{2020_cryptographic_obfuscation_a_survey} that we deem not ready for use in practice yet.

This survey includes papers that present (i) novel SPs in scope; (ii) reverse engineering tools and techniques with which those SPs in scope are attacked, identified, detected, located, circumvented, deobfuscated, analyzed, undone, bypassed, worked-around, etc.; (iii) practical or theoretical methods or models to reason about, evaluate, and deploy those SPs and attacks; (iv) as well as surveys on them.
Pure position papers, non-peer-reviewed papers, blog posts, books, and posters are out of scope.
\Cref{tab:criteria} lists all our exclusion criteria.

\begin{table}
    \caption{Inclusion and exclusion criteria. We first compiled a list of \totalBeforeSelection{} papers meeting at least one inclusion criterion. We then excluded papers meeting at least one exclusion criterion, ending up with \totalPapers{} papers.} \label{tab:criteria}
    \begin{tabular}{r p{13cm}}
        \toprule
        \multicolumn{2}{l}{\textbf{Inclusion criteria}}                                                                                                                                                                                                                                                           \\
        \midrule
        1 & Obfusc*\big \vert deobfusc* in paper title published between 2016--2022 and listed in ACM Digital Library, IEEE Xplore, SpringerLink or published at the USENIX Security or NDSS symposia                                                                                                             \\
        2 & Paper is cited in~\cite{2016_protecting_software_through_obfuscation_can_it_keep_pace_with_progress_in_code_analysis,2018_diversification_and_obfuscation_techniques_for_software_security_a_systematic_literature_review,2021_measuring_software_obfuscation_quality_a_systematic_literature_review} \\
        3 & Offensive or defensive paper on software protection or malware analysis published before 2023 by any of a self-curated list of authors\footnotemark{}                                                                                                                                                 \\
        \toprule
        \multicolumn{2}{l}{\textbf{Exclusion criteria}}                                                                                                                                                                                                                                                           \\
        \midrule
        1 & Not peer-reviewed                                                                                                                                                                                                                                                                                     \\
        2 & Posters, books, pure position papers                                                                                                                                                                                                                                                                  \\
        3 & Targets obfuscation of something else than software, such as location or energy meter data                                                                                                                                                                                                            \\
        4 & Topic is a form of cryptographic obfuscation                                                                                                                                                                                                                                                          \\
        5 & Main focus is special-purpose obfuscation such as white-box cryptography                                                                                                                                                                                                                              \\
        6 & Protection aims for mitigating vulnerability exploits, not on confidentiality of software                                                                                                                                                                                                             \\
        7 & Focus of paper is software integrity protection rather than confidentiality                                                                                                                                                                                                                           \\
        8 & Focus is on steganography instead of software confidentiality                                                                                                                                                                                                                                         \\
        9 & Evaluated or analyzed techniques do not include protections in scope                                                                                                                                                                                                                                  \\
        \bottomrule
    \end{tabular}
\end{table}

\subsection{Paper Retrieval and Selection Process}\label{sec:paper_selection}
The starting point of our study are papers on the topic of obfuscation gathered from three online libraries.
In the ACM Digital Library, IEEE Xplore, and SpringerLink we searched for paper titles including \enquote{obfuscation}, \enquote{obfuscate}, \enquote{obfuscating}, and other variants thereof, including variants of \enquote{deobfuscation}, from 2016 to 2022.
From those years, we also added papers based on their title from the USENIX Security and Network and Distributed System Security (NDSS) symposia, two top security conferences that are not included in the aforementioned libraries.
For papers older than 2016, we limited the selection to papers referenced in three well-known surveys related to this survey~\cite{2016_protecting_software_through_obfuscation_can_it_keep_pace_with_progress_in_code_analysis,2018_diversification_and_obfuscation_techniques_for_software_security_a_systematic_literature_review,2021_measuring_software_obfuscation_quality_a_systematic_literature_review}.
In addition, a self-curated list\footnotemark[\value{footnote}]{} of researchers working in the area of SP was used to manually identify additional papers.
\footnotetext{C.~Basile, A.~Pretschner, A.~Lakhotia, A.~Francillon, B.~Wyseur, C.~Henderson, C.~Collberg, C.~Thomborson, C.~Maurice, C.~Mulliner, D.~Boneh, J.~Davidson, L.~Goubin, M.~Videau, M.~Bailey, M.~Franz, M.~Dala Preda, M.~Ahmadvand, N.~Stakhanova, N.~Provos, N.~Eyrolles, R.~Giacobazzi, S.~Debray, S.~Banescu, S.~Bardin, S.~Katzenbeisser, T.~McDonald, and Y.~Gu.}
\Cref{tab:criteria} lists these inclusion criteria.
The preliminary list obtained with the described, rather mechanistic retrieval process totaled \totalBeforeSelection{} papers. Many of those were false positives, however, so after the initial selection we manually filtered out papers that are out of scope as determined by the listed exclusion criteria.
Eventually, \totalPapers{} (\percentTotalInScope{}\%) remained for in-depth analysis.
This is hence by far the largest literature study on SP to date.
By comparison, the aforementioned surveys only contain 203, 367, and 98 cited works.

\begin{figure}
    \centering
    \includegraphics[width=0.80\textwidth]{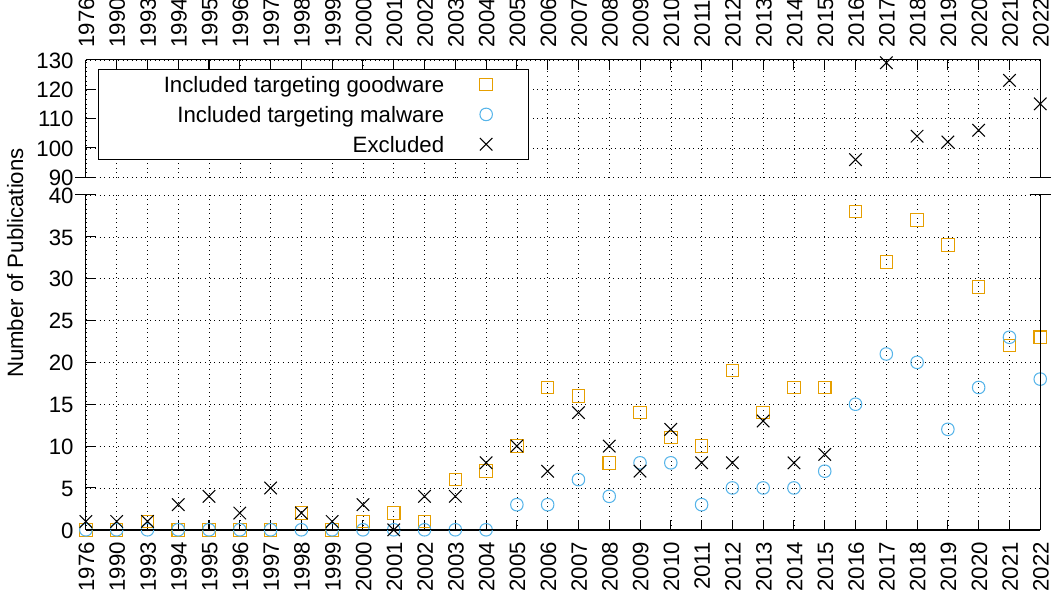}
    \caption{Evolution of included goodware and malware papers, as well as excluded papers.}\label{fig:exclude}
    \Description[Number of ations excluded and included per year.]{TODO add a long Description for visually challenged readers.}
\end{figure}

\Cref{fig:exclude} depicts the publications included and excluded per publication year. About 60\%~$\hat{=}$~343/\totalPapers{} of the papers are from the 2016--2022 period.
This is in line with our ambition to shed a light on currently used evaluation methods in the domain of SP.
We deem the inclusion of the other 40\%~$\hat{=}$~230/\totalPapers{} of older papers important, however, because it enables us to identify potential historic trends and evolutions in how SPs are being evaluated.

\subsection{Information Collection}
For each paper, we collected \numberAspects{} aspects ranging from the contribution area (obfuscation, deobfuscation, analysis, etc.), over types of software targeted, and types of performed measurements and used sample sets, to the discussed and deployed protection and analysis methods and tools.
We always use our own terms and definitions for all categories, which sometimes diverge from how individual authors of surveyed papers used the same terms.
Our definitions are listed in the supplemental material in \cref{sec:protection_definitions,sec:analysis_definitions,sec:measurements_definitions}.
Overall, \numberDatapoints{} individual data points were gathered.
We partitioned the papers among all authors of this survey.
Care was taken not to finalize any paper's data collection before discussing any corner cases or doubts that surfaced and before obtaining consensus, i.e., agreeing unanimously on how to classify them.
All classifications and information were collected in a spreadsheet.
At times, additional categories or evaluation aspects were added and definitions of existing categories were updated or refined.
At those occasions, potentially impacted, already reviewed papers were revisited.

We further conducted multiple rounds of double checking in the spreadsheet for consistency and performed various plausibility checks on the data to further improve its quality. %

\subsection{Top-level Paper Categorization}\label{sec:top_level_categories}

\changed{We categorized all papers non-exclusively using the top-level categories listed in \cref{tab:all_papers_short}.
    In doing so, we consider malware detection and malware classification papers to be analysis papers. %
    Papers labeled as analysis papers are in addition labeled as deobfuscation papers if the presented analysis reveals the information necessary to deobfuscate protected code fragments.
    Examples are control flow analyses that identify opaque predicate-based bogus control flow as unrealizable, or techniques that reveal which library APIs are called at obfuscated call sites.

    If a paper's contribution is some analysis or framework to assess the strength of an obfuscation, we consider this an obfuscation contribution, as it advances our knowledge about obfuscations and evaluation methods thereof. We only consider it an analysis contribution if it also advances knowledge about analysis techniques. Detection algorithms, such as for library code and cryptographic primitive detection, are framed by some authors as analysis research, and by others as deobfuscation research. We consider them analysis papers because they classify software components similarly to what malware classification does. We also label them as deobfuscation papers, however, because in the eyes of an attacker they summarize complex, obfuscated low-level code instances into abstract concepts that are void of any obfuscation.%
}
\setlength{\tabcolsep}{2pt}

\begin{table}
    \caption{Categorization of the \totalPapers{} surveyed papers. Categories are non-exclusive.}\label{tab:all_papers_short}
    \small
    \begin{tabular}{l r p{9.5cm}}
        \toprule
        \textbf{Perspective} &                                                                 &                                                                               \\
        \midrule
        Obfuscation          & \totalObfuscation{}~$\hat{=}$~\percentObfuscation{}\%           & novel SPs, deployment and evaluation methods, insights into SP practicee      \\
        Deobfuscation        & \totalDeobfuscation{}~$\hat{=}$~\percentDeobfuscation{}\%       & novel tools/methods for obtaining deobfuscated software representations       \\
        Analysis             & \totalAnalysis{}~$\hat{=}$~\percentAnalysis{}\%                 & other analysis and (malware) classification techniques, evaluation thereof    \\
        \midrule
        \textbf{Application} &                                                                 &                                                                               \\
        \midrule
        Diversification      & \totalDiversification{}~$\hat{=}$~\percentDiversification{}\%   & diversification of SPs or as a SP                                             \\
        Tamperproofing       & \totalTamperproofing{}~$\hat{=}$~\percentTamperproofing{}\%     & use of SPs to strengthen tamperproofing of code, or attacks on them           \\
        Watermarking         & \totalWatermarking{}~$\hat{=}$~\percentWatermarking{}\%         & use of SPs for building watermarks, or attacks on them                        \\
        \midrule
        \textbf{Target}      &                                                                 &                                                                               \\
        \midrule
        Malware              & \totalMalware{}~$\hat{=}$~\percentMalware{}\%                   & paper focuses on malware and/or includes malware evaluation samples           \\
        Goodware             & \totalGoodware{}~$\hat{=}$~\percentGoodware{}\%                 & all papers not categorized as malware according to above criterion            \\
        Malware only         & \totalMalwareOnly{}~$\hat{=}$~\percentMalwareOnly{}\%           & malware paper focusing on techniques that are only useful for malware         \\
        Mobile               & \totalMobile{}~$\hat{=}$~\percentMobile{}\%                     & paper targets Android or iOS apps                                             \\
        \midrule
        \textbf{Nature}      &                                                                 &                                                                               \\
        \midrule
        Survey               & \totalSurveys{}~$\hat{=}$~\percentSurveys{}\%                   & literature studies, surveys, meta-analysis papers                             \\
        Theoretical          & \totalTheory{}~$\hat{=}$~\percentTheory{}\%                     & theoretical approaches, e.g., abstract interpretation for strength evaluation \\
        Data science         & \totalMachinelearning{}~$\hat{=}$~\percentMachinelearning{}\ \% & presented approaches rely on machine learning or artificial intelligence      \\
        Human Experiment     & \totalHumanExperiments{}~$\hat{=}$~\percentHumanExperiments{}\% & paper presents experiments with humans performing tasks                       \\
        \bottomrule
    \end{tabular}
\end{table}

\begin{figure}
    \centering
    \resizebox{0.92\columnwidth}{!}
    {
        \input{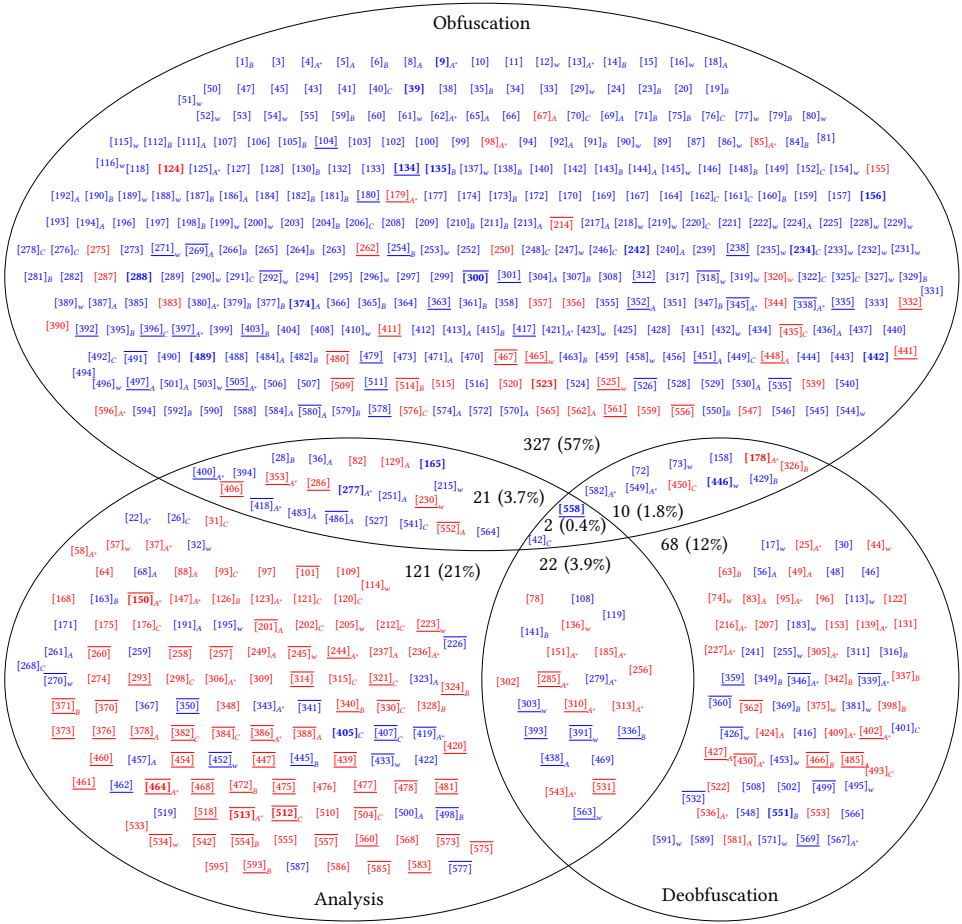}
    }
    \caption{Venn diagram of the \totalPapers{} surveyed \textcolor{blue}{goodware} and \textcolor{red}{malware} papers. Papers targeting \underline{\scriptsize{mobile platforms}} are underlined. Overlined papers use \textoverline{\scriptsize{data science}} methods such as machine learning and deep learning. Bold references are \textbf{surveys and literature and meta analyses}. A subscript index indicates the CORE ranking of the paper's publication venue, ranging from $A^*$, $A$, $B$, to $C$, and $w$ for workshop. An absent index denotes that the venue is national, new, or unranked.}\label{fig:venn}
    \Description[TODO add description for visually challenged readers.]{TODO add a long Description for visually challenged readers.}
\end{figure}

\changed{Finally, all papers not categorized as \textit{malware} using the criterion formulated in \cref{tab:all_papers_short} are categorized as \textit{goodware} papers.
    The goodware category hence includes papers that do not specifically target malware and that do not use malware samples, but of which the content can still apply to both benign and malicious software.

    \bemph{Our survey includes many more defensive papers than offensive ones}. This can be seen in the dominance of the blue goodware papers in the obfuscation part and the dominance of the red malware papers in the analysis part of \cref{fig:venn}.
    In the context of malware, defensive papers focus on analysis, detection, and deobfuscation techniques to safeguard systems from malware.
    In the context of goodware, by contrast, defensive papers focus on obfuscation to protect the confidentiality of software assets against reverse engineering.
}

\changed{
    \bemph{Our survey includes few multi-perspective papers, and even fewer papers (1.2\%~$\hat{=}$~7/\totalPapers{}) that contribute novel techniques and at once evaluate the impact of countermeasures that adversaries might adopt}.
    \Cref{tab:multiperspectives} lists the various ways in which multi-perspective papers contribute.
    This lack of multi-perspectivism is important because SP is an arms race between defenders and attackers, a.k.a.\ a cat and mouse game.
    In case evaluating a countermeasure against one's own contribution requires a considerable research or engineering effort, it is acceptable to publish the initial contribution and the countermeasure separately.
    However, we think it occurs much more often that one can evaluate the impact of (small) adaptations by adversaries with relatively little effort.
    In such cases, it is warranted for reviewers and readers to expect and demand such an evaluation in the papers.
    This is all the more important because a crucial aspect of MATE attackers' modus operandi is to find and exploit the attack-path-of-least-resistance.
    Whenever some novel SP is proposed and evaluated that impacts some path-of-least-resistance, it is important to evaluate whether there are no trivially similar, alternative paths-of-least-resistance remain unimpacted.
    A best practice for SP researchers that make a cat or mouse move is therefore to look ahead and evaluate at least the smallest next moves that can be anticipated.
    This best practice expectation was explicitly raised at the aforementioned Dagstuhl seminar.

    \bemph{Our survey confirms that the literature is lacking with respect to multi-perspectivism and direct countermeasure evaluation.}
}

\begin{table}
    \caption{Multiperspective papers and how they combine contributions in (de)obfuscation and analysis.\label{tab:multiperspectives}. Papers were marked with an * contribute novel techniques and at once evaluate the impact of countermeasures.}
    \scriptsize
\begin{tabular}{p{2.3cm} p{11.2cm}}
        \toprule
        \multicolumn{2}{l}{\textbf{Obfuscation, Deobfuscation, and Analysis: \percentAnalysisDeobfuscationObfuscationOnly{}\%~$\hat{=}$~\totalAnalysisDeobfuscationObfuscationOnly{}/\totalPapers{} papers}}                                                                                                                                                                                                                                                                                                                                                                                                                                                                                                                                                                                                                                                                                                                                                                                                                                \\
\bottomrule
    \end{tabular}
\end{table}

\begin{figure}
    \includegraphics[width=10cm]{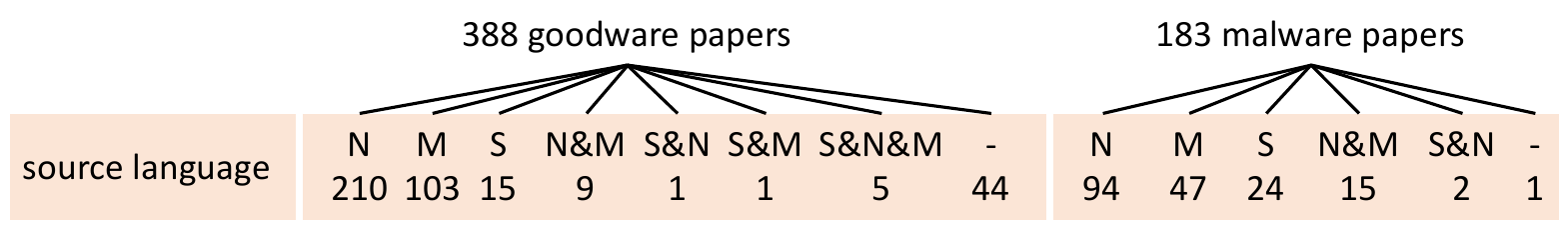}
    \caption{Targeted programming languages: \underline{M}anaged, \underline{N}atively compiled, and \underline{S}cript languages. \changed{Language-agnostic papers are marked with a dash (-).}}\label{fig:languages_graph_source}
    \Description[Targeted programming languages]{TODO add a long Description for visually challenged readers.}
\end{figure}

\changed{
    We also categorized the papers according to the categories of programming languages they target, i.e, the languages on which the paper authors demonstrate or evaluate their contributions.
    \Cref{fig:languages_graph_source} shows the results.
    In line with existing work~\cite{2016_a_survey_on_aims_and_environments_of_diversification_and_obfuscation_in_software_security}, we considered four categories:
    \begin{itemize}
        \item \emph{Native languages} such as C/C++ are compiled to and distributed as native code binaries.
        \item \emph{Managed languages} such as Java or C\# are typically compiled to and distributed as bytecode to be executed in a managed environment, for which enough symbolic information needs to be included, e.g., to support garbage collection, type correctness verification, and reflection.
        \item \emph{Scripts}, e.g., in JavaScript or PHP, are distributed as source code mostly for web applications.
        \item No \emph{domain specific languages} are in scope as we exclude special-purpose obfuscation.
    \end{itemize}
}

\changed{\bemph{A significant number of papers (44 goodware + 1 malware papers) are language-agnostic.}
    They do not target any particular type of programming language, either because they are surveys or theoretical papers, or because they present practical techniques in a language-independent way.

    \bemph{For goodware, by far the most targeted languages are native languages at 54\%~$\hat{=}$~210/388. Managed languages are considerably less popular at 27\%~$\hat{=}$~103/388.} The latter mostly target Java and C\#.
    Their lower popularity is no surprise.
    The run-time management of these programs requires that the distributed bytecode adheres to stricter rules than required in native code, and that it is accompanied by quite some symbolic information. Protection tools for managed software have hence much less freedom to operate and to generate unconventional code than native language protection tools. The semantic gap between the managed language source code and the corresponding (protected) bytecode is hence also smaller than the semantic gap between native language source code and its corresponding (protected) assembly code. For these reasons (protected) managed bytecode is typically an easier target for reverse engineers than (protected) native code. Software developers caring for their assets and considering SP are hence incentivized to opt for a natively compiled language rather than for a managed language, and when they have the freedom to make that choice (given their other industrial requirements), they often do so in practice. It then also follows that programs written in native languages are more interesting targets for SP.

    \bemph{Only 2.3\%~$\hat{=}$~9/388 of the papers study the combination of managed and natively compiled software. Given the market importance of the Android platform, we find this surprisingly few:}
    only 1.3\%~$\hat{=}$~5/388 of the papers explicitly target Android applications that contain both Java bytecode and native code~\cite{2020_a_large_scale_study_on_the_adoption_of_anti_debugging_and_anti_tampering_protections_in_android_apps,2019_robust_detection_of_obfuscated_strings_in_android_apps,2017_on_the_effectiveness_of_code_reuse_based_android_application_obfuscation,2017_an_anti_reverse_engineering_technique_using_native_code_and_obfuscator_llvm_for_android_applications,2022_apkdiff_matching_android_app_versions_based_on_class_structure}.
    We find this case important, because the aforementioned choice to embed the security-sensitive assets in the native libraries in Android applications does not fully void the need to also protect the Java part, in particular its interfaces to the native parts. If those are not protected, they can provide trivially exploitable attack vectors.

    \bemph{Scripting languages are clearly the least popular targets of goodware protection papers}. The reason is of course that they are in general even easier to reverse engineer than bytecode of managed languages.
    Only 3.9\%~$\hat{=}$~15/388 of the goodware papers specifically target them; 13 papers target JavaScript, two target PHP.
    JavaScript is hence clearly the most popular script language for obfuscation research.

    Of the 1.3\%~$\hat{=}$~5/388 goodware papers that target all three types of programming languages, four are surveys that cover obfuscations for all types~\cite{2020_layered_obfuscation_a_taxonomy_of_software_obfuscation_techniques_for_layered_security,2018_diversification_and_obfuscation_techniques_for_software_security_a_systematic_literature_review,2017_a_tutorial_on_software_obfuscation,2016_protecting_software_through_obfuscation_can_it_keep_pace_with_progress_in_code_analysis}, while one presents an obfuscation tool for all three types of languages~\cite{2019_design_and_implementation_of_obfuscating_tool_for_software_code_protection}.

    Finally, one goodware paper discusses not closely related techniques applicable to script code and techniques applicable to native code~\cite{2021_a_comprehensive_solution_for_obfuscation_detection_and_removal_based_on_comparative_analysis_of_deobfuscation_tools} and one paper focuses (mostly) on identifier renaming which is mostly applicable to both script and managed languages~\cite{2015_from_obfuscation_to_comprehension}.

    Overall, the prevalence of the three types of languages in our survey is similar to that reported in a previous survey~\cite{2018_diversification_and_obfuscation_techniques_for_software_security_a_systematic_literature_review}. For each type, the fractions we report are within 5\% of theirs.

    \bemph{Of the malware papers, 51\%~$\hat{=}$~94/183 target native malware. This is typically Windows malware. Next, 26\%~$\hat{=}$~47/183 of the malware papers focus on code written in managed languages, mostly (Android) Java.}
    Two papers target pure Java malware~\cite{2016_detection_of_obfuscation_in_java_malware,2017_partial_evaluation_of_string_obfuscations_for_java_malware_detection}, and 45 papers target the Java part of Android apps.
    In addition, 8.2\%~$\hat{=}$~15/183 consider both the Java and the native part in Android apps.
    The 13\%~$\hat{=}$~24/183 of the papers targeting script malware focus on JavaScript (11), PowerShell (5), PHP (3), Visual Basic (3), PDF scripts (1), and shell scripts (1).
    Finally, two papers discuss malware analyses for both script and native malware~\cite{2012_a_survey_on_automated_dynamic_malware_analysis_techniques_and_tools,2022_a_systematic_approach_for_evading_antiviruses_using_malware_obfuscation}.

    \bemph{Finally, script languages are much more popular in malware papers than in goodware papers, with 14\%~$\hat{=}$~26/183 of the malware papers targeting script languages}, while only 5.7\%~$\hat{=}$~22/388 of the goodware papers do so. Scripts language papers are then also the only type of language for which malware papers dominate the goodware papers, with a ratio of 26:22. For native and managed languages, the ratios are 111:226 and 61:118 respectively. This is in line with what we discussed above for goodware defenders being incentivized not to use managed and script languages if they care about their assets. In the case of malware, by contrast, the omnipresence of scripts in online applications and their exploitation in the real-world have effectively raised the attention of researchers, resulting in more focus on them in literature as well.
}

\subsection{Quality of Venues}

We assessed the quality of the \totalPapers{} papers' publication venues using the Australian CORE conferences and journal rankings~\cite{core}.
The results are as follows: A* \percentVenueAs{}\%~$\hat{=}$~\venueAs/\totalPapers{}, A \percentVenueA{}\%~$\hat{=}$~\venueA/\totalPapers{}, B \percentVenueB{}\%~$\hat{=}$~\venueB/\totalPapers{}, C \percentVenueC{}\%~$\hat{=}$~\venueC/\totalPapers{}, unranked \percentVenueUnranked{}\%~$\hat{=}$~\venueUnranked/\totalPapers{}, and workshop \percentVenueWorkshop{}\%~$\hat{=}$~\venueWorkshop/\totalPapers{}.
In addition, \percentVenueNational{}\%~$\hat{=}$~\venueNational/\totalPapers{} of the surveyed papers were published at national-only venues, and \percentVenueNew{}\%~$\hat{=}$~\venueNew/\totalPapers{} at new venues not (yet) ranked by CORE\@.
\changed{Importantly, these rankings} are based on the latest available CORE databases (2020 for journals and 2021 for conferences) but venue rankings might change over time. The ranking of a venue may hence have been different at the time of publication.

A detailed study of the numbers revealed that while the distribution of papers over differently ranked venues varies from one year to another, we observed no major trends over the years. In particular, the distributions for the years 2016--2022 did not differ substantially from the distribution in the years up to 2015. In other words, the different methods we used for selecting papers up to 2015 (based on existing surveys and authors) and from 2016 onwards (additionally based on title keywords in online databases) did not have a noticeable impact on the quality of the papers included from those years as judged by the ranking of their publication venue.

\changed{\bemph{The most important finding from our data is that papers published at top-tier venues (A*) are predominantly from the malware category.}}
Although only \percentMalware{}\% of all surveyed papers fall into the malware category, about 58\% of all A* papers are from this category. Goodware obfuscation papers are instead mostly published at specialized workshops.

\changed{\bemph{Overall, the number of goodware SP papers (\ref{tab:recommendation_matrix})
        at top-tier security venues is low}: USENIX Security\@: 5, ACM CCS\@: 4, IEEE S\&P\@: 1, NDSS\@: 1.}
The cause of this can only be speculated.
However, it coincides with the subjective feeling among many goodware SP researchers, including many participants of the mentioned Dagstuhl seminar, that it is difficult to get goodware SP papers published at top-tier venues.
Part of the reason could be that the evaluations and validations of the proposed techniques are less convincing to reviewers, e.g., because the used methodologies are ad hoc rather than standardized, and because the goals and provided security guarantees are much fuzzier than in other domains such as cryptography.
With this survey, we shed more light on the used evaluations, such that the SP community can progress from a subjective feeling towards more objective observations.

\section{Sample Sets}\label{sec:sample_sets}
\changed{In the SP community, no consensus on the use of particular samples for the evaluation of SPs and code analysis methods exists~\cite{2019_software_protection_decision_support_and_evaluation_methodologies_dagstuhl_seminar_19331}. Instead, used samples vary widely both in complexity of individual samples and overall sample set size. This results in a strikingly large number of distinct samples used in the SP literature,
in stark contrast to other fields in computer science, such as in machine learning and computer vision with the omni-present data sets MNIST, CIFAR10, CIFAR100, ImageNet; in compilers with the SPEC, parsec, and DaCapo benchmark suites; in circuit design with ITC'99, MCNC'91, ISCAS'89 and ISCAS'85; and even in hardware obfuscation, where Trust Hub offers a set of labeled hardware obfuscation benchmarks~\cite{2018_development_and_evaluation_of_hardware_obfuscation_benchmarks}.
}

This section presents the results of a fine-grained analysis of samples used in SP research.
Methodologically, we included only samples used for evaluating solutions presented in a paper.
Short demo programs or fragments of programs explaining an approach were omitted from the analysis.
Overall, \haveSamplesetPercentage{}\%~$\hat{=}$~\haveSampleset{}/\totalPapers{} of all papers have specified a sample set for evaluation.
Among the other papers that have no sample set, there are 8.4\%~$\hat{=}$~48/\totalPapers{} of which this is to be expected, because they are surveys, theoretic papers, and papers focusing specifically on SP tool features that do not need to be evaluated on samples, such as tool transparency for maintainers and users~\cite{2019_epona_and_the_obfuscation_paradox_transparent_for_users_and_developers_a_pain_for_reversers}.

\changed{\bemph{In the remaining 7.0\%~$\hat{=}$~40/\totalPapers{} papers, which are all goodware papers and of which the vast majority present novel obfuscation techniques claimed to be practical, no samples are used for evaluating the contributions.
This lack of evaluation obviously is major shortcoming of a significant number of papers.}}
However, our data in \cref{sec:sample-no-sample} in the supplemental material indicates a downward trend.

\begin{figure}[t]
    \centering
    \includegraphics[width=0.9\textwidth]{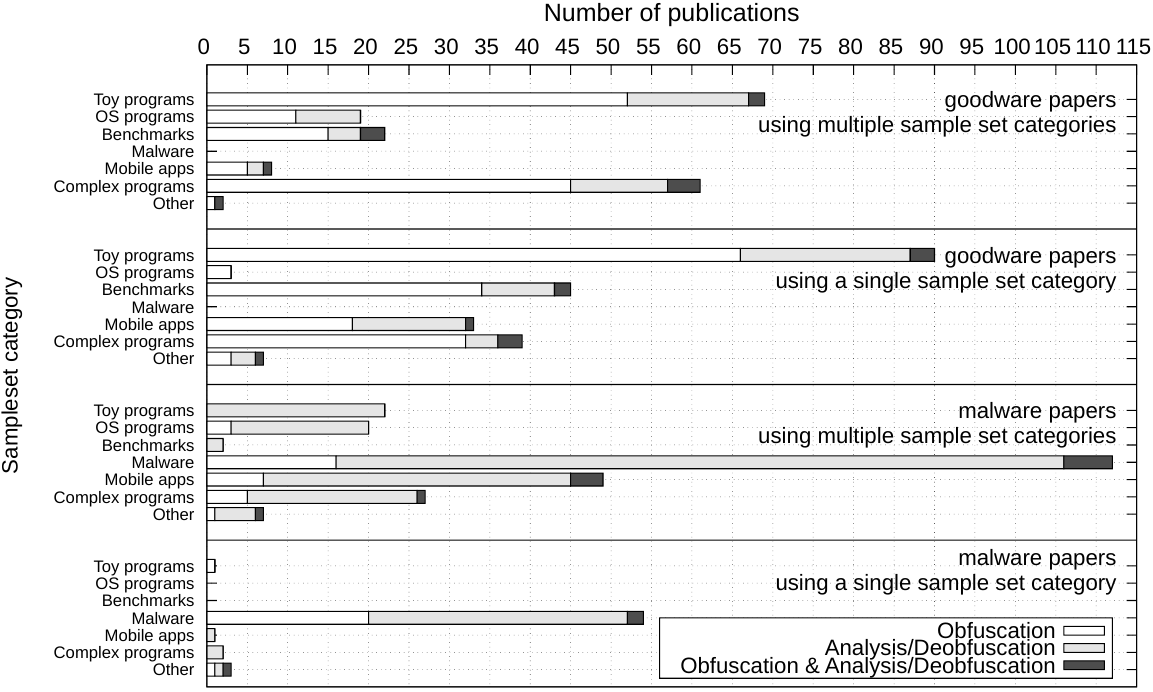}
    \caption{Number of papers per category of sample set. The categories are non-exclusive.}\label{fig:sample_types}
    \Description[Number of papers per category of sample set.]{TODO add a long Description for visually challenged readers.}
\end{figure}

\subsection{Sample Categories}
First, we analyzed which types of samples were used in the papers, including whether samples were reused, and we classified samples into distinct categories \textit{benchmarks}, \textit{malware}, \textit{toy programs}, \textit{complex programs}, \textit{oS programs}, \textit{mobile apps}, \textit{other}, and \textit{unknown}.
\Cref{fig:sample_types} shows the distribution of sample set categories in the surveyed papers for each of four paper types: goodware and malware papers that experiment with either one category of samples or multiple categories. %
Notice that some malware papers rely exclusively on non-malware samples.
The (valid) reason is that these papers focus on techniques aimed at malware, but which are only evaluated on benign samples.
This happens, e.g., for papers presenting repackaging detection techniques.

\paragraph{Toy programs}
\changed{Toy programs are used in a strikingly high number of evaluations. 
Goodware papers stand out in this regard, with \samplesetsNomwToyPercentage{}\%~$\hat{=}$~\samplesetsNomwToy{}/\samplesetsNomwTotal{} of their sample sets containing toy programs, and \samplesetsNomwToyOnlyPercentage{}\%~$\hat{=}$~\samplesetsNomwToyOnly{}/\samplesetsNomwToy{} of them only using toy programs for evaluation. Of all papers with samples, \samplesetsToyPercentage{}\%~$\hat{=}$~\samplesetsToy{}/\haveSampleset{} include at least one toy program.

We define toy programs as small, mostly single-function programs that are not standardized, i.e., not from a known benchmark suite.
They are often written for the sole purpose of performing evaluations. They include well-known algorithms for sorting and searching (\samplesSortingprogsPercentage{}\%~$\hat{=}$~\samplesSortingprogs{}/\samplesetsToy{});
encoding, encrypting, and hashing (\samplesEncprogsPercentage{}\%~$\hat{=}$~\samplesEncprogs{}/\samplesetsToy{});
and math functions such as Fibonacci and prime number generators and basic string operations (\samplesMathprogsPercentage{}\%~$\hat{=}$~\samplesMathprogs{}/\samplesetsToy{}).
Toy programs also include synthesized programs generated by tools such as Tigress and csmith (\samplesGeneratedprogsPercentage{}\%~$\hat{=}$~\samplesGeneratedprogs{}/\samplesetsToy{}), and simple programs with a special structure such as nested loops or a certain way of branching (\samplesTestingprogsPercentage{}\%~$\hat{=}$~\samplesTestingprogs{}/\samplesetsToy{}).

The use of toy programs does not always imply a threat to validity.
Local protections such as mixed Boolean-arithmetic (MBA) can be evaluated locally within their function, isolated from the rest of a program.
Thus even simple samples can be the foundation for meaningful evaluations.
However, we observe a more ambivalent picture, in which complex, non-local protections are also often evaluated with toy programs only.
For example, \samplesetsToyOnlyVirtualizationPercentage{}\%~$\hat{=}$~\samplesetsToyOnlyVirtualization{}/\samplesetsToyOnly{} of all toy-program-only papers deal with virtualization-based protection, which is not a local protection in real-world usage. Furthermore, nine out of ten goodware papers that introduce techniques based on symbolic execution include toy programs in their sample set. Five even do so exclusively (the remaining paper uses a single unknown sample).
It is known that \changed{symbolic execution} becomes impractical as program sizes increase due to the challenge of path explosion, so when evaluations are done with small programs only, no statement about the applicability to real programs can be made.

\bemph{The observed relience on toy programs, in particular in goodware papers, threats the external validity of much research, showing it remains at the lowest technology readiness levels (TRLs). The relience on non-standardized toy programs of which the implementations are rarely published, also hinders the reproducibility and interpretation of presented research results.}
For many categories of toy programs, there exist myriads of different implementations that differ significantly in terms of complexity and structure, such as when algorithms can be implemented with or without recursion. It should be a strict requirement that evaluation toy programs are made available as artifacts.

\paragraph{OS programs}
Around \samplesetsOsPercentage{}\%~$\hat{=}$~\samplesetsOs{}/\haveSampleset{} of all sample sets include OS programs, i.e., programs that can be attributed to an operating system.
For Windows systems, these are bundled tools such as the calculator or notepad.
For Linux, we included tools that are an integral part of a typical Linux distributions, such as the GNU core utilities.
Almost all identified samples in this category have in common their comparatively low complexity and small binary size.

Samples of the GNU core utilities occur in \samplesCoreutilsPercentage{}\%~$\hat{=}$~\samplesCoreutils{}/\samplesetsOs{} papers that rely on OS programs.
Following the UNIX philosophy of small single-task utilities, the median lines of code (LoC) of the 126 utilities is 339, with the biggest program (\texttt{ls}) being 5626 LoC. Most real-world software that requires SP is magnitudes bigger. Core utilities hence are not representative.

\bemph{Lacking representativeness, the use of Linux OS programs is as doubtful as that of toy programs.}

The situation is different for the \samplesWindowsosprogsPercentage{}\%~$\hat{=}$~\samplesWindowsosprogs{}/\samplesetsOs{} papers with Windows samples, which are typically more complex. For example, a Windows~7 calc.exe binary is almost a megabyte large. %
Another major difference between Linux and Windows OS samples is the code format on which obfuscations can be applied. 
Since their source code is not publicly available, Windows samples only have SPs applied on binary code, while Linux OS programs can be obfuscated at various build stages.%

\paragraph{Benchmarks}
Of all sample sets, \samplesetsBenchmarksPercentage{}\%~$\hat{=}$~\samplesetsBenchmarks{}/\haveSampleset{} contain samples from benchmark suites. Almost all of them are originally intended for performance evaluations.
The SPEC suites are by far the most often used, at \samplesSpecPercentage{}\%~$\hat{=}$~\samplesSpec{}/\samplesetsBenchmarks{}.
They include real programs with realistic inputs.
SPEC CPU 2017, e.g., includes the GCC compiler, a weather forecasting program, a perl interpreter, and an x264 video compressor.
The surveyed papers use samples from only two out of SPEC's current offering of 25 suites: the SPEC CPU suite (including SPECint to measure integer arithmetic performance) for binary code and the SPECjvm suite for Java bytecode evaluations.
The second most used suite is MiBench, at \samplesMibenchPercentage{}\%~$\hat{=}$~\samplesMibench{}/\samplesetsBenchmarks{}.
This embedded benchmark suite's programs vary from single-function quicksort implementations to complex programs such as Ghostscript~\cite{guthaus2001mibench}.
All other suites, such as DaCapo, SciMark, and the CompCert benchmarks are used by less than five papers.

We identified only benchmark suite specifically compiled for SP research, from the Technical University of Munich~\cite{tum_benchmark_git}.
It includes basic algorithms, other small toy programs, and programs automatically created by Tigress.
Despite this benchmark suite being presented at an A-ranked conference in December 2016, only \samplesTumPercentage{}\%~$\hat{=}$~\samplesTum{}/\samplesetsBenchmarks{} papers from the surveyed corpus use it~\cite{2022_cadecff_compiler_agnostic_deobfuscator_of_control_flow_flattening,2020_similarity_features_for_the_evaluation_of_obfuscation_effectiveness,2020_semantics_aware_obfuscation_scheme_prediction_for_binary,2019_deobfuscating_android_native_binary_code,2017_predicting_the_resilience_of_obfuscated_code_against_symbolic_execution_attacks_via_machine_learning,2016_code_obfuscation_against_symbolic_execution_attacks}, including four papers originating from other research institutions (with disjoint authors). This is surprisingly low, in particular considering that \samplesetsToyAfterSeventeen{} papers published in the period 2018--2022, when everyone had long had the opportunity to learn about this obfuscation benchmark suite, still rely on other, non-standardized toy programs.

The widespread use of samples from performance benchmark suites is unsurprising, as they serve well for measuring the performance cost of SPs. However, the fact that 45 goodware papers rely solely on sample programs from performance-oriented benchmarks and from a benchmark suite consisting only of toy programs, clearly constitutes a benchmark representativeness problem.

\bemph{In conclusion, the community is missing opportunities to standardize their evaluation methodologies, and there is the lack of standardized, representative benchmarks for goodware SP research.}

\paragraph{Malware}
Of all sample sets, \samplesetsMalwarePercentage{}\%~$\hat{=}$~\samplesetsMalware{}/\haveSampleset{} include malware samples.
Sample sets containing malware samples are hence on average the largest sets. The reason is straightforward:

\bemph{Access to malware samples is easy with several publicly available malware repositories.} The Drebin~\cite{drebin_download} dataset with 5,560 samples from the years 2010--2012 is used most often in \samplesetsDrebinPercentage{}\%~$\hat{=}$~\samplesetsDrebin{}/\samplesetsMalware{} of the papers, followed by MalGenome~\cite{malgenomeproject} (1,200 samples from the years 2010--2011) with \samplesetsMalgenomePercentage{}\%~$\hat{=}$~\samplesetsMalgenome{}/\samplesetsMalware{} and Contagio~\cite{contagio_blogspot} with \samplesetsContagioPercentage{}\%~$\hat{=}$~\samplesetsContagio{}/\samplesetsMalware{}.
Other used repositories are VirusShare.com~\cite{virusshare}, in which almost 50M malware samples are freely available, at \samplesetsVirussharePercentage{}\%~$\hat{=}$~\samplesetsVirusshare{}/\samplesetsMalware{}; and AndroZoo~\cite{androzoo} with 20M samples of goodware and malware Android APKs, at \samplesetsAndrozooPercentage{}\%~$\hat{=}$~\samplesetsAndrozoo{}/\samplesetsMalware{}.

\paragraph{Mobile apps}
\samplesetsMobileappsPercentage{}\%~$\hat{=}$~\samplesetsMobileapps{}/\haveSampleset{} sample sets include mobile app samples.
Similar to malware, mobile app samples also contribute to significantly larger sample sets than those from other sample categories. 

\bemph{There is a vast difference in the use of Android samples and iOS apps.} 
The easy, direct availability of Android mobile apps enables their use as evaluation samples. These apps can be batch-downloaded from various app stores in the form of APK files~\cite{geiger2018datasets}. This is done for goodware research, but also for malware detection/classification research, in which sample sets need to contain malicious as well as benign samples. 
By contrast, access to iOS app samples is much more restricted: there exists only one official app store for iOS from which apps can be obtained, and apps from the store are DRM protected and encrypted.
Sharing sample sets with iOS apps is therefore both technically and legally challenging.
Only two SP papers include iOS samples~\cite{2018_protecting_million_user_ios_apps_with_obfuscation_motivations_pitfalls_and_experience,2018_software_protection_on_the_go_a_large_scale_empirical_study_on_mobile_app_obfuscation}.

\paragraph{Complex programs}
Complex programs are included in \samplesetsComplexPercentage{}\%~$\hat{=}$~\samplesetsComplex{}/\haveSampleset{} samples set.
In this category, we collect all samples which are more complex than toy programs and do not fall into any of the other categories.
These programs include well-known programs from various domains:
compression or archiving programs (\samplesComprprogsPercentage{}\%~$\hat{=}$~\samplesComprprogs{}/\samplesetsComplex{} papers), games (\samplesGamesPercentage{}\%~$\hat{=}$~\samplesGames{}/\samplesetsComplex{}),
chat and instant messaging applications (\samplesMessagingprogsPercentage{}\%~$\hat{=}$~\samplesMessagingprogs{}/\samplesetsComplex{}),
media processing software (\samplesMediaprogsPercentage{}\%~$\hat{=}$~\samplesMediaprogs{}/\samplesetsComplex{}), webservers (\samplesWebserversPercentage{}\%~$\hat{=}$~\samplesWebservers{}/\samplesetsComplex{}),
browsers, (\samplesBrowsersPercentage{}\%~$\hat{=}$~\samplesBrowsers{}/\samplesetsComplex{}), and
databases (\samplesDatabaseprogsPercentage{}\%~$\hat{=}$~\samplesDatabaseprogs{}/\samplesetsComplex{}).
Some papers also used less known programs from public repositories (e.g., GitHub) in their sample set (\samplesPublicreposPercentage{}\%~$\hat{=}$~\samplesPublicrepos{}/\samplesetsComplex{}).

\bemph{We found no evidence of considerable re-use of complex sample. Moreover, the vast majority of papers do not mention the samples' specific version. Hence reproducibility is severely limited.}

A notable exception is observed in the \percentGoodwareHumanExperiments{}\%~$\hat{=}$~\totalGoodwareHumanExperiments{}/\totalGoodware{} goodware papers that present human experiments. Six of those do use the same two samples: a simple race car game and a chat client~\cite{2008_towards_experimental_evaluation_of_code_obfuscation_techniques,
    2009_the_effectiveness_of_source_code_obfuscation_an_experimental_assessment,
    2014_a_family_of_experiments_to_assess_the_effectiveness_and_efficiency_of_source_code_obfuscation_techniques,
    2014_an_other_exercise_in_measuring_the_strength_of_source_code_obfuscation,
    2018_programming_experience_might_not_help_in_comprehending_obfuscated_source_code_efficiently,
    2020_creative_manual_code_obfuscation_as_a_countermeasure_against_software_reverse_engineering}. These papers span 12 years, with the latter three cited papers being from different authors than the first three.

\paragraph{Other}
This category unites samples that are not executable programs, which are used in \samplesetsOtherPercentage{}\%~$\hat{=}$~\samplesetsOther{}/\haveSampleset{} of all samples sets.
It includes office documents, obfuscated expressions such as MBA expressions, malicious URLs, etc.

\bemph{In this category of samples, we do observe more reuse: MBA expression sample sets are shared and reused frequently}~\cite{2017_syntia_synthesizing_the_semantics_of_obfuscated_code,2021_search_based_local_black_box_deobfuscation_understand_improve_and_mitigate,2021_software_obfuscation_with_non_linear_mixed_boolean_arithmetic_expressions,xu2021boosting,2022_efficient_deobfuscation_of_linear_mixed_boolean_arithmetic_expressions,2022_program_synthesis_based_simplification_of_mba_obfuscated_malware_with_restart_strategies,2022_program_synthesis_based_simplification_of_mba_obfuscated_malware_with_restart_strategies,2020_qsynth_a_program_synthesis_based_approach_for_binary_code_deobfuscation}.
}

\paragraph{Unknown}
Finally, we included a category {unknown}, which are included in \samplesetsUnknownPercentage{}\%~$\hat{=}$~\samplesetsUnknown{}/\haveSampleset{} samples sets. We use this category for papers that clearly evaluate their contributions on executable software, but that lack specificity to enable us to categorize that software. For example, some papers only state they use Windows applications, or applications downloaded from GitHub.%

\subsection{Sample Set Sizes}
For each paper, we also counted two sample set sizes. The first is the \textit{total sample set size}, which is the total number of different samples used in the paper's evaluation, including variations of the same program, document, expression, etc. The second is the \textit{original sample set size}. This is the number of original programs/documents/expressions from which variations were generated and evaluated. In goodware research, for example, the original samples are typically the vanilla, unprotected programs. The total sample set then contains those originals, plus a number of protected versions of each of them. When authors downloaded samples from app stores or malware repositories, we counted each downloaded sample as an original sample, even if they belong to the same malware family or if they happen to be repackaged versions.

\changed{\bemph{The total sample set sizes range from a single sample to over two million.
Large sample sets over 4000 samples consist predominantly of the categories \textit{malware} and \textit{mobile apps}.}}
In addition, there exist a few large sample sets from samples of the \textit{other} category, such as large sets of MBA expressions.
Few papers use large sample sets from the remaining categories. For instance, \citeauthor{2016_metadata_recovery_from_obfuscated_programs_using_machine_learning}~\cite{2016_metadata_recovery_from_obfuscated_programs_using_machine_learning} have automatically generated a large number of toy samples with the help of the \textit{Tigress} obfuscator to create a sample set of over 11,000 samples.

\changed{\bemph{By contrast, one third of all evaluations (36\%) feature small total sample sets of at most 20 samples.}}
Eleven evaluations are based on only a single sample, which originated from one of four sample categories: \textit{complex programs} (4), \textit{toy programs} (3), \textit{malware} (3), and \textit{unknown} (1).

\begin{figure}[t]
    \centering
    \includegraphics[width=0.99\textwidth]{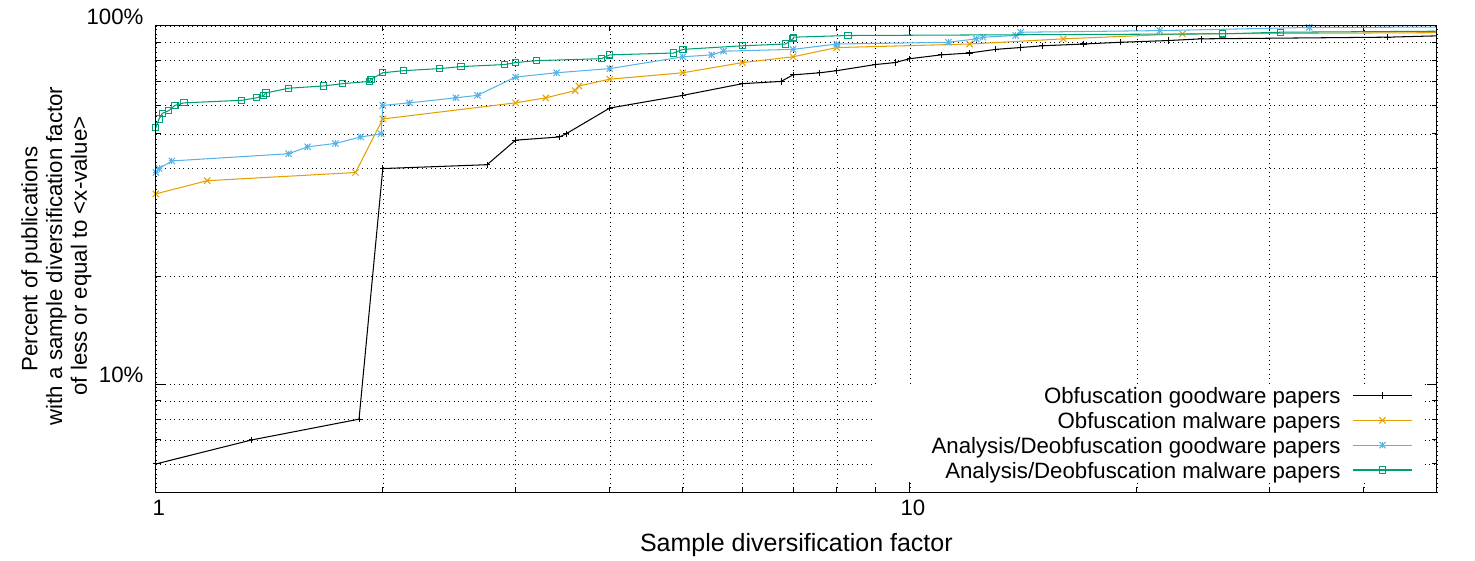}
    \caption{Cumulative graph of ratio between total and original sample set sizes. The x-axis displays the sample diversification factor, the y-axis the percentage of papers lower or equal to the current x-value.}\label{fig:samplediversification}
    \Description[TODO]{TODO add a long Description for visually challenged readers.}
\end{figure}

We further analyzed the ratio between the total sample set size and the original sample set size (i.e., how many variations of the original samples were generated).
\Cref{fig:samplediversification} shows the ratios between the two sample set sizes for all four combinations of goodware/malware and obfuscation/deobfuscation-analysis papers.
Unsurprisingly, a majority (51\%) of the malware analysis papers has a diversification factor of 1: these paper only evaluate original malware samples and original benign apps, but no variations of either of them, to evaluate malware detection rates.
This number is in line with \MalwareOnlyOfMalwarePercentage{}~$\hat{=}$\totalMalwareOnly{}/\totalMalware{} of the malware papers being malware only papers.
Many malware papers, in particular the ones focusing on techniques that can also be used on goodware, also evaluate their techniques on samples they built themselves, and then evaluate more than one version, such as an unprotected one and a protected one.
That is why the green line gradually rises from 51\% to 100\%.
Noteworthy is the big step on the black curve at two for goodware obfuscation papers.
9\% of the papers feature a diversification factor of less than 2, meaning that they do not even consider perform measurements on two versions (unprotected vs.\ protected) of all of their samples.
An additional 31\% of the goodware obfuscation papers only reaches a diversification factor of 2, which in almost all cases corresponds to one unprotected and one protected version for each sample.

\changed{\bemph{This means that 40\% of the goodware obfuscation papers do not evaluate diversified protection at all.}}
In other words, they do not evaluate the sensitivity of protection strength or costs to configuration parameters or to random seeds used in stochastic approaches, such as the parameters or seeds that might determine precisely where a particular protection is injected, which might be in a hot (i.e., frequently executed) part of a program or in a cold part, or the parameters that determine how sparsely or densely some transformation is deployed.
We find this a major shortcoming, that plagues too many papers in this research domain.

\changed{\bemph{For malware obfuscation research, where authors typically also generate their own samples, the situation is even worse, with close to 70\% of those papers not evaluating diversified protection.}}

\subsection{Correlation with Publication Venue}
\label{sub:venue_ranking-sample_set_size}
We analyzed the relation between the quality of the evaluations (number of samples, number of different sample categories, sample complexity and diversification) and the rankings of the venues at which the works where published.
For malware papers, we did not make interesting observations.

\changed{\bemph{For goodware papers, we did observe clear correlations between publication venue quality and sample set quality and size.}}
In this category, the total number of samples used in an evaluation of obfuscations correlates with the ranking of the venue.
The median sample set sizes go from 51 for A* down to 21 for C-rated venues.
\Cref{sub:venue_ranking-sample_set_size-supp} in the supplemental material analyses this relation in more detail.
In addition, we observed that the share of toy programs is significantly higher in evaluations of goodware publications in the C, unranked, and workshop categories (34--48\%) compared to B-ranked or better (20--25\%) venues.

\changed{\bemph{The lack of diversification, in particular in papers that present novel obfuscation, is not limited to lower-rated publication venues. To the contrary: the problem appeared almost as frequently in A*/A papers as it occurred in other papers.}}

\subsection{Identified challenges}
In our analysis of sample sets, we identified three relevant challenges.
\begin{itemize}
    \item \textbf{Lack of real programs in the evaluations:} \changed{Simple toy samples are omnipresent in SP research. Samples from other categories such as the GNU core utilities are also of low complexity. Such samples seem unsuitable for measuring the strength of a SP or its applicability in real software, which typically is much more complex.} The only exception are local protections such as MBA which can be evaluated independently of the program which they are intended to protect. However, our data clearly shows that even for non-local protections and analysis methods, often low-complexity samples are used. In top publication venues, this issue occurs less frequently, but still frequently enough to be worrisome.
    \item \textbf{Limited sample availability:} \changed{Depending on the sample set category}, reproducibility in SP research is most often low, including in top publication venues. In contrast to other research domains such as machine learning, where large datasets are publicly available, in SP research no commonly-used sample sets exit. This makes it difficult to reproduce results and to compare different approaches. For toy programs, many different implementations are used (rather than the ones available in a benchmark suite~\cite{2016_code_obfuscation_against_symbolic_execution_attacks}), and of complex programs the used versions are most often not specified, all of which makes it difficult to compare, interpret, and reproduce results. For OS programs, we observe more reuse, primarily due to the easy availability of the GNU core utilities. Another exception is recent work on MBA, where reuse of expressions exists (e.g., the public sample set of Syntia~\cite{2017_syntia_synthesizing_the_semantics_of_obfuscated_code}).
    \item \textbf{Limited diversification:} Evaluations by and large neglect diversification of the protection deployment. This issue plays in particular in goodware papers that present novel obfuscations, including in those published in A*/A venues.
\end{itemize}

To tackle these challenges, we advocate for a collaborative compilation of open sample sets within the research community.
A publicly available set of samples that takes into account the different motivations and goals for protections would assist researchers in evaluating new protections, contribute to the validity of research, and improve the comparability and reproducibility of research results.
It should furthermore be of sufficient size to increase its credibility for A*-rated conferences.
(If necessary, authors can then still pick a small subset of this sample set to do exploratory research.)
Moreover, the standardized sample set should come with guidelines on how to diversify the deployment of protections on the samples, e.g., by specifying small, medium, and large sets of functions in the samples that are candidates for obfuscation, similar to how the performance benchmark suites include testing, training, and references inputs.

\section{Sample Treatment}\label{sec:treatment}

The protected samples discussed above have been compiled and built using certain compilation tools. They feature combinations of protections that were deployed by protection tools that transformed one or more representations of those samples. In addition, the last phase of the samples' treatment is the application of code analysis, deobfuscation, and classification techniques on them. This section analyzes the representations of the samples that were used to deploy protections, the tools that were used to do so, the protections that were deployed with those tools, and the analysis methods used to reverse a protection and to reduce its protection potency and/or resilience.

\subsection{Protection Code Representation}\label{sec:obfuscation_representation}
\changed{For categorizing papers according to the code representation (i.e., format) on which the studied protection transformations are applied, we used similar categories as in \cref{sec:top_level_categories} for the types of programming languages.
  First, \emph{Source code} can be rewritten to inject protections, e.g., with the C-to-C rewriter Tigress.
  Secondly, \emph{Native code} in the form of assembler of binary object code can be transformed, e.g., with link-time binary rewriters such as Diablo.
  Finally, protections can be deployed on \emph{Intermediate code} formats.
  These more symbolic formats can be either compiler intermediate representations such as LLVM's bitcode, or the bytecode distribution formats of managed language, such as Java bytecode or the C Intermediate Language (CIL).
}
\begin{figure}[t]
  \includegraphics[width=12cm]{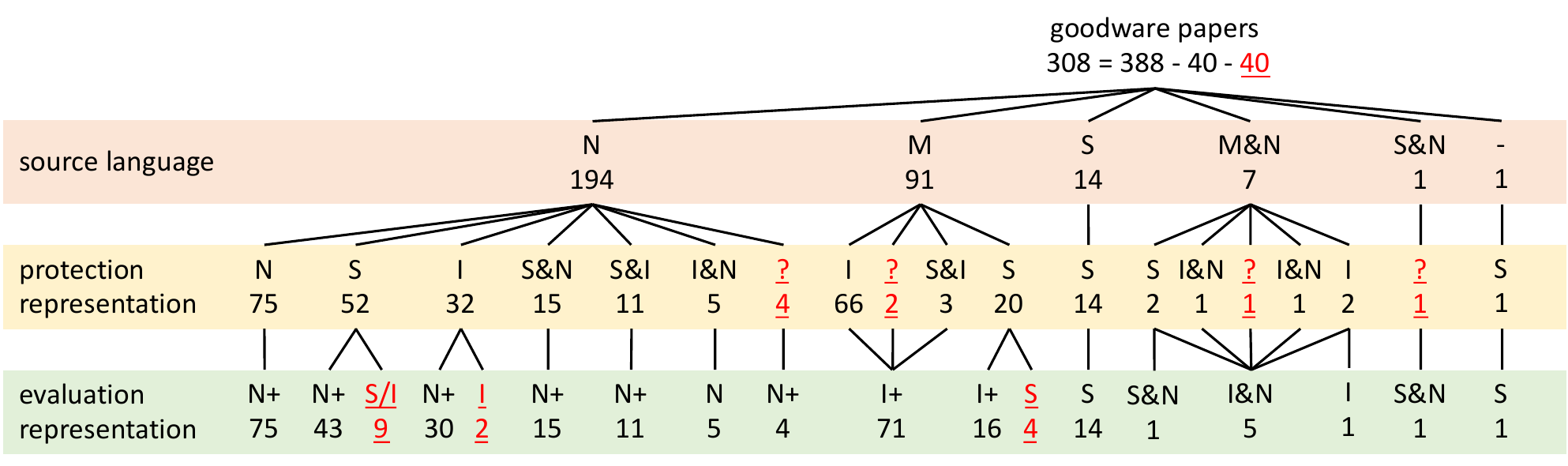}\\
  \vspace{0.5cm}
  \includegraphics[width=9.5cm]{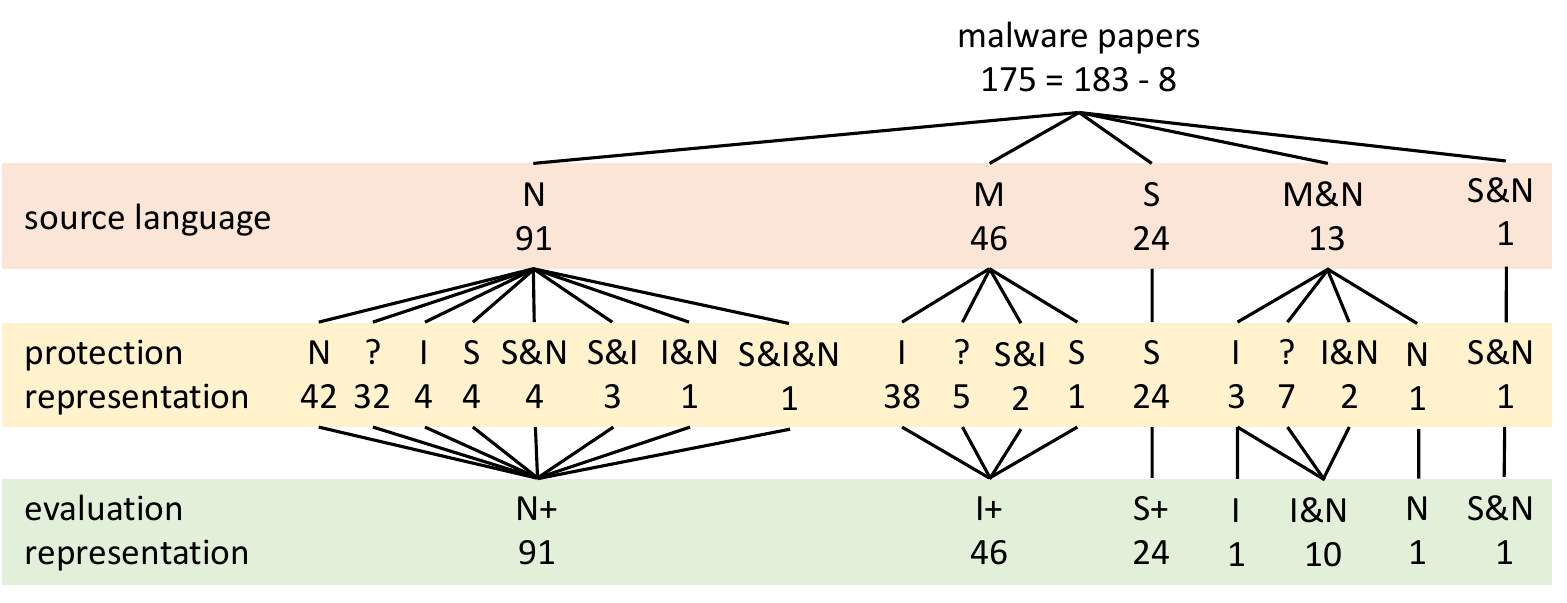}
  \caption{Sample protection and evaluation languages for goodware papers (top) and malware papers (bottom). M~=~\underline{M}anaged, N~=~\underline{N}ative, S~=~\underline{S}cript, I~=~\underline{I}ntermediate formats, ?~=~unknown, +~=~and more, \&~=~and, /~=~or, -~=~ language-agnostic. \changed{Red and underlined data indicate cases that we consider problematic.}}\label{fig:protection_languages_graph}
  \Description[Sample languages and representations for goodware and malware publications.]{TODO add a long Description for visually challenged readers.}
\end{figure}

\changed{
  For the \percentHaveSamplesetGoodware{}\%~$\hat{=}$~\haveSamplesetGoodware{}/\totalGoodware{} goodware and \percentHaveSamplesetMalware{}~$\hat{=}$~\haveSamplesetMalware{}/\totalMalware{} malware papers that report measurements on protected samples generated by transforming some code format, the middle bar in \cref{fig:protection_languages_graph} displays how many papers apply protections on the different formats, per the different programming language types. This section discusses only the most striking observations.

  First, 10\%~$\hat{=}$~40/\totalGoodware{} of the goodware papers are not considering any specific language or protection code representation.
  These include theoretical papers (e.g., on using  abstract interpretation to model protection strength), surveys, and a few papers on aspects of SP tools that do not need to be evaluated on samples, such as making a tool's transformation transparent for maintainers and users~\cite{2019_epona_and_the_obfuscation_paradox_transparent_for_users_and_developers_a_pain_for_reversers}.

  \bemph{Another 10\%~$\hat{=}$~40/\totalGoodware{} of the goodware papers lack an evaluation on samples, despite the fact that 36 (90\%) of them do present novel obfuscation techniques their authors claim to be practical.}
  This is a rather negative result that clearly shows how the field of goodware software protection often fails to maintain the highest standards.
  In malware research, that problem does not occur.
  The 4.4\%~$\hat{=}$~8/\totalMalware{} of malware papers that do not feature an evaluation on protected samples are all surveys and theoretical papers.

  \bemph{\percentHaveSamplesetGoodwareNoObflang{}\%~$\hat{=}$~\haveSamplesetGoodwareNoObflang{}/\haveSamplesetGoodware{} of the goodware papers fail to clarify which code format is being transformed for their evaluation, and hence lack this critical information for reproducibility.}
  On the malware side, \percentHaveSamplesetMalwareNoObflang{}\%~$\hat{=}$~\haveSamplesetMalwareNoObflang{}/\haveSamplesetMalware{} of the papers similarly do not report a representation used to obfuscate the samples.
  This is to be expected and okay, however, as malware papers typically use real-world malware samples which can be found online in various datasets for reproducing the research results, even if their provenance is not documented.

  \bemph{Most importantly, goodware papers targeting natively compiled languages understudy the composition and layering of protections applied at different code representations, i.e., at different levels of abstraction.}\footnote{Multiple protections to be composed/layered if they are deployed together in at least one sample. Papers never applying multiple protections together on at least a sample are hence not considered to be composing or layering protections.}
  Transformations on more abstract program representations such as source code early on in the build process can protect more abstract properties (e.g., invariants) and can be selected based on more abstract program features such as type information.
  Moreover, because the injected protections' code is then compiled together with the application code, it can be integrated more stealthily.
  The freedom to transform code is then restricted, however, by the conventions to which the higher-level representations need to adhere (language or IR specification compliance).
  Complementary thereto, obfuscating transformations applied on lower-level code formats can completely break the conventional mapping between higher-level software engineering constructs and lower-level software artifacts to mitigate decompilation and human comprehension and to prevent, e.g., that disassemblers produce correct CFGs.
  The detailed analysis presented below shows that such complementarity has hardly been researched despite its importance~\cite{2008_a_compiler_based_infrastructure_for_software_protection}.

  First, only \percentHaveSamplesetGoodwareSourcenativeObfsn{}\%~$\hat{=}$~\haveSamplesetGoodwareSourcenativeObfsn{}/\haveSamplesetGoodwareSourcenativeMeasuresomething{} goodware papers targeting natively compiled languages consider both source-level and binary-level transformations (S\&N).
  Among them are two papers that only compare the strength of their source-level transformation against other binary-rewriting-based alternatives~\cite{2022_loki_hardening_code_obfuscation_against_automated_attacks,2016_translingual_obfuscation}.
  Among the five papers that actually compose source-level and native-level transformations, one paper evaluates return-oriented-programming obfuscation through binary rewriting on programs already obfuscated at source-level by Tigress~\cite{2021_hiding_in_the_particles_when_return_oriented_programming_meets_program_obfuscation}.
  The other four papers originate from the ASPIRE research project in which both source-level and binary-level transformations are composed on the same samples~\cite{2020_code_renewability_for_native_software_protection,2019_a_meta_model_for_software_protections_and_reverse_engineering_attacks,2017_towards_optimally_hiding_protected_assets_in_software_applications,2016_tightly_coupled_self_debugging_software_protection}.
  The remaining eight papers all have a deobfuscation/analysis perspective and rely on samples that are either protected with source-level obfuscators such as Tigress or with binary-level tools such as VMProtect or Themida, but never with both.
  We thus identified only two research efforts that layered source-level and native-level obfuscations on C/C++ goodware.

  All \percentHaveSamplesetGoodwareSourcenativeObfsm{}\%~$\hat{=}$\haveSamplesetGoodwareSourcenativeObfsm{}/\haveSamplesetGoodwareSourcenativeMeasuresomething{} papers that consider both source-to-source rewriting and compile-time transformations (S \& I) for native languages use Tigress and Obfuscator-LLVM to generate samples~\cite{2022_fine_grained_obfuscation_scheme_recognition_on_binary_code,2021_machine_learning_classification_of_obfuscation_using_image_visualization,2021_input_output_example_guided_data_deobfuscation_on_binary,2020_semantics_aware_obfuscation_scheme_prediction_for_binary,2020_power_profiling_and_analysis_of_code_obfuscation_for_embedded_devices,2020_a_framework_for_evaluation_of_software_obfuscation_tools_for_embedded_devices,2019_fine_grained_static_detection_of_obfuscation_transforms_using_ensemble_learning_and_semantic_reasoning,2019_defeating_opaque_predicates_statically_through_machine_learning_and_binary_analysis,2019_asm2vec_boosting_static_representation_robustness_for_binary_clone_search_against_code_obfuscation_and_compiler_optimization,2018_combining_obfuscation_and_optimizations_in_the_real_world,2016_code_obfuscation_against_symbolic_execution_attacks}.
  Only two of those eleven papers compose protections of the two tools on the same samples~\cite{2018_combining_obfuscation_and_optimizations_in_the_real_world,2021_machine_learning_classification_of_obfuscation_using_image_visualization}.

  Of the \percentHaveSamplesetGoodwareSourcenativeObfmn{}\%~$\hat{=}$~\haveSamplesetGoodwareSourcenativeObfmn{}/\haveSamplesetGoodwareSourcenativeMeasuresomething{} papers that consider compiler-level and binary-level obfuscation, only two actually compose them:
  One composes compiler-based control flow obfuscation with assembly-level code layout randomization~\cite{2005_control_flow_based_obfuscation}, and one combines a high-level obfuscation based on multithreading implemented in a compiler with packing applied at the binary level~\cite{2021_model_of_execution_trace_obfuscation_between_threads}.

  \bemph{So overall, only 4.6\%~$\hat{=}$~9/\haveSamplesetGoodwareSourcenativeMeasuresomething{} of papers on native goodware from only five projects/teams explicitly target tool flows that deploy protections at multiple software abstraction levels. This gap between published research results and best practices in industry~\cite{2008_a_compiler_based_infrastructure_for_software_protection} is a major shortcoming of this field.}
}

\subsection{Deployed Protections}\label{sec:protections}
\changed{
  For each of the papers, we collected the types of protective transformations and features thereof such as relying on aliasing, that were deployed or theorized about beyond merely discussing them as related work.
  We classified them in \numberProtections{} different classes, of which short descriptions are provided as supplemental material in \cref{sec:protection_definitions}, where we also discuss some subtleties of our information gathering process.
  Our classification was derived from previous work by \citeauthor{1997_a_taxonomy_of_obfuscating_transformations}~\cite{1997_a_taxonomy_of_obfuscating_transformations}, and from three surveys~\cite{2016_protecting_software_through_obfuscation_can_it_keep_pace_with_progress_in_code_analysis,2018_diversification_and_obfuscation_techniques_for_software_security_a_systematic_literature_review,2021_measuring_software_obfuscation_quality_a_systematic_literature_review}.

  For each protection, we distinguish between theoretical considerations as found in surveys or qualitative security analyses on the one hand and practical implementations that actually get deployed on samples on the other hand.
  In the remainder of this section, we only consider the practical implementations.
  For the interested reader, \cref{sec:protection_crosstabs} in the supplemental material presents a number of tables with detailed counts of occurrences of protection combinations.

  \bemph{The popularity (i.e., the relative deployment frequency) of the considered obfuscation classes differs between malware and goodware and between different types of programming languages.}

  Overall, data encoding/encryption (\CTtotalDENC{}) is the most frequently used SP technique, followed by identifier renaming (\CTtotalIREN{}), junk code insertion (\CTtotalJCI), opaque predicates (\CTtotalOP{}), code diversity (\CTtotalCDIV{}), and packing/encryption (\CTtotalENC{}).
  Some often-studied SPs such as data encoding/encryption, packing/encryption, junk code insertion, and also repackaging are more popular for malware than for goodware.
  Vice versa, opaque predicates are relatively more frequently researched in goodware than in malware, as are control flow flattening, class-based transformations, data flow transformations, aliasing-based techniques, MBA, code mobility, loop transformations, server-side execution and hardware-supported obfuscation.
  This is in line with the distinct goals for which SPs are used: for preventing reverse engineering, code comprehension, and tampering in goodware; and for preventing detection and classification of malware.

  Few protections are popular for all types of programming languages:
  Only data encoding/encryption, junk code insertion, code diversity, packing/encryption, control flow transformation, and function transformations are all deployed in 10\% or more of the papers on each type of language.
  For the interested reader, \cref{sec:protection_crosstabs} in the supplemental material presents a more extensive analysis of which protections are popular for which languages, complementary to the above.

  \bemph{Aliasing is rarely exploited (explicitly) in the deployed protections, even though this is an SP feature that is often cited as increasing the resilience of protections due to alias analysis being an intractable problem~\cite{1998_manufacturing_cheap_resilient_and_stealthy_opaque_constructs}.}

  Only \percentPracticalPapersUsingAliasing{}\%~$\hat{=}$~\practicalPapersUsingAliasing{}/\practicalPapers{} of papers that deploy protections exploit aliasing to complicate analyses and to make analysis results less precise.
  This feature is much more popular in managed languages than in natively compiled, let alone script languages.
  The reason is that the computation of precise call graphs for Java bytecode depends heavily on type inference and points-to set analysis.
  Various transformations have been presented to increase points-to set sizes and to hamper type inference and call graph reconstruction.
  Most script languages are dynamically typed, and no C++-specific obfuscations exist. Hence, transformations that aim for points-to set increases are not researched for those types of languages.
  What remains is the use of conventional pointer aliasing to hamper data and control flow analysis, but very few papers target this.

  \bemph{For most protections, we identified a balance between papers that present protections from an obfuscation perspective vs.\ papers that consider and evaluate them from an analysis/deobfuscation perspective, meaning that comparable attention is given to those protections from the two perspectives.}
  However, we also found that some protections are much more likely to be studied from the perspective of protection than deobfuscation or analysis, and vice versa.
  The biggest imbalance towards a protection perspective can be seen for hardware-assisted protection (\practicalPapersObfUsingHardware{}:\practicalPapersAnaDeobfUsingHardware{}).
  While ten papers present and implement new hardware-assisted protection techniques, none target deobfuscation or analysis of this type of protection.
  Other techniques with a big imbalance towards the obfuscation perspective are code mobility (\practicalPapersObfUsingCodemob{}:\practicalPapersAnaDeobfUsingCodemob{}), aliasing (\practicalPapersObfUsingALI{}:\practicalPapersAnaDeobfUsingALI{}), server side execution (\practicalPapersObfUsingSSE{}:\practicalPapersAnaDeobfUsingSSE{}), data-to-code conversion (\practicalPapersObfUsingDtC{}:\practicalPapersAnaDeobfUsingDtC{}), and class-based transformations (\practicalPapersObfUsingCBT{}:\practicalPapersAnaDeobfUsingCBT{}).
  Large imbalances towards deobfuscation and analysis occur less often.
  The biggest one is observed for repackaging (\practicalPapersObfUsingRPA{}:\practicalPapersAnaDeobfUsingRPA{}).
  Some of the imbalances can be explained by taking into account that the number of defensive papers (\totalDefensive{}) in our survey is much higher than the number of offensive papers (\totalOffensive{}), and the fact that certain protections are used relatively more frequently for malware than for goodware or vice versa.
  Notable exceptions are virtualization, which is included \defensiveVIR{} times in defensive studies and \offensiveVIR{} times in offensive studies, and MBA is included \defensiveMBA{} times in defensive research, and \offensiveMBA{} times in offensive research.
  We conjecture that this is due to the fact that these obfuscations were at some point considered among the strongest ones in the past and widely deployed in practice, which made them both relevant and challenging cases for (academic) offensive research.

  \bemph{Few papers explicitly layer protections on each other. SP research, in particular for goodware protection, is clearly lacking in this regard.}
  While quite some papers consider and evaluate multiple classes of protections, most of them deploy the different protections on different samples, e.g., to compare the protections' strength, rather than evaluating their combined/layered deployment within the same samples.
  For both goodware and malware, about one quarter of the papers only deploy a single protection.
  Moreover, about \percentHaveLessThanFourProtectionsXGoodware{}\%~$\hat{=}$~\haveLessThanFourProtectionsXGoodware{}/\totalGoodware{} of all goodware papers and about \percentHaveLessThanFourProtectionsXMalware{}\%~$\hat{=}$~\haveLessThanFourProtectionsXMalware{}/\totalMalware{} of all malware papers deploy three or fewer protections, and about \percentHaveLessThanSixProtectionsX{}\%~$\hat{=}$~\haveLessThanSixProtectionsX{}/\totalPapers{} deploy five or fewer protections, on mostly disjunct sets of samples as discussed.
  On average, few protections are hence considered in the surveyed papers.
  Slightly more are considered in papers from A*/A-rated venues, but the difference to lower-rated venues is small.
  This is the case despite the common wisdom that strong protection can only be obtained if multiple protections are combined and layered.
  For example, commercial tools such as Cloakware~\cite{cloakware} and Dexguard~\cite{dexguard} support and explicitly recommend the combination of multiple different protection techniques in so-called protection recipes in their user manuals and whitepapers in order to mitigate various attack strategies.

  \bemph{Only in a small minority of the papers, contributions get evaluated in protected samples that are representative of real-world SP usage.}
  This lack of mature evaluation is a shortcoming for deobfuscation and analysis papers that are evaluated on unrealistically weak protected samples, as well as for obfuscation papers, many of which apparently do not evaluate the complementarity of their contributions over existing ones.
  In particular, in goodware research, where the authors create protected samples themselves and thus can provide clear descriptions of the deployed protections, the numbers are appalling.
  Obviously, not all papers that limit the evaluation to a single or few protections are problematic.
  In particular, for all fundamental research at low technology readiness levels (TRLs), it can be perfectly reasonable to evaluate techniques on unrealistically simple use cases.
  Our numbers do indicate, however, that the field of goodware SP is running behind in demonstrating that the presented techniques are capable of achieving a high TRL.

  Two protections stand out as being deployed in isolation in analysis/deobfuscation papers: virtualization for goodware and packing/encryption for malware.
  As for the latter, this is mostly due to the papers not containing all ground-truth / provenance information:
  Many malware papers mention that their real-world malware samples are packed, either because they specifically target packed malware or because an unpacking step is included in their tool flow, but do not mention any other techniques that are deployed on those samples because the authors do not know which ones have been applied and because their analysis is (sufficiently) orthogonal to other SPs.

  As for the high frequency with which virtualization is studied in isolation in goodware analysis/deobfuscation papers, there are certainly a number of authors that simply have opted to target programs protected with only virtualization.
  But to some extent, authors also were forced to do so: on natively compiled code, the early virtualization tools, both academic (e.g., based on Strata) and commercial, transformed binary code.
  For the commercial ones, such as VM protect, it is not obvious which additional SPs they deploy on top of virtualization, and hence it is hard to report on those in papers.
  Moreover, binary rewriting tools are not easily combined with other obfuscation tools to layer SPs.
  Up to 2015, this definitely impacted research.

  From 2015 onwards, this changed.
  Of the goodware papers on natively compiled code, \percentHaveSamplesetGoodwareSourcenativeMeasuresomethingDeployVirtUseTigress{}\%~$\hat{=}$~\haveSamplesetGoodwareSourcenativeMeasuresomethingDeployVirt{}/\haveSamplesetGoodwareSourcenativeMeasuresomething{} deploy virtualization, of which \percentHaveSamplesetGoodwareSourcenativeMeasuresomethingDeployVirtUseTigress{}\%~$\hat{=}~$\haveSamplesetGoodwareSourcenativeMeasuresomethingDeployVirtUseTigress{}/\haveSamplesetGoodwareSourcenativeMeasuresomethingDeployVirt{} use Tigress, the source-rewriting obfuscator that allows one to compose virtualization with a range of other SPs and became available around 2015.
  Of those \haveSamplesetGoodwareSourcenativeMeasuresomethingDeployVirtUseTigress{}~papers, only two are virtualization-only papers, one of which is a Systematization of Knowledge paper on automated virtualization deobfuscation~\cite{2021_sok_automatic_deobfuscation_of_virtualization_protected_applications}.
  In other words, once the necessary tool was available to compose virtualization with other SPs, researchers effectively started doing so.

  \bemph{This example of how tool availability improved the research methodology stresses the need for reusable tools in SP research}, an aspect that will be discussed further in \cref{sec:used_tools}.
}

\subsection{Employed Analysis Methods}\label{sec:analysis_methods}
\changed{Similar to how we analyzed the papers' deployment of protections, we classified their use of code analysis methods. These are used by MATE attackers and malware analysts in practice and can hence be used by researchers to evaluate their obfuscations' practical strength.
  In total, we considered \numberAnalyses{} different methods and features thereof, which were again derived from previous work~\cite{1997_a_taxonomy_of_obfuscating_transformations,2016_protecting_software_through_obfuscation_can_it_keep_pace_with_progress_in_code_analysis,2018_diversification_and_obfuscation_techniques_for_software_security_a_systematic_literature_review,2021_measuring_software_obfuscation_quality_a_systematic_literature_review}.
  A list of all analysis methods and short descriptions is provided as supplemental material in \cref{sec:analysis_definitions}.
  In the collected data, we again differentiate between theoretical discussions and actual deployment on samples.
  Here, we only report on the latter.

  \bemph{A major observation is that more than half (\percentPracticalPapersGoodwareObfuscationHaveNoAnalysis{}\%~$\hat{=}$~\practicalPapersGoodwareObfuscationHaveNoAnalysis{}/\practicalPapersGoodwareObfuscation{}) of the goodware obfuscation papers that deploy protections on samples do not evaluate those protections' strength by assessing how analysis methods fare against them.}
  Moreover, \percentPracticalPapersGoodwareObfuscationHaveOneAnalysis{}\%~$\hat{=}$~\practicalPapersGoodwareObfuscationHaveOneAnalysis{}/\practicalPapersGoodwareObfuscation{} only evaluate one analysis method, leaving only \percentPracticalPapersGoodwareObfuscationHaveMoreThanOneAnalysis{}\%~$\hat{=}$~\practicalPapersGoodwareObfuscationHaveMoreThanOneAnalysis{}/\practicalPapersGoodwareObfuscation{} that evaluate more than one method.
  These numbers are inflated by lower-quality papers of lower-quality publication venues, but the observation is certainly not limited to such papers:
  \percentPracticalPapersGoodwareObfuscationHaveNoAnalysisAandAstar{}\%~$\hat{=}$~\practicalPapersGoodwareObfuscationHaveNoAnalysisAandAstar{}/\practicalPapersGoodwareObfuscationAAndAstar{} of the goodware obfuscation papers published in A*/A rated venues do not deploy any analysis methods.
  This is strikingly low, given that (layered) SP is all about mitigating the attack paths (plural) of least resistance, including attacks on the deployed protections.
  While it is perfectly acceptable for authors presenting a new SP to aim to mitigate only one existing attack strategy, we do think it deserves a recommendation to evaluate the impact of the SP on multiple attacks, in particular attacks on the new SP itself.
  In other words, beyond the potency of an obfuscation to hamper one attack strategy, also its resilience against other attacks~\cite{2019_understanding_the_behaviour_of_hackers_while_performing_attack_tasks_in_a_professional_setting_and_in_a_public_challenge} should be evaluated.
  This is clearly not done in papers in which no or only one analysis method is evaluated.

  In the obfuscation malware papers, the numbers are better, with \percentPracticalPapersMalwareObfuscationHaveAnalysis{}\%~$\hat{=}$~\practicalPapersMalwareObfuscationHaveAnalysis{}/\practicalPapersMalwareObfuscation{} of them reporting on at least one analysis method, but still \percentPracticalPapersMalwareObfuscationHaveOneAnalysis{}\%~$\hat{=}$\practicalPapersMalwareObfuscationHaveOneAnalysis{}/\practicalPapersMalwareObfuscation{} of them consider only one such method.

  In the analysis/deobfuscation papers, there, of course, is no issue of papers not evaluating analyses.
  The average number of analysis methods per paper (with at least one analysis method) is also significantly higher in those  papers (\averageAnaPerAnaOrDeobfWithAna{}) than in obfuscation papers (\averageAnaPerObfWithAna{}).

  Overall, control flow analysis (\totalCFANAonlyX{}), pattern matching (\totalPATMonlyX{}), tracing (\totalTRACEonlyX{}), statistical analysis (\totalSTATonlyX{}), and machine learning (\totalMLANAonlyX{}) are the top five  methods, the latter being a feature rather than an analysis.

  \bemph{Similar to what we noticed for deployed protections, we observe significant differences between the types of analysis methods used in obfuscation papers and in deobfuscation/analysis papers.}
  For example, pattern matching is used in \anaDeobfPATMonlyX{} analysis/deobfuscation papers.
  This clearly indicates that in some scenarios, this rather generic technique is considered a highly relevant alternative to more targeted analyses.
  In stark contrast, only \obfPATMonlyX{} papers that introduce new protections evaluate their strength against pattern matching. Similar observations can be made for statistical analysis (\obfSTATonlyX{}:\anaDeobfSTATonlyX{}) and machine learning (\obfMLANAonlyX{}:\anaDeobfMLANAonlyX{}). To some extent, these effects are driven by some analyses' predominance in  malware research. When we exclude malware papers from the analysis, the imbalance between obfuscation and deobfuscation/analysis papers is less pronounced. However, these observations still add to our concern as to whether protections are sufficiently being evaluated against path-of-least-resistance attack and analysis methods.

  As for pattern matching, in goodware papers the ratio is still \obfPATMonlyXonlyGW{}:\anaDeobfPATMonlyXonlyGW{}. \citeauthor{2016_protecting_software_through_obfuscation_can_it_keep_pace_with_progress_in_code_analysis}~\cite{2016_protecting_software_through_obfuscation_can_it_keep_pace_with_progress_in_code_analysis} previously observed that pattern matching is easily mitigated, even with very simple obfuscations. However, the scope of their observation is limited to the use of pattern matching to identify artifacts/assets in the original programs, excluding its use to identify deployed protections. In other words, they observed that even simple obfuscations can be \emph{potent} against pattern matching but did not discuss the \emph{resilience} of obfuscations against pattern matching-based deobfuscation. Our observation that pattern matching is among the more popular techniques in analysis/deobfuscation goodware papers confirms it is considered a viable technique to attack SPs, which makes it all the worse that pattern matching is hardly being evaluated in obfuscation goodware papers.

  \bemph{This confirms our earlier point on defensive goodware research lacking in evaluations of resilience.}

  A second observation by \citeauthor{2016_protecting_software_through_obfuscation_can_it_keep_pace_with_progress_in_code_analysis}~\cite{2016_protecting_software_through_obfuscation_can_it_keep_pace_with_progress_in_code_analysis} was that dynamic analyses are significantly more effective against dynamic protections than static analyses.
  They defined static protections as ones that execute exactly the code present in the static binary, while dynamic protections perform additional code transformations at run time (e.g., unpacking).
  Following up on their observation and to verify whether it still applies to our more recent data, we analyzed how virtualization, the most popular dynamic protection method in the surveyed papers, is being targeted by analyses.
  Our findings indicate that against virtualization, indeed, predominantly dynamic analyses such as trace-based techniques and symbolic execution are used, sometimes supplemented by human analysis.
  However, our data also reveals that static analysis techniques like lifting, control flow analysis, data flow analysis, normalization, and slicing are also frequently utilized in the analysis and deobfuscation of virtualization-based protections.
  A crosstab analysis (for which the data is available in the supplemental material in \cref{sec:suppl_analysis_deployment}) provides a robust explanation, namely that these static techniques are often used in conjunction with the identified dynamic techniques.
  For instance, symbolic execution is often combined with lifting, normalization, and data flow analysis.

  \bemph{We, therefore, conclude that the claim by \citeauthor{2016_protecting_software_through_obfuscation_can_it_keep_pace_with_progress_in_code_analysis}~\cite{2016_protecting_software_through_obfuscation_can_it_keep_pace_with_progress_in_code_analysis} about dynamic analyses techniques in the face of dynamic protections should be refined somewhat. Dynamic analyses, indeed, are stronger, but they seem to complement static analyses rather than replacing them.}

}

\subsection{Used Tools}
\label{sec:used_tools}

\newcommand{\totalIdaProPlugins}{21}
\newcommand{\totalBinDiffOnly}{7}
\newcommand{\totalIdaProPluginsNotJustBinDiff}{14} %
\newcommand{\totalgccbuild}{52}
\newcommand{\totalvsbuild}{14}
\newcommand{\totalclangbuild}{16}
\newcommand{\totalllvm}{73}
\newcommand{\totalllvmprotect}{57}
\newcommand{\totalllvmbuild}{71}
\newcommand{\totalllvmprotectsole}{22} %
\newcommand{\totaltigress}{37}
\newcommand{\ollvmextended}{7}
\newcommand{\virustotalpctversion}{8.3}
\newcommand{\totalvirustotal}{24}
\newcommand{\versionvirustotal}{2}
\newcommand{\idapctversion}{23}
\newcommand{\idatotal}{63}
\newcommand{\versionida}{14}
\newcommand{\tigresspctversion}{23}
\newcommand{\tigresstotal}{34}
\newcommand{\versiontigress}{8}
\newcommand{\tigressrecentpctversion}{28}
\newcommand{\tigressrecenttotal}{18}
\newcommand{\versiontigressrecent}{5}
\newcommand{\tigressConnectionWithAuthors}{3}
\newcommand{\ollvmConnectionWithAuthors}{1}

\changed{As already hinted above, the existence and availability of tools have an impact on how authors treat their samples.
  This section discusses the tool use in papers with sample sets in more detail, covering tools used for building, protecting, and analyzing samples. %
  Our analysis includes tools that are newly introduced in a paper, as well as tools that are re-used as is or built upon.
  For example, when a work deploys Obfuscator-LLVM (OLLVM) on samples, we count this towards OLLVM as well as towards the LLVM framework on which OLLVM builds.
  A caveat is that not all authors are specific and complete in describing the tools they build on. For the interested readers, \Cref{sec:suppl_tools} in the supplemental material provides additional information on the tools, including recommendations regarding their use for evaluating SPs.
}

\begin{table}[t]
\renewcommand{\arraystretch}{0.9}
\setlength\tabcolsep{2.4pt}
\footnotesize
\caption{The \totalToolsMoreThanFive{} tools used by more than five surveyed papers in their experimental evaluation. The columns A~=~\underline{A}nalysis, B~=~\underline{B}uild, and P~=~\underline{P}rotection list how many papers use a tool for which purposes. T~=~\underline{T}otal gives the number of publications using the tool. The \$~column indicates whether the tool is: \protect\priority{100}~=~free to use, \protect\priority{50}~=~has a demo version, \protect\priority{0}~=~paid version only. The symbols in column O describe the access to the source-code and mean \protect\priority{100}~=~open-source, \protect\priority{25}~=~upon request, \protect\priority{0}~=~closed-source. The F column describes how adjustable the tool is, meaning \protect\priority{100}~=~adaptable/extendable, \protect\priority{50}~=~plug-ins, \protect\priority{25}~=~configurable, \protect\priority{0}~=~rigid. Column History~=~Usage per year since 2012 (left to right).}\label{tab:tools}
\begin{subtable}[t]{.5\linewidth}
\centering
\begin{tabular}[t]{ p{0.25ex} l c c c c c c c c }
\multicolumn{2}{l}{\textbf{Tool}} & \textbf{History} & \textbf{\$} & \textbf{O} & \textbf{F} & \textbf{B} & \textbf{A} & \textbf{P} & \textbf{T} \\\toprule
\multicolumn{2}{l}{\textbf{LLVM}} & \begin{sparkline}{10} 
\sparkspike 1.0 0.9 
\sparkspike 0.9 0.9 
\sparkspike 0.8 1.2 
\sparkspike 0.7 0.9 
\sparkspike 0.6 0.8 
\sparkspike 0.5 0.8 
\sparkspike 0.4 0.6 
\sparkspike 0.3 0.8 
\sparkspike 0.2 0.4 
\sparkspike 0.1 0 
\sparkspike 0.0 0 
\spark 0.0 0 0.1 0 0.2 0 0.3 0 0.4 0 0.5 0 0.6 0 0.7 0 0.8 0 0.9 0 1.0 0 / 
\end{sparkline} & \priority{100} & \priority{100} & \priority{100} & 71 & 22 & 57 & 73 \\
\end{tabular}
 \end{subtable}%
 \begin{subtable}[t]{.5\linewidth}
     \centering
     \begin{tabular}[t]{ p{0.25ex} l c c c c c c c c }
\multicolumn{2}{l}{\textbf{Tool}} & \textbf{History} & \textbf{\$} & \textbf{O} & \textbf{F} & \textbf{B} & \textbf{A} & \textbf{P} & \textbf{T} \\\toprule
\multicolumn{2}{l}{\textbf{Themida}} & \begin{sparkline}{10} 
\sparkspike 1.0 0.2 
\sparkspike 0.9 0.2 
\sparkspike 0.8 0.2 
\sparkspike 0.7 0 
\sparkspike 0.6 0.3 
\sparkspike 0.5 0.1 
\sparkspike 0.4 0.1 
\sparkspike 0.3 0.1 
\sparkspike 0.2 0.1 
\sparkspike 0.1 0 
\sparkspike 0.0 0 
\spark 0.0 0 0.1 0 0.2 0 0.3 0 0.4 0 0.5 0 0.6 0 0.7 0 0.8 0 0.9 0 1.0 0 / 
\end{sparkline} & \priority{50} & \priority{0} & \priority{0} & --- & --- & 14 & 14 \\
 & \multicolumn{9}{>{\raggedright}p{6.4cm}}{\tiny GW:\cite{2022_loki_hardening_code_obfuscation_against_automated_attacks, 2022_chosen_instruction_attack_against_commercial_code_virtualization_obfuscators, 2020_vcf_virtual_code_folding_to_enhance_virtualization_obfuscation, 2020_symsem_symbolic_execution_with_time_stamps_for_deobfuscation, 2018_vodka_virtualization_obfuscation_using_dynamic_key_approach, 2018_identifying_input_dependent_jumps_from_obfuscated_execution_using_dynamic_data_flow_graphs, 2016_comparing_the_effectiveness_of_commercial_obfuscators_against_mate_attacks, 2014_a_framework_for_understanding_dynamic_anti_analysis_defenses} --- MW:\cite{2021_obfuscation_resilient_executable_payload_extraction_from_packed_malware, 2021_an_assessment_of_obfuscated_ransomware_detection_and_prevention_methods, 2018_vmhunt_a_verifiable_approach_to_partially_virtualized_binary_code_simplification, 2017_a_framework_for_recognition_and_confronting_of_obfuscated_malwares_based_on_memory_dumping_and_filter_drivers, 2015_symbolic_execution_of_obfuscated_code, 2009_automatic_reverse_engineering_of_malware_emulators}}\normalsize \\
\end{tabular}
 \end{subtable}%
 \end{table}

\changed{
  \bemph{In general, authors do not share or reuse research artifacts much.}
  At \totalTools{}, we counted more tools than papers, and \percentToolsOnce{}\%~$\hat{=}$~\totalToolsOnce{}/\totalTools{} of them are used in only one paper.
  Only the \percentToolsMoreThanFive{}\%~$\hat{=}$~\totalToolsMoreThanFive{}/\totalTools{} listed in \Cref{tab:tools} are used in at least five papers.
  We discuss several interesting aspects in more detail below.
}

\paragraph{Tool Popularity}
\changed{LLVM (\totalllvmbuild{} papers) and gcc (\totalgccbuild{}) are by far the most popular build tools. Clang (\totalclangbuild{}), Visual Studio (\totalvsbuild{}), and ACTC (7). Beyond those, the only additional compilers being mentioned are CompCert~\cite{2019_formal_verification_of_a_program_obfuscation_based_on_mixed_boolean_arithmetic_expressions,2016_formal_verification_of_control_flow_graph_flattening,2022_obfuscation_resilient_semantic_functionality_identification_through_program_simulation}, g++\cite{2008_implementation_of_an_obfuscation_tool_for_c_c_source_code_protection_on_the_xscale_architecture,2016_deviation_based_obfuscation_resilient_program_equivalence_checking_with_application_to_software_plagiarism_detection}, and tinycc and TenDRA~\cite{2022_obfuscation_resilient_semantic_functionality_identification_through_program_simulation}.

  \bemph{Too few authors mention the compilation tools they use to build their samples.}
  Only 126 papers provide this information, of which 111 target natively compiled code.
  So 45\%~$\hat{=}$~91/202 of the goodware papers that target at least natively compiled code do not mention which compiler they used to generate samples, let alone how their compiler was configured, e.g., with respect to the optimization level.
  Given how strongly optimized code can differ from non-optimized code, this lack of information in so many papers is astonishing and an important issue.

  Unlike gcc, which is only used once to deploy protections~\cite{2005_control_flow_based_obfuscation}, LLVM is frequently used for this purpose (\totalllvmprotect{} papers).
  Its more modular compilation pipeline design enables obfuscating IR transformations at various stages. The popular OLLVM obfuscator builds in this, (\totalToolsOLLVM{}), and so do other obfuscators developed on top of LLVM (\totalllvmprotectsole{}).
  While the clang compiler typically serves as a front-end for LLVM IR, it is also used to inject protections with source-to-source rewriting~\cite{2020_code_renewability_for_native_software_protection}.

  In addition, LLVM is used in the popular KLEE dynamic symbolic execution engine (16 papers). This demonstrates that core software analysis and transformation infrastructure can be re-used for building, protecting, and analyzing software. Other multi-purpose tools are the Java bytecode rewriting framework Soot (e.g., in the Dava decompiler (\totalToolsDava{} papers)), and Visual Studio and Eclipse, which are used for building programs as well as for dynamic analysis (debugging). Eclipse's code refactoring support is also used to protect programs~\cite{2008_an_inter_classes_obfuscation_method_for_java_program,2010_theory_and_practice_of_program_obfuscation}. Finally, beyond being used for compilation, gcc is used to diversify software (e.g., by compiling at multiple optimization levels), and as an attack tool to undo source-level obfuscations~\cite{2014_measuring_the_robustness_of_source_program_obfuscation_studying_the_impact_of_compiler_optimizations_on_the_obfuscation_of_c_programs,2021_search_based_local_black_box_deobfuscation_understand_improve_and_mitigate,2019_how_to_kill_symbolic_deobfuscation_for_free_or_unleashing_the_potential_of_path_oriented_protections,2021_program_protection_through_software_based_hardware_abstraction}.

  \bemph{LLVM and Tigress are currently the only popular protection tools.} The \emph{History} data in \cref{tab:tools} shows that other SP tools are declining in popularity: only a few recent papers still use ProGuard after an initial peak of activity; a similar fate has befallen Zelix Klassmaster, Sandmark, and Diablo.

  \bemph{\Cref{tab:tools} hints for two reasons for tool popularity: their origin and their low cost.}
  As for the cost, the only commercial SP tools used at least five times (VMProtect, Themida, Code Virtualizer, DashO, Allatori, and Zelix Klassmaster) all offer a demo version.
  All other SP tools in \cref{tab:tools} are freely available. Ten of them come from academia: LLVM, OLLVM, Tigress, Diablo, Soot, clang, KLEE, Sandmark, Obfuscapk, and DroidChameleon.
  Relatively absent are the \emph{advanced} commercial offerings from, e.g., Irdeto, Digital Ai (formerly known as Arxan), and GuardSquare.
  The latter offers the free but limited-functionality ProGuard and the commercial DexGuard.
  ProGuard is used in \totalToolsProGuard{} papers, DexGuard in only \totalToolsDexGuard{}.
  Of the other tools commonly used to protect software, VMProtect and Code Virtualizer only offer specific obfuscations, namely virtualization.
  Themida virtualizes and encrypts code fragments, and the injected execution engine embeds several techniques to prevent analysis, but it performs no further transformations of the original application code.

  The popularity of academic protection tools raises a question:
  Is their popularity due to their widespread adoption across the research community, or are their own authors prolific publishers?

  \bemph{A detailed analysis found that Tigress and OLLVM have seen widespread take-up in the community:} only \tigressConnectionWithAuthors{} out of \totaltigress{} Tigress papers have a connection with Tigress' developers, %
  and only \ollvmConnectionWithAuthors{} out of \totalToolsOLLVM{} OLLVM papers have a connection with the OLLVM authors.
  This is in stark contrast to Diablo, the third most popular academic protection tool, for which 18 out of 24 papers originate from within the team that developed Diablo.
  Another binary rewriter, PLTO, suffers a similar fate:
  It has only \totalToolsPlto{} uses in our dataset, %
  of which two originate from its own research group.
  We conjecture that this lack of reuse of (post-)link-time rewriters like Diablo and PLTO is, to a large, degree due to the difficulty of using and maintaining such binary-rewriting tools.

  \bemph{We conclude with the major problem that goodware research involves very little deployment of commercial-grade protection tools, even in papers in A*/A publication venues.} It implies that the strength of commercial SP software is obscured by a lack of publicly documented evaluations thereof.
  The root cause of this problem is, of course, that commercial protection tool vendors do not allow independent researchers to evaluate their tools and publish about it, e.g., by means of restrictions in their end-user license agreements.

  On the analysis tool side, by contrast, the most popular tool is commercial: the interactive binary disassembler IDA Pro.
  Being relatively cheap and having a free demo version might contribute to its popularity.
  The competition that has surfaced in recent years with Binary Ninja (released in 2016) and Ghidra (released in 2019) has not yet made much of a dent in IDA Pro's popularity:
  Ghidra was only used \totalToolsGhidra{} time in the surveyed papers, Binary Ninja \totalToolsBinaryninja{} times.
  The most popular academic binary code analysis tool is angr.
  In 2016, this tool has been put forward as a unifying binary analysis framework~\cite{angr}.
  In practice, however, in the period 2018--2022, well after the publication of angr (2016~\cite{angr}), IDA Pro was used in \totalToolsIdaProAfterTwentySeventeen{} papers, angr in \totalToolsAngrAfterTwentySeventeen{}, Triton in \totalToolsTritonAfterTwentySeventeen{}, Pin in \totalToolsPinAfterTwentySeventeen{}, etc.

  \bemph{So clearly, many different analysis tools and frameworks keep being used. There is no sign of overall unification or standardization in their use for evaluating the strength of SPs on native binaries.}

  When angr is used, it is most often for \changed{symbolic execution} of native code.
  KLEE is another tool for \changed{symbolic execution} of natively compiled code that has remained almost as popular as angr, with \totalToolsKleeAfterTwentySeventeen{}~deployments in the years 2018--2022.
  KLEE operates on LLVM IR code, which is sometimes obtained by lifting binary code~\cite{2020_similarity_features_for_the_evaluation_of_obfuscation_effectiveness,2020_on_the_automatic_analysis_of_the_practical_resistance_of_obfuscating_transformations}, but almost always by compiling source code.
  Fortunately, most authors seem to understand that one can question how accurately such KLEE results reflect real-world attacks on binary code, and at least from 2018 onwards, they all complement their KLEE results with experimental data obtained with angr, IDA Pro, or other binary analysis tools.
}

\paragraph{Tool Versions and Configurations}
\changed{Some tools have been in use for a long time.
  Some did not evolve significantly over that time, such as OLLVM, which saw only one release.
  Other tools evolved considerably, in which case reproducibility requires that the used versions are reported.

  \bemph{We observed that all too many papers lack in the reporting of used versions and configuration.}
  For example, for IDA Pro, \idapctversion{}\%~$\hat{=}$~\versionida{}/\idatotal{} papers specified the version number; for the source-to-source obfuscator Tigress, \tigresspctversion{}\%~$\hat{=}$~\versiontigress{}/\totaltigress{} did so.
  For Tigress, the situation is improving over time, as \tigressrecentpctversion{}\%~$\hat{=}$~\versiontigressrecent{}/\tigressrecenttotal{} papers report a version in the period 2020--2022.
  All papers using Tigress mention at least which SPs were deployed with it, but in many cases, the authors omit the used configuration options, of which Tigress offers a wide variety.
  Some mention they used default configurations, but as these evolve over time, that is insufficient for reproducing the research and for interpreting the results.

  The worst tool with respect to version reporting is VirusTotal. Users provide software samples to this online service, which are then scanned by a set of malware detection engines. Both the deployed engines and their versions have evolved over the years.
  In quite some papers, VirusTotal is used to measure the extent to which concrete obfuscations hinder malware detection.
  To interpret the result in those papers, it is paramount to know which version of VirusTotal was used, i.e., when the service was accessed.
  However, only \virustotalpctversion{}\%~$\hat{=}$~\versionvirustotal{}/\totalvirustotal{} of the papers using VirusTotal do so.
}

\paragraph{Tool Flexibility}
\changed{
  Some tools offer their users more flexibility than others.
  This does not imply, however, that this flexibility is exploited in research.
  In fact, we observed quite the contrary.

  \bemph{SP researchers mostly use tools as is, even the flexible ones, rather than customizing them the way attackers do.}
  Consider the Tigress obfuscator, of which the developers share its source code with colleagues in academia on demand.
  Still, we observed that no outsider papers (i.e., not involving members of Christian Collberg's team behind Tigress) discuss or evaluate extensions of improvements of Tigress' transformations.
  At most, other authors combine Tigress protections with their own or with other existing tools.
  Despite OLLVM being open source, we made similar observations, albeit to a lesser degree: \ollvmextended{}/\totalToolsOLLVM{} OLLVM papers papers extend OLLVM.

  As another example, consider IDA Pro. This closed-source tool is extensible through a plug-in and scripting interface.
  Like all disassemblers that are by and large developed to reverse engineer non-obfuscated binaries, IDA Pro can easily be thwarted with control flow obfuscations~\cite{2003_obfuscation_of_executable_code_to_improve_resistance_to_static_disassembly,2021_obfuscated_integration_of_software_protections}.
  It has also been shown, however, that it is fairly easy to improve those tools' handling of obfuscated code with custom extensions~\cite{2021_obfuscated_integration_of_software_protections}.
  That is also how professional penetration testers customize them~\cite{2019_understanding_the_behaviour_of_hackers_while_performing_attack_tasks_in_a_professional_setting_and_in_a_public_challenge,2017_how_professional_hackers_understand_protected_code_while_performing_attack_tasks,2014_a_family_of_experiments_to_assess_the_effectiveness_and_efficiency_of_source_code_obfuscation_techniques}.
  However, of the authors using IDA Pro, only \percentTotalToolsIdaProPlugins{}\%~$\hat{=}$~\totalToolsIdaProPlugins{}/\totalToolsIdaPro{} extended its functionality with custom plug-ins or reused ones.
  Of those, \percentTotalToolsIdaProOnlyBindiff{}\%~$\hat{=}$~\totalToolsIdaProOnlyBindiff{}/\totalToolsIdaPro{} only extend the functionality with BinDiff, a special-purpose analysis tool to compare binaries.
  BinDiff merely builds on IDA Pro's disassembly to match code across different binaries rather than improve the disassembly of IDA Pro itself.
  This leaves only \percentTotalToolsIdaProNotJustBindiff{}\%~$\hat{=}$~\totalToolsIdaProNotJustBindiff{}/\totalToolsIdaPro{} papers extending or customizing IDA Pro in one way or another.
  Notably, \percentTotalToolsIdaProDis{}\%~$\hat{=}$~\totalToolsIdaProDis{}/\totalToolsIdaPro{} of the papers that use IDA Pro and that specifically target disassembly and CFG reconstruction, only two papers extend IDA Pro's functionality with regards to disassembly~\cite{2017_backward_bounded_dse_targeting_infeasibility_questions_on_obfuscated_codes,2021_obfuscated_integration_of_software_protections}.
  Others, such as the seminal static disassemblers thwarting paper~\cite{2003_obfuscation_of_executable_code_to_improve_resistance_to_static_disassembly}, make no such attempt.

  \bemph{We consider this discrepancy between published SP research and industrial SP practice a major shortcoming of the research in this field.}
  It is linked to our earlier remark in \cref{sec:protections} about SP being a cat and mouse game in which researchers fail all too often in anticipating even the simplest custom extensions of the functionality they propose to thwart with a new SP.
}

\subsection{Identified challenges}
In our analysis of sample sets, we identified six relevant challenges.

\begin{itemize}
  \item \textbf{Gap between analysis methods used in analysis papers and evaluations of newly proposed protections:} \changed{The used analysis methods differ significantly between analysis/deobfuscation papers and obfuscation papers.}
        While several simple analysis methods are frequently used in analysis/deobfuscation papers, they are much less common in obfuscation papers. This is the case even in A*/A papers.
        This leaves open the question of whether appropriate analysis methods are chosen for the evaluation of newly presented protections.
  \item \textbf{Gap between tools used in published research and the commercial, industrial state-of-the-art:} Our analysis of tools used in the surveyed papers across all publication venue rankings identified a concerning absence of commercial state-of-the-art protection tools.
        While in analysis papers, well-known commercial tools (e.g., IDA Pro) are heavily used, commercial protection tools such as Irdeto, Arxan, and GuardSquare are rarely applied, even in papers in top publication venues. And even the used analysis tools are most often not exploited in the way practitioners do so, e.g., by means of plug-ins.
        These are obvious threats to the validity of research, as the effectiveness of proposed protections is not compared against the commercial state-of-the-art and real-world practices.
  \item \textbf{Limited exploration of obfuscation combinations:} Our analysis reveals that in the examined papers, protections are rarely combined; a significant majority investigates protections in isolation rather than layered or combined with other protections, even in the papers in A*/A-rated venues. This limited exploration of multi-layered/combined SPs in research may not fully capture the complexity of protections deployed in real-world applications, which often utilize a combination of techniques.
  \item \textbf{Evaluations with no analysis method} In the goodware obfuscation category, too many papers (even in A*/A venues) lack an evaluation \changed{with} analysis methods from an attacker's toolbox, meaning they do not evaluate real-world potency and resilience.
  \item \textbf{Evaluations with just a single analysis method:} Still in the goodware obfuscation category, but only in the B, C, and lower-rated publication venues, even the papers that do evaluate the proposed SPs with analysis methods, still heavily lean towards using a single analysis method for evaluation. This measurement methodology sharply contrasts with real-world scenarios, where attackers typically employ a multitude of analysis methods in various combinations to undo, bypass, and work-around protections. It also implies that either potency or resilience is evaluated, but not both.
  \item \textbf{Lack of specificity in experimental setups:} All too many papers, especially in lower-rated publication venues, do not specify which versions and configurations of protection tools, analysis tools, and build tools were used.
\end{itemize}

To overcome the latter challenges, we simply invite researchers to provide more details in the description of their experimental setups.
Regarding the use of commercial protection tools, we invite the vendors to offer their tools at reduced prices to academics, with the permission to evaluate their tools and to publish their results.
With regards to the combination of obfuscations, we invite all academic tool developers to open up their tools to enable others to build on them and experiment more freely with them, and we invite all researchers to exploit the already available flexibility that tools such as Tigress and OLLVM offer to compose protections.

As for the insufficient use of appropriate analysis methods, we do not think we can ask researchers to deploy more tools to evaluate their SP contributions. Instead, more work is needed to ease the reuse of analysis tools.  We will come back to that in the next section, which analyzes which forms of measurements are reported in the papers, including measurements with analysis tools.

\section{Measurements}\label{sec:measurements}
This section analyzes how papers measure their contributions in the form of new protections and/or analysis/deobfuscation methods.
We first focus on the evaluation representation, i.e., the software format on which measurements and other evaluations are performed as reported.
Next, we analyze which aspects of strength and cost are measured and what measurements are used thereto.

\subsection{Evaluation Representation}
For all papers performing measurements on software, we recorded the \textbf{evaluation representation}, i.e., the program format from which the experimental evaluations in the paper start and from which measurements are performed.

\changed{
    \bemph{In most papers, most experimental evaluations are performed on software in the format in which it is being distributed to customers, i.e., the format on which real-world attackers get their hands.}

    The bottom bars in \cref{fig:protection_languages_graph} show the representations of samples on which the papers report performed measurements.%
    Some papers include additional measurements on a second, higher-level representation to obtain ground truth information or to evaluate the applicability of protections. Furthermore, some papers, such as those presenting MBA-deobfuscation techniques~\cite{2021_mba_blast_unveiling_and_simplifying_mixed_boolean_arithmetic_obfuscation,2016_defeating_mba_based_obfuscation}, evaluate their tools on MBA expressions in their source code format.

    On top of the 80 goodware papers that do not report any experimental evaluation, 4.9\%~$\hat{=}$~15/309 goodware papers fail to perform an experimental evaluation on software in the format on which attackers get their hands, as shown by the red underlined numbers in the bottom bar of \cref{fig:protection_languages_graph}.

    This poses no problem for papers focusing on the obfuscation of source code of natively compiled languages (e.g.,~\cite{2021_deepobfuscode_source_code_obfuscation_through_sequence_to_sequence_networks,2016_assessment_of_source_code_obfuscation_techniques,2018_common_program_similarity_metric_method_for_anti_obfuscation}), for surveys, and for papers that evaluate decision tool support rather than protection or analysis strength (e.g.,~\cite{2016_towards_automatic_risk_analysis_and_mitigation_of_software_applications}).
    For other papers, it does pinpoint a problem.

    \bemph{We observed that for most papers that have an experimental evaluation but lack one on the format on which attackers get their hands, that lack coincides with a lack of layered protection.}
    This coincidence makes sense:
    \changed{If no additional protection is deployed to widen the semantic gap between a higher-level representation on which the evaluated protection is deployed and a lower-level representation on which attackers get their hands, i.e., if there is a one-to-one and hence reversible mapping from higher-level obfuscation code constructs to lower-level ones, strength measurements on the richer, more abstract format can serve as reasonable approximations of the protection's real strength against attackers that only get their hands on the lower-level representation. In other words, it is fine for researchers to use ground-truth information available on the higher-level representations to estimate practical strength when attackers can easily reconstruct the used information. }
    For example, the practical potency of anti-slicing obfuscations can be evaluated accurately on static slices in C source code~\cite{2019_analysis_of_obfuscated_code_with_program_slicing,2007_metrics_based_evaluation_of_slicing_obfuscations,2007_slicing_obfuscations_design_correctness_and_evaluation} if no additional obfuscation has been applied that prevents the reconstruction of those slices from assembly code.
    Similarly, the practical potency of control flow or data flow obfuscations can be measured on Java source code instead of on bytecode~\cite{2020_creative_manual_code_obfuscation_as_a_countermeasure_against_software_reverse_engineering,2011_suggesting_potency_measures_for_obfuscated_arrays_and_usage_of_source_code_obfuscators_for_intellectual_property_protection_of_java_products} if no anti-decompilation obfuscations have been applied that prevent reconstruction of the obfuscated flows from the bytecode.

    However, we still do consider performing evaluations of protections in isolation a threat to their real-world relevance and validity because in practice only layered protections provide real protection~\cite{2008_a_compiler_based_infrastructure_for_software_protection}, and in practically useful compositions one or more layers do exploit the semantic gap between higher-level and lower-level software formats, such as source code vs.\ native code, in order to hamper code understanding and reconstruction of the original code structure.
    Even if such additional obfuscations are deployed, it is often unclear to what extent metrics computed on higher-level representations are representative of practical protection strength.
    For example, in the mentioned example of anti-slicing obfuscations it is not clear whether a potency metric computed on slices that are statically and accurately computed on source code (i.e., using the ground truth information that defenders have access to) provides an upper bound or a lower bound on the practical potency in cases where additional obfuscations prevent the accurate static computation of slices by attackers starting from the binary code.

    \bemph{We hence consider protections getting evaluated in isolation a bigger threat to validity than evaluations happening on a higher-level representation than the one attackers can attack in practice.}

    Unfortunately, that threat to validity is not limited to the relatively few papers that completely lack any evaluation on the appropriate format.
    Further analysis revealed that of all goodware papers performing at least some evaluation on samples written in natively compiled languages, 32\%~$\hat{=}$~64/(194+7+1) actually only measure overhead (compilation time, run time, memory consumption, and power consumption) and applicability (i.e., to what percentage of a program is some transformation applicable).
    They do not measure any potency, resilience, stealth\changed{,} or other form of protection strength on actual binaries.
    Another 8.9\%~$\hat{=}$~18/(194+7+1) papers do measure strength-related features, but they do so on other representations.

    \bemph{We consider this lack of adequate strength measurements a major shortcoming and lack of maturity in obfuscation research for natively compiled goodware.}
}

\subsection{Measurement Aspects and Categories}

\changed{We identified ten \emph{protection strength measurement categories} that are used to evaluate three strength aspects of SPs, namely potency\footnote{\label{footnote_potency}\changed{The original definitions of potency and resilience by Collberg et al.~\cite{1997_a_taxonomy_of_obfuscating_transformations,1998_manufacturing_cheap_resilient_and_stealthy_opaque_constructs} unnecessarily aimed to discriminate between human and automated analysis, in an unsatisfactory manner. Collberg's revised definition of potency fixed this by unifying the two aspects under the umbrella of potency 2.0, which he then defined as a protection's effect of making at least one analysis harder to perform, and no analysis easier~\cite{cc_book}. We find this revised definition insufficiently discriminative, however, as it does not discriminate between analyses of the protected assets and analyses of the protections. For this reason, we refine Collberg's potency 2.0, and complement it with a revised definition of resilience.}}, resilience\textsuperscript{\ref{footnote_potency}}, and stealth. We define the potency of an SP as the extent to which it can prevent, confuse, or hamper some analysis of the asset the SP protects. With this definition, SPs can have different potencies, each of which is specific to the considered analysis. Importantly, that can be a manual human analysis such as code comprehension, but also an automated analysis such as disassembling or malware detection. Our definition of resilience of an SP then is its capability to resist targeted attacks on the SP itself (instead of on the protected asset) to undo it or otherwise nullify or reduce its impact. Finally, our definition of stealth concerns the ability of an SP to remain unidentified, unrelated to whether or not the SP helps in hiding the presence of the asset protected by the SP. Note that there hence exists no one-to-one mapping between the measurement categories and the three aspects of protection strength. For example, precision metrics can be used to evaluate stealth in obfuscation classification papers, to evaluate potency in malware detection papers, and to evaluate resilience in goodware deobfuscation papers.

In addition to the strength measurement categories, we identified six \emph{cost measurement categories} that are used to measure the cost of protections. \Cref{tab:measurements} lists all measurement categories, with counts of their occurrence in different (non-exclusive) types of papers. A complete list of all categories, including descriptions, is provided in the supplemental material in \cref{sec:measurements_definitions}.
}

\begin{table}[t]
    \caption{Number of papers that report measurements for different strength and cost criteria of SPs.}\label{tab:measurements}
    \small
    \begin{tabular}{rccccc}
                                                                     & Deobf./Ana.                                      & Deobf./Ana.                                   & Obfuscation                                           & Obfuscation                                        &                                          \\
                                                                     & Goodware                                         & Malware                                       & Goodware                                              & Malware                                            & Total                                    \\
        \toprule
        Papers with protection implementation                        & \papersNonMalwareProtectionsXAnadeobfX{}         & \papersMalwareProtectionsXAnadeobfX{}         & \papersNonMalwareProtectionsXObfX{}                   & \papersMalwareProtectionsXObfX{}                 & \papersProtectionsX{}                    \\
        \midrule
        \multicolumn{1}{r}{\textbf{Aspect of Protection Measured}}   &                                                  &                                               &                                                       &                                                    &                                          \\
        Costs                                                        & \costsAnaDeobfPapersNonMalwareX{}                & \costsAnaDeobfPapersMalwareX{}                & \costsProtectionPapersNonMalwareX{}                   & \costsProtectionPapersMalwareX{}                   & \costsAllCategoriesX{}                   \\
        Potency                                                      & \potencyAnaDeobfPapersNonMalwareX{}              & \potencyAnaDeobfPapersMalwareX{}              & \potencyProtectionPapersNonMalwareX{}                 & \potencyProtectionPapersMalwareX{}                 & \potencyAllCategoriesX{}                 \\
        Resilience                                                   & \resilienceAnaDeobfPapersNonMalwareX{}           & \resilienceAnaDeobfPapersMalwareX{}           & \resilienceProtectionPapersNonMalwareX{}              & \resilienceProtectionPapersMalwareX{}              & \resilienceAllCategoriesX{}              \\
        Stealth                                                      & \stealthAnaDeobfPapersNonMalwareX{}              & \stealthAnaDeobfPapersMalwareX{}              & \stealthProtectionPapersNonMalwareX{}                 & \stealthProtectionPapersMalwareX{}                 & \stealthAllCategoriesX{}                 \\
        \midrule
        \multicolumn{1}{r}{\textbf{Cost Measurement Categories}}     &                                                  &                                               &                                                       &                                                    &                                          \\
        Static program size                                          & \sizeAnaDeobfPapersNonMalwareX{}                 & \sizeAnaDeobfPapersMalwareX{}                 & \sizeProtectionPapersNonMalwareX{}                    & \sizeProtectionPapersMalwareX{}                    & \sizeAllCategoriesX{}                    \\
        Execution time                                               & \performanceOverheadTimeDeobfPapersNonMalwareX{} & \performanceOverheadTimeDeobfPapersMalwareX{} & \performanceOverheadTimeProtectionPapersNonMalwareX{} & \performanceOverheadTimeProtectionPapersMalwareX{} & \performanceOverheadTimeAllCategoriesX{} \\
        Compilation/protection time                                  & \compilationTimeDeobfPapersNonMalwareX{}         & \compilationTimeDeobfPapersMalwareX{}         & \compilationTimeProtectionPapersNonMalwareX{}         & \compilationTimeProtectionPapersMalwareX{}         & \compilationTimeAllCategoriesX{}         \\
        Dynamic memory consumption                                   & \memoryUsageDeobfPapersNonMalwareX{}             & \memoryUsageDeobfPapersMalwareX{}             & \memoryUsageProtectionPapersNonMalwareX{}             & \memoryUsageProtectionPapersMalwareX{}             & \memoryUsageAllCategoriesX{}             \\
        Dynamic power consumption                                    & \powerConsumptionDeobfPapersNonMalwareX{}        & \powerConsumptionDeobfPapersMalwareX{}        & \powerConsumptionProtectionPapersNonMalwareX{}        & \powerConsumptionProtectionPapersMalwareX{}        & \powerConsumptionAllCategoriesX{}        \\
        Other costs                                                  & \otherCostsDeobfPapersNonMalwareX{}              & \otherCostsDeobfPapersMalwareX{}              & \otherCostsProtectionPapersNonMalwareX{}              & \otherCostsProtectionPapersMalwareX{}              & \otherCostsAllCategoriesX{}              \\
        \midrule
        \multicolumn{1}{r}{\textbf{Strength Measurement Categories}} &                                                  &                                               &                                                       &                                                    &                                          \\
\emph{Papers with 0 strength measurements}                          & \emph{\zeroStrengthMeasurementsGWanadeobfX{}}           & \emph{\zeroStrengthMeasurementsMWanadeobfX{}}        & \emph{\zeroStrengthMeasurementsGWobfX{}}                     & \emph{\zeroStrengthMeasurementsMWobfX{}}                  & \emph{\zeroStrengthMeasurementsAllX{}}          \\
        Other precision, recall, F-score, \ldots{}                   & \precisionAnaDeobfPapersNonMalwareX{}            & \precisionAnaDeobfPapersMalwareX{}            & \precisionProtectionPapersNonMalwareX{}               & \precisionProtectionPapersMalwareX{}               & \precisionAllCategoriesX{}               \\
        Automated attack/analysis time                               & \attackAnaOverheadTimeDeobfPapersNonMalwareX{}   & \attackAnaOverheadTimeDeobfPapersMalwareX{}   & \attackAnaOverheadTimeProtectionPapersNonMalwareX{}   & \attackAnaOverheadTimeProtectionPapersMalwareX{}   & \attackAnaOverheadTimeAllCategoriesX{}   \\
        Malware detection precision                                  & \malwareDetPrecAnaDeobfPapersNonMalwareX{}       & \malwareDetPrecAnaDeobfPapersMalwareX{}       & \malwareDetPrecProtectionPapersNonMalwareX{}          & \malwareDetPrecProtectionPapersMalwareX{}          & \malwareDetPrecAllCategoriesX{}          \\
        Code complexity                                              & \codeComplexityDeobfPapersNonMalwareX{}          & \codeComplexityDeobfPapersMalwareX{}          & \codeComplexityProtectionPapersNonMalwareX{}          & \codeComplexityProtectionPapersMalwareX{}          & \codeComplexityAllCategoriesX{}          \\
        Deltas / similarity                                          & \similarityAnaDeobfPapersNonMalwareX{}           & \similarityAnaDeobfPapersMalwareX{}           & \similarityProtectionPapersNonMalwareX{}              & \similarityProtectionPapersMalwareX{}              & \similarityAllCategoriesX{}              \\
        Applicability (Code Coverage)                                & \applicabilityDeobfPapersNonMalwareX{}           & \applicabilityDeobfPapersMalwareX{}           & \applicabilityProtectionPapersNonMalwareX{}           & \applicabilityProtectionPapersMalwareX{}           & \applicabilityAllCategoriesX{}           \\
        Human analysis (effort, success rate)                        & \manualAnaDeobfPapersNonMalwareX{}               & \manualAnaDeobfPapersMalwareX{}               & \manualAnaProtectionPapersNonMalwareX{}               & \manualAnaProtectionPapersMalwareX{}               & \manualAnaAllCategoriesX{}               \\
        Opcode distribution                                          & \opcodeDistributionDeobfPapersNonMalwareX{}      & \opcodeDistributionDeobfPapersMalwareX{}      & \opcodeDistributionProtectionPapersNonMalwareX{}      & \opcodeDistributionProtectionPapersMalwareX{}      & \opcodeDistributionAllCategoriesX{}      \\
        Entropy                                                      & \entropyAnaDeobfPapersNonMalwareX{}              & \entropyAnaDeobfPapersMalwareX{}              & \entropyProtectionPapersNonMalwareX{}                 & \entropyProtectionPapersMalwareX{}                 & \entropyAllCategoriesX{}                 \\
    \end{tabular}
\end{table}

\changed{\bemph{A first observation is that the costs of applying SPs are much more frequently measured than their strengths.} This is particularly the case in obfuscation goodware papers, where cost measurements (79\%~$\hat{=}$~197/248) are reported twice as often as potency measurements (39\%~$\hat{=}$~97/248), the most popular measured aspect of strength.
The vast majority of the cost measurements focus on static program size and execution time, which is not unexpected.

    \bemph{Still, the other 21\% of the obfuscation goodware papers fail to report the costs of the protections they discuss.}
    This lack of cost evaluation is unfortunately not limited to lower-rated publication venues: also in A*/A obfuscation goodware papers, we observed that (20\%~$\hat{=}$~8/41) of the papers that do feature protection implementations lack cost measurements.

    Of the SP strength measurements, stealth is clearly the least frequently evaluated (15\%~$\hat{=}$~73/495).

    In the deobfuscation/analysis types of papers, the balance between potency and resilience measurements matches the balances that can be observed in \cref{fig:venn} for goodware and malware papers: analysis papers typically measure potency, while deobfuscation papers measure resilience.

    \bemph{In obfuscation papers, potency is measured much more often than resilience.}
    For obfuscation malware papers, this is to be expected, as malware obfuscation aims by and large at preventing detection, not at preventing deobfuscation.
    For goodware obfuscation papers, the fact that potency measurements (39\%~$\hat{=}$~97/248) are more than twice as popular than resilience measurements (16\%~$\hat{=}$~40/248) is in line with our earlier observations (see \cref{sec:top_level_categories,sec:used_tools}) that few researchers study how attackers might adapt their strategies to the SPs they propose.
    Among researchers that publish in A*/A venues, this appears not to be the case, however: 37\%~$\hat{=}$~18/49 A*/A of the goodware obfuscation papers present potency measurements, while only slightly less (33\%~$\hat{=}$~16/49) present resilience measurements.
    Only 10\%~$\hat{=}$~5/49 present both.

    In defensive malware papers, analysis time (46\%~$\hat{=}$~63/137) and malware detection precision (36\%~$\hat{=}$~49/137) and other precision metrics (36\%~$\hat{=}$~50/137) are by far the most popular measurement categories.
    Delta/similarity measurements (9.5\%~$\hat{=}$~13/137) is the only remaining category that is somewhat popular in those papers.
    In offensive malware research, malware detection precision is also by far the most popular measurement category (64\%~$\hat{=}$~30/47).
    The dominance of these measurement categories for malware research is not surprising.

    In goodware papers, many more strength measurement categories have some popularity.
    In offensive goodware papers, precision (46\%~$\hat{=}$~40/87) and automated attack/analysis time (40\%~$\hat{=}$~35/87) are the \changed{dominant} measurement categories.
    In defensive, obfuscation goodware papers, code complexity is the most popular measurement, being reported in 19\%~$\hat{=}$48/248 of the papers.
    The only other category measured in more than 10\% of the obfuscation goodware papers is precision (13\%~$\hat{=}$~31/248).

    \bemph{Perhaps the most striking result is that 25\%~$\hat{=}$~125/495 of the papers with protection implementations report no strength measurements.} This is mostly due to 37\%~$\hat{=}$~92/248 of the obfuscation goodware papers reporting no strength measurements on the obfuscations they present, an astonishingly high number. While there is a correlation between this number and the quality of the publication venue, this problem is certainly not limited to lower-rated venues: 22\%~$\hat{=}$9/41 of the A*/A obfuscation goodware papers that feature implemented obfuscations report no strength measurements. This trend of lacking adequate strength measurements in obfuscation goodware papers also shows in the average number of such measurements per paper: \averageStrengthMeasurementsGWobfX{} overall, and still only \averageStrengthMeasurementsGWobfXAAstar{} in A*/A papers.

    We do not know for sure what the reason behind this lack of strength measurements is.
    We see two possible reasons.
    First, authors of goodware obfuscation might not consider such measurements relevant.
    We doubt that this is the case.
    Secondly, performing strength measurements might be considered too complex and/or too time-consuming to be worthwhile.
    We can think of several reasons why authors might think so.
    First, for many types of SPs, there is no consensus on which metrics to use.
    Theoretical complexity metrics from the field of software engineering often suffer from issues when used on obfuscated software, and practical metrics of the performance of concrete tools are most often ad hoc.
    Moreover, using real-world tools such as the IDA Pro disassembler (and its plug-in capabilities) or symbolic execution engines is rather difficult.
    Secondly, the relevant measurements can require experiments mimicking attack strategies involving multiple analyses.
    This is the case, e.g., for \enquote{binary} protections that completely prevent certain types of analysis tools (such as anti-debuggers, or code mobility preventing static disassembly).
    Measuring the impact of such protections requires comparing attack strategies with those tools to alternative attack strategies without those tools rather than evaluating the performance of individual analyses.

    While tools/frameworks have been proposed in the past to ease the use of reverse engineering tools in research~\cite{2022_argon_a_toolbase_for_evaluating_software_protection_techniques_against_symbolic_execution_attacks,angr,2015_a_framework_for_measuring_software_obfuscation_resilience_against_automated_attacks,2020_a_framework_for_evaluation_of_software_obfuscation_tools_for_embedded_devices} and to move into the direction of a commonly accepted set of criteria~\cite{2007_program_obfuscation_a_quantitative_approach}, our analysis of the literature in this section and in \cref{sec:used_tools} on the use of tools reveals that further work is needed to convince authors to include more strength measurements and to facilitate the use of tools to obtain relevant measurements.
}

\subsection{Identified challenges}
In our analysis of measurements, we identified one major challenge.

\begin{itemize}
    \item \textbf{Focus on cost measurements, lack of strength measurement:} In the surveyed papers, the most often measured aspects of protections are costs.
          This is unsurprising, as cost measurements and their interpretation are rather straightforward.
          By contrast, the lack of established methods for evaluating the strength of protections is clearly evident in the limited number of strength measurements in the surveyed papers. This is particularly the case in obfuscation goodware papers, including those from top venues with A*/A ranking.
\end{itemize}

Compared to the challenges related to samples and treatments, which might be solvable by a community consensus on shared sample sets and by more sharing of analysis tools, the strength measurement challenges are probably much more difficult to address as the weaknesses of the evaluations arise from both the choice of measurement types and tools used.
Clearly, it is not useful to define a standardized methodology for all evaluations, as different protections have very different motivations and goals, and thus the choice of appropriate measurements can also differ significantly.
Still, we strongly believe these challenges need to be addressed. The development of a flexible and easily re-usable protected software analysis toolbox, to which researchers can contribute their own measurement techniques and scripts, could help to improve the situation. A large community effort is needed to advance beyond the fragmental past efforts~\cite{2022_argon_a_toolbase_for_evaluating_software_protection_techniques_against_symbolic_execution_attacks,angr,2015_a_framework_for_measuring_software_obfuscation_resilience_against_automated_attacks,2020_a_framework_for_evaluation_of_software_obfuscation_tools_for_embedded_devices}.

\section{Experiments with Human Subjects}\label{sec:human}

In the large body of literature studied in this survey, few papers present experiments in which the performance of human subjects deploying SP or attacks on protected software is evaluated.
\Cref{tab:human} summarizes our findings on these papers.
A more complete \cref{tab:human_full} can be found in \Cref{sec:suppl_human} in the supplemental material.
Notice that this list excludes papers that merely discuss how a human performs some analysis or uses some tool.
In the papers listed in \cref{tab:human}, the performance of the humans is analyzed and evaluated to gather knowledge about the strength of SPs.

\begin{table}
    \caption{Papers reporting experiments with human subjects performing MATE protection and attack tasks. \enquote{Subjects}~indicates how many subjects participated of various levels of expertise: \protect\priority{00} bachelor and master students, \protect\priority{25} PhD and postgraduate students that are not experts in SP or reverse engineering, \protect\priority{50} students and amateurs with considerable experience in SP or reverse engineering, \protect\priority{75} professional programmers, and \protect\priority{100} professional security experts and pen testers. \enquote{Samples}~indicates the handled type of samples. Asterisks mark samples that are real-world programs rather than just \enquote{Complex} programs somewhere between toy and real-world programs . \changed{\enquote{\# Protections} indicates the number of protections (if any) composed in different samples. Commas separate different samples; \enquote{Vanilla}~means unprotected.} \enquote{Language}~indicates the targeted type of programming language. \enquote{Format}~indicates the format from which the software was reverse engineered. \enquote{Time}~indicates how long the experiments lasted. Question marks indicate the information is not available.}
\label{tab:human}
\small
\begin{tabular}{c p{1cm}  p{1.7cm} p{2cm} p{2.2cm} l l r}
\textbf{Paper} & \textbf{Year} & \textbf{Subjects} & \textbf{Samples} & \textbf{\# Protections} & \textbf{Language} & \textbf{Format} & \textbf{Time} \\ \midrule
\cite{2021_input_output_example_guided_data_deobfuscation_on_binary} & 2021 & 5\priorityss{50} & Toy  & 1,1 & Native & Native & 12h \\ %
\cite{2021_experimental_evaluation_of_the_obfuscation_techniques_against_reverse_engineering} & 2021 & 14\priorityss{75} & Toy  & Vanilla,2 & Managed & Intermediate & ? \\ %
\cite{2020_empirical_assessment_of_the_effort_needed_to_attack_programs_protected_with_client_server_code_splitting} & 2020 & 87\priorityss{0} & Complex  & Vanilla,1 & Native & Source & 2h \\ %
\cite{2020_creative_manual_code_obfuscation_as_a_countermeasure_against_software_reverse_engineering} & 2020 & 22\priorityss{50}\priorityss{100}\priorityss{0} & Complex  & Vanilla, ? & Managed & Source & 1h \\ %
\cite{2019_understanding_the_behaviour_of_hackers_while_performing_attack_tasks_in_a_professional_setting_and_in_a_public_challenge} & 2019 & 6\priorityss{100} 1\priorityss{50} & *Mobile, Toy & 9,8,7,3,2,1,1,1 & Native & Native & 30d \\ %
\cite{2019_the_impact_of_control_flow_obfuscation_technique_on_software_protection_against_human_attacks} & 2019 & 14\priorityss{75} & Toy  & Vanilla, 2 & Managed & Intermediate & ? \\ %
\cite{2019_resilient_user_side_android_application_repackaging_and_tampering_detection_using_cryptographically_obfuscated_logic_bombs} & 2019 & 4\priorityss{75} & Mobile & 1 & Managed & Intermediate & 40h \\ %
\cite{2018_programming_experience_might_not_help_in_comprehending_obfuscated_source_code_efficiently} & 2018 & 2\priorityss{25} 64\priorityss{0} & Complex  & Vanilla,1,1 & Managed & Intermediate & 1.5h \\ %
\cite{2018_enhance_virtual_machine_based_code_obfuscation_security_through_dynamic_bytecode_scheduling} & 2018 & 2\priorityss{25} 13\priorityss{0} & Complex, Toy & 2 & Native & Native & 72h \\ %
\cite{2018_a_large_scale_investigation_of_obfuscation_use_in_google_play} & 2018 & 63\priorityss{75} & Mobile & ? & Managed & Intermediate & ? \\ %
\cite{2017_stochastic_optimization_of_program_obfuscation} & 2017 & 10\priorityss{25} 10\priorityss{0} & Complex  & ? & Script & Source & ? \\ %
\cite{2017_how_professional_hackers_understand_protected_code_while_performing_attack_tasks} & 2017 & 6\priorityss{100} & *Mobile & 9,8,7 & Native & Native & 30d \\ %
\cite{2016_towards_better_program_obfuscation_optimization_via_language_models} & 2016 & 20\priorityss{75} & Complex  & ? & Script & Source & ? \\ %
\cite{2016_comparing_the_effectiveness_of_commercial_obfuscators_against_mate_attacks} & 2016 & 1\priorityss{75} & Complex, Toy & 3,2,2 & Native & Native & ? \\ %
\cite{2016_assessment_of_source_code_obfuscation_techniques} & 2016 & 1\priorityss{25} 14\priorityss{0} & Complex  & Vanilla,1 & Native & Source & 3.5h \\ %
\cite{2014_an_other_exercise_in_measuring_the_strength_of_source_code_obfuscation} & 2014 & 12\priorityss{0} & Complex  & Vanilla,1,1,2 & Managed & Source & 1h \\ %
\cite{2014_a_family_of_experiments_to_assess_the_effectiveness_and_efficiency_of_source_code_obfuscation_techniques} & 2014 & 22\priorityss{25} 52\priorityss{0} & Complex  & Vanilla,1,1 & Managed & Intermediate & 4h \\ %
\cite{2009_visualizing_compiled_executables_for_malware_analysis} & 2009 & 6\priorityss{50} & Malware & 1 & Native & Native & ? \\ %
\cite{2009_the_effectiveness_of_source_code_obfuscation_an_experimental_assessment} & 2009 & 22\priorityss{25} 10\priorityss{0} & Complex  & Vanilla,1 & Managed & Intermediate & 4h \\ %
\cite{2008_towards_experimental_evaluation_of_code_obfuscation_techniques} & 2008 & 8\priorityss{0} & Complex  & Vanilla,1 & Managed & Intermediate & 4h \\ %
\cite{2007_security_strength_measurement_for_dongle_protected_software} & 2007 & 5\priorityss{100} & *Complex  & ?+5 & Native & Native & 80m \\ %
\end{tabular}
\end{table}

The first observation is that few validations of SPs have been performed involving human subjects.
Important to know, in~\cite{2019_understanding_the_behaviour_of_hackers_while_performing_attack_tasks_in_a_professional_setting_and_in_a_public_challenge} \citeauthor{2019_understanding_the_behaviour_of_hackers_while_performing_attack_tasks_in_a_professional_setting_and_in_a_public_challenge} report on a superset of the experiments reported in~\cite{2017_how_professional_hackers_understand_protected_code_while_performing_attack_tasks}.
Together with~\cite{2016_comparing_the_effectiveness_of_commercial_obfuscators_against_mate_attacks} and~\cite{2007_security_strength_measurement_for_dongle_protected_software}, these are the only experiments in which more than two protections are layered on top of each other.
Moreover, only three experiments involved professional SP experts~\cite{2019_understanding_the_behaviour_of_hackers_while_performing_attack_tasks_in_a_professional_setting_and_in_a_public_challenge,2017_how_professional_hackers_understand_protected_code_while_performing_attack_tasks,2020_creative_manual_code_obfuscation_as_a_countermeasure_against_software_reverse_engineering,2007_security_strength_measurement_for_dongle_protected_software}, and only five papers have experiments lasting at least one working day~\cite{2021_input_output_example_guided_data_deobfuscation_on_binary,2019_understanding_the_behaviour_of_hackers_while_performing_attack_tasks_in_a_professional_setting_and_in_a_public_challenge,2017_how_professional_hackers_understand_protected_code_while_performing_attack_tasks,2019_resilient_user_side_android_application_repackaging_and_tampering_detection_using_cryptographically_obfuscated_logic_bombs,2018_enhance_virtual_machine_based_code_obfuscation_security_through_dynamic_bytecode_scheduling}.
While it is understandable that few research groups have the budget to hire professionals for multiple days, these results do indicate that few experiments have been performed that are representative of real-world MATE attacks.
For all the other experiments, it is also understandable that they are performed on toy programs or small \enquote{complex} programs. This enables the student participants, which can in practice not be required or demanded to participate to longer running experiments, to finish their assignments in the experiments' limited time frames.

\changed{\bemph{It is an open question whether results from such short running experiments with non-expert subjects and toy program lacking SP layering can be extrapolated to real-world attack scenarios.}}

\section{Recommendations for Goodware Obfuscation Research}\label{sec:recommendations}
\changed{
Previous sections identified a number of evaluation methodology challenges that the field of software obfuscation is facing. Most of those challenges concern goodware obfuscation/deobfuscation/analysis research. This section follows up on that by formulating a number of recommendations for improving the evaluations reported in future research papers.

Importantly, we explicitly do not propose moving in the direction of one universal evaluation methodology for SPs and countermeasures. In goodware SP, the different types of obfuscations are deployed to counter different attack strategies. Hence the obfuscations should be evaluated differently: with different forms of measurements, and by evaluating their impact on different analysis tools from the attacker's toolbox. In short, different evaluation methodologies can and should be used for different SPs. What all evaluations should have in common, such as including an evaluation of the potency and of the resilience, using concrete attack tools, on code obfuscated with state-of-the-art tools and layered protections, is precisely captured in the our recommendations.

\paragraph{Multiperspectivism} When presenting a new obfuscation/analysis contribution that aims to counter some existing analysis/obfuscations, do not only evaluate how the contribution fares against those existing analyses/obfuscations as is. Also evaluate how it fares against adversaries that adapt their strategy. In other words, try to think as your adversary, at least considering what their immediate reactions might be. For example, for a new deobfuscation technique, propose some potential countermeasure obfuscations~\cite{2021_search_based_local_black_box_deobfuscation_understand_improve_and_mitigate}, and when evaluating novel obfuscations to defeat the function reconstruction heuristics of commercial disassemblers, do so using plug-ins that to some degree override the default heuristics with ones that aim at defeating your new obfuscations~\cite{2021_obfuscated_integration_of_software_protections}.

\paragraph{Complete strength evaluation} When presenting a new SP, evaluate its potency, i.e., its strength in terms of protecting the assets it is supposed to protect; its resilience, i.e., its strength for resisting attacks on the SP itself rather than on the protected asset; and, if relevant, its stealth, i.e., how easy it is to detect the presence and configuration of the SP in a program. Ideally, you evaluate these aspects with respect to multiple analysis techniques, including static and dynamic ones that are known, from the scientific literature and other sources, to be deployed by real-world MATE attackers. Pattern matching is a prime example of a popular analysis to be considered. Provide convincing arguments for your choice of analyses, and why you exclude other popular, capable analyses, if any. If you, for some reason, cannot include experimental results with samples for some relevant analysis, at least present a theoretical security assessment thereof. %

\paragraph{Layered SP deployment} Deploy multiple, layered SPs on your samples, similar to how they are deployed in the real world. For a novel SP, evaluate its marginal value when combined with existing (popular) SPs, rather than its value when used in isolation. For managed and script languages, at least identifier renaming and string encryption should be included in the composed protections. Natively compiled software samples should, at the very least, be stripped.

\paragraph{Concrete attacks evaluation} To evaluate an SP, do not solely rely on complexity metrics computed on ground truth data. Instead, measure the SP's actual impact on analyses executed with concrete analysis tools. For example, evaluate the impact on their run times and on the precision of their results. If the tools are flexible and support plug-ins, as is the case for many disassemblers, consider using plug-ins (available online) rather than only the base tool. If possible, preference should be given to tools that attackers might use in the real world. For example, when evaluating how symbolic execution performs on protected code, tools that operate on binary code such as BINSEC/SE~\cite{2016_binsec_se_a_dynamic_symbolic_execution_toolkit_for_binary_level_analysis} or angr~\cite{angr} are to be preferred over the use of KLEE on IR produced with LLVM. Alternatively, KLEE can, of course, also be deployed on IR lifted from binaries~\cite{2019_saturn_software_deobfuscation_framework_based_on_llvm}. The survey by Schrittwieser et al.\ on obfuscations vs.\ analysis techniques~\cite{2016_protecting_software_through_obfuscation_can_it_keep_pace_with_progress_in_code_analysis} and the data presented in the supplemental work \Cref{fig:crosstab_anaprot} can help researchers to select the most relevant analyses for evaluating a new obfuscation's strengths, or to select the best obfuscations to stress-test a novel analysis technique.

\paragraph{State-of-the-art SP tools} The most advanced, commercial SP tools such as the offerings by Irdeto, Digital Ai, and GuardSquare are unavailable to most if not all researchers. So instead, evaluate analyses and deobfuscation techniques on samples generated with what comes closest to state-of-the-art SP tools. Several tools are available for free or affordable prizes, including Tigress, VMProtect, and Themida. Some older tools such as OLLVM and ProGuard cannot be considered today's state of the art. In addition, configure the tools properly such that your deployment is representative of real-word deployment. This can require quite some effort, e.g., to decide which SPs are combined and layered on which parts of the sample programs.

\paragraph{Setup specificity} For all tools used to build, protect, and analyse samples, specify the used versions and configurations. If you use a default configuration, specify what that entails. This is particularly important for tools of which the default settings evolve over time, such as Tigress where the default configurations of individual SPs evolved with different releases.

\paragraph{Tool availability} Make your research tools available as artifacts for reuse and reproducibility.

\paragraph{Sample complexity} Include in your data set a sufficient number of sample programs of sufficient complexity to be representative of real-world use cases. Which concrete complexity metrics are relevant and which levels of complexity should be included depends entirely on the claims for which the evaluation is supposed to provide evidence. For example, when claiming that some deobfuscation technique can fail even on the simplest of samples, experiments with simple samples suffice~\cite{2022_chosen_instruction_attack_against_commercial_code_virtualization_obfuscators}. By contrast, if a technique requires identifying the relevant fragments in program traces, large programs that generate long traces should be included~\cite{2017_syntia_synthesizing_the_semantics_of_obfuscated_code}. In addition, the samples build process should be representative of how real-world software is being built. For example, do not include native binaries compiled at \texttt{-O0}, i.e., without any compiler optimizations enabled.%

\paragraph{Sample availability} Make the samples in your dataset available as artifacts for reuse and reproduction by others. As for commercial programs, make sure you mention the exact versions.

\paragraph{Sample diversification} When performing an evaluation on protected samples that you generate yourself with some configurable obfuscation and/or with an obfuscation of which the behavior is randomized, include multiple obfuscated versions for varying configurations and random seeds and deploy accepted statistical techniques to aggregate the obtained measurements. For measuring the performance overhead, include multiple obfuscated samples where the protection has been deployed on a range of program points with varying degrees of execution frequencies.

\vspace{0.2cm}

These recommendations are concrete and realistic, as evidenced by the fact that there exist quite a few papers that already implemented them in the past. To help readers find good examples, \Cref{tab:recommendation_matrix} lists the extent to which A* goodware papers in our survey implement our recommendations. The sample diversification column lists multiple numbers for papers with multiple, distinct subsets of samples with varying levels of diversification. Each reported number is the rounded ratio between the total sample set size of such a subset and its original sample set size. In the state-of-the-art SP tools column, we marked Javascript minifiers such as Google's Closure compiler that do not deploy more advanced obfuscations with \priority{50}. Apart from that, it should be clear that our judgments are subjective: they are our interpretation of our recommendations and of the papers. Still, we hope this table can provide guidance for future research in software obfuscation.

For inspiration as to how to instantiate some of these generic recommendations for concrete SPs or analyses, we refer to the report of the Dagstuhl seminar that motivated us for this survey~\cite{2019_software_protection_decision_support_and_evaluation_methodologies_dagstuhl_seminar_19331}. The report's Section 4 lists concrete recommendations for research into anti-disassembly SPs and into trace-based analysis techniques. In addition, \Cref{sec:suppl_tools} in the supplemental material contains concrete recommendations for the usage of some of the most popular research tools. Finally, \Cref{sec:suppl_human} of the supplemental material presents some potential sources of inspiration for selecting relevant combinations of layered protections.
}

\begin{table}[t]
\small %
\caption{Goodware papers published in A* venues, scored for their implementation of the recommendations from Section~\ref{sec:recommendations}. \protect\priority{100} means well done, \protect\priority{0} not so; n/a means the recommendation is not applicable.}\label{tab:recommendation_matrix}
\centering
\begin{tabular}[t]{ c c c c c c c c c r c c c c c l l }
    \textbf{Paper}  &  \rotatebox{90}{\textbf{Obfuscation}}  &  \rotatebox{90}{\textbf{Deobfuscation}}  &  \rotatebox{90}{\textbf{Analysis}}  &  \rotatebox{90}{\textbf{Multiperspectivism}}  &  \rotatebox{90}{\textbf{Complete strength eval.}}  &  \rotatebox{90}{\textbf{Layered SP deployment}}  &  \rotatebox{90}{\textbf{Concrete attacks eval.}}  &  \rotatebox{90}{\textbf{State-of-the-art SP tools}} &  \rotatebox{90}{\textbf{Setup specificity}}  &  \rotatebox{90}{\textbf{Tool availability}}  &  \rotatebox{90}{\textbf{Sample Complexity}}  &  \rotatebox{90}{\textbf{Sample availability}}  &  \rotatebox{90}{\textbf{Sample diversification}}  &  \textbf{Year}  &  \textbf{Venue}  & \textbf{Authors} \\
\toprule
\cite{2022_loki_hardening_code_obfuscation_against_automated_attacks} & \checkmark{} & \checkmark{} &  & \priority{100} & \priority{50} & \priority{100} & \priority{100} & \priority{100} & \priority{100} & \priority{100} & \priority{100} & \priority{100} & 175 & 2022 & Usenix Security & \citeauthor{2022_loki_hardening_code_obfuscation_against_automated_attacks}\\
\cite{2022_chosen_instruction_attack_against_commercial_code_virtualization_obfuscators} &  & \checkmark{} &  & \priority{0} & n/a & \priority{100} & n/a & \priority{100} & \priority{100} & \priority{100} & \priority{50} & \priority{100} & 3 & 2022 & NDSS & \citeauthor{2022_chosen_instruction_attack_against_commercial_code_virtualization_obfuscators}\\
\cite{2021_search_based_local_black_box_deobfuscation_understand_improve_and_mitigate} & \checkmark{} & \checkmark{} &  & \priority{100} & \priority{50} & \priority{100} & \priority{100} & \priority{100} & \priority{0} & \priority{100} & \priority{100} & \priority{100} & 1; 2; 3 & 2021 & ACM CCS & \citeauthor{2021_search_based_local_black_box_deobfuscation_understand_improve_and_mitigate}\\
\cite{2020_ui_obfuscation_and_its_effects_on_automated_ui_analysis_for_android_apps} & \checkmark{} &  &  & \priority{0} & \priority{50} & \priority{100} & \priority{100} & \priority{0} & \priority{0} & \priority{100} & \priority{100} & \priority{100} & 10 & 2020 & IEEE/ACM ACE & \citeauthor{2020_ui_obfuscation_and_its_effects_on_automated_ui_analysis_for_android_apps}\\
\cite{2019_cfhider_control_flow_obfuscation_with_intel_sgx} & \checkmark{} &  &  & \priority{0} & \priority{50} & \priority{0} & \priority{0} & \priority{0} & \priority{0} & \priority{100} & \priority{50} & \priority{50} & 2; 4 & 2019 & IEEE INFOCOM & \citeauthor{2019_cfhider_control_flow_obfuscation_with_intel_sgx}\\
\cite{2019_asm2vec_boosting_static_representation_robustness_for_binary_clone_search_against_code_obfuscation_and_compiler_optimization} &  &  & \checkmark{} & \priority{0} & n/a & \priority{100} & n/a & \priority{100} & \priority{100} & \priority{100} & \priority{100} & \priority{100} & 211 & 2019 & IEEE S\&P & \citeauthor{2019_asm2vec_boosting_static_representation_robustness_for_binary_clone_search_against_code_obfuscation_and_compiler_optimization}\\
\cite{2019_anything_to_hide_studying_minified_and_obfuscated_code_in_the_web} &  &  & \checkmark{} & \priority{100} & n/a & \priority{100} & n/a & \priority{50} & n/a & \priority{100} & \priority{100} & \priority{0} & 1; 12 & 2019 & WWW Conf. & \citeauthor{2019_anything_to_hide_studying_minified_and_obfuscated_code_in_the_web}\\
\cite{2018_software_protection_on_the_go_a_large_scale_empirical_study_on_mobile_app_obfuscation} & \checkmark{} &  &  & \priority{100} & \priority{50} & \priority{100} & \priority{100} & \priority{0} & n/a & \priority{0} & \priority{100} & \priority{0} & n/a & 2018 & ACM/IEE ICSE & \citeauthor{2018_software_protection_on_the_go_a_large_scale_empirical_study_on_mobile_app_obfuscation}\\
\cite{2018_protecting_million_user_ios_apps_with_obfuscation_motivations_pitfalls_and_experience} & \checkmark{} &  &  & \priority{0} & \priority{50} & \priority{100} & \priority{50} & \priority{0} & \priority{0} & \priority{0} & \priority{50} & \priority{0} & 2 & 2018 & ACM/IEE ICSE & \citeauthor{2018_protecting_million_user_ios_apps_with_obfuscation_motivations_pitfalls_and_experience}\\
\cite{2018_hybrid_obfuscation_to_protect_against_disclosure_attacks_on_embedded_microprocessors} & \checkmark{} &  &  & \priority{50} & \priority{100} & \priority{50} & \priority{0} & \priority{0} & \priority{0} & \priority{0} & \priority{0} & \priority{100} & 6 & 2018 & IEEE ToC & \citeauthor{2018_hybrid_obfuscation_to_protect_against_disclosure_attacks_on_embedded_microprocessors}\\
\cite{2017_syntia_synthesizing_the_semantics_of_obfuscated_code} &  & \checkmark{} &  & \priority{0} & n/a & \priority{100} & n/a & \priority{100} & \priority{0} & \priority{100} & \priority{50} & \priority{100} & 1; 2 & 2017 & Usenix Security & \citeauthor{2017_syntia_synthesizing_the_semantics_of_obfuscated_code}\\
\cite{2017_stochastic_optimization_of_program_obfuscation} & \checkmark{} &  &  & \priority{0} & \priority{100} & \priority{0}\footnote{While the focus of~\cite{2017_stochastic_optimization_of_program_obfuscation} is the determination of optimal combinations and orderings of protections to be composed and layered, the paper lacks strong evidence that actual combinations were deployed in its evaluation.} & n/a & \priority{50} & \priority{0} & \priority{0} & \priority{100} & \priority{50} & 3 & 2017 & ACM/IEE ICSE & \citeauthor{2017_stochastic_optimization_of_program_obfuscation}\\
\cite{2017_semantics_based_obfuscation_resilient_binary_code_similarity_comparison_with_applications_to_software_and_algorithm_plagiarism_detection} &  &  & \checkmark{} & \priority{0} & n/a & \priority{100} & n/a & \priority{0} & \priority{0} & \priority{0} & \priority{100} & \priority{100} & 1; 5; 6 & 2017 & ACM FSE & \citeauthor{2017_semantics_based_obfuscation_resilient_binary_code_similarity_comparison_with_applications_to_software_and_algorithm_plagiarism_detection}\\
\cite{2017_recovering_clear_natural_identifiers_from_obfuscated_js_names} &  & \checkmark{} &  & \priority{0} & n/a & \priority{0} & n/a & \priority{0} & n/a & \priority{100} & \priority{100} & \priority{0} & 1 & 2017 & ACM FSE & \citeauthor{2017_recovering_clear_natural_identifiers_from_obfuscated_js_names}\\
\cite{2017_predicting_the_resilience_of_obfuscated_code_against_symbolic_execution_attacks_via_machine_learning} & \checkmark{} &  &  & \priority{0} & \priority{50} & \priority{100} & \priority{50} & \priority{100} & \priority{100} & \priority{100} & \priority{50} & \priority{100} & 6 & 2017 & Usenix Security & \citeauthor{2017_predicting_the_resilience_of_obfuscated_code_against_symbolic_execution_attacks_via_machine_learning}\\
\cite{2016_reliable_third_party_library_detection_in_android_and_its_security_applications} &  & \checkmark{} & \checkmark{} & \priority{0} & n/a & \priority{0} & n/a & \priority{0} & \priority{0} & \priority{100} & \priority{100} & \priority{0} & 1 & 2016 & ACM CCS & \citeauthor{2016_reliable_third_party_library_detection_in_android_and_its_security_applications}\\
\cite{2010_mimimorphism_a_new_approach_to_binary_code_obfuscation} & \checkmark{} &  &  & \priority{0} & \priority{50} & \priority{0} & \priority{50} & \priority{0} & n/a & \priority{0} & \priority{0} & \priority{0} & 201 & 2010 & ACM CCS & \citeauthor{2010_mimimorphism_a_new_approach_to_binary_code_obfuscation}\\
\cite{2007_binary_obfuscation_using_signals} & \checkmark{} &  &  & \priority{0} & \priority{100} & \priority{50} & \priority{50} & \priority{0} & \priority{100} & \priority{0} & \priority{100} & \priority{100} & 2 & 2007 & Usenix Security & \citeauthor{2007_binary_obfuscation_using_signals}\\
\cite{2004_static_disassembly_of_obfuscated_binaries} &  &  & \checkmark{} & \priority{0} & n/a & \priority{50} & n/a & \priority{0} & \priority{100} & \priority{0} & \priority{100} & \priority{100} & 1 & 2004 & Usenix Security & \citeauthor{2004_static_disassembly_of_obfuscated_binaries}\\
\cite{2003_obfuscation_of_executable_code_to_improve_resistance_to_static_disassembly} & \checkmark{} &  &  & \priority{0} & \priority{50} & \priority{50} & \priority{50} & \priority{0} & \priority{0} & \priority{0} & \priority{100} & \priority{100} & 6 & 2003 & ACM CCS & \citeauthor{2003_obfuscation_of_executable_code_to_improve_resistance_to_static_disassembly}\\
\cite{1998_manufacturing_cheap_resilient_and_stealthy_opaque_constructs} & \checkmark{} &  &  & \priority{0} & \priority{100} & \priority{0} & \priority{0} & \priority{0} & n/a & n/a & \priority{0} & n/a & n/a & 1998 & ACM POPL & \citeauthor{1998_manufacturing_cheap_resilient_and_stealthy_opaque_constructs}\\
\bottomrule
     \end{tabular}
 \end{table}

\section{Related Work}\label{sec:related_work}

While ours is the largest survey in the field of SP to date, it is not the first.
We identified \totalSurveysIncluded{} in-scope surveys and several non-survey related works, which we classify in this section. A complete list of these papers including short descriptions is available in the supplemental material \cref{tab:relwork}.

We identified several literature surveys~\cite{2016_a_survey_on_aims_and_environments_of_diversification_and_obfuscation_in_software_security,2016_protecting_software_through_obfuscation_can_it_keep_pace_with_progress_in_code_analysis,2017_on_secure_and_usable_program_obfuscation_a_survey,2018_a_systematic_study_on_static_control_flow_obfuscation_techniques_in_java,2018_diversification_and_obfuscation_techniques_for_software_security_a_systematic_literature_review,2020_a_survey_of_android_malware_detection_with_deep_neural_models,2021_a_survey_of_android_malware_static_detection_technology_based_on_machine_learning,android_app_forensics_survey,2021_measuring_software_obfuscation_quality_a_systematic_literature_review}, which --- similar to our work --- provide a broad overview of SP research from different perspectives.
Another category of publications similar to our work are taxonomies, introductions, tutorials and other theoretical publications~\cite{2002_watermarking_tamper_proofing_and_obfuscation_tools_for_software_protection,2010_assure_high_quality_code_using_refactoring_and_obfuscation_techniques,2010_theory_and_practice_of_program_obfuscation,2011_a_taxonomy_of_self_modifying_code_for_obfuscation,2012_cloud_protection_by_obfuscation_techniques_and_metrics,2016_a_study_review_on_code_obfuscation,2017_a_tutorial_on_software_obfuscation,2018_tutorial_an_overview_of_malware_detection_and_evasion_techniques,2019_a_taxonomy_of_software_integrity_protection_techniques,2019_obfuscated_code_quality_measurement,2020_layered_obfuscation_a_taxonomy_of_software_obfuscation_techniques_for_layered_security,2021_a_survey_of_binary_code_similarity}.
These papers do meta descriptions such as categorization or classification of SP techniques, attack scenarios, methods for measuring the strength of SPs.
Some papers survey available tools and techniques for SP.
Besides some early works, which look at the SP domain in general~\cite{2002_watermarking_tamper_proofing_and_obfuscation_tools_for_software_protection,2005_code_obfuscation_literature_survey} there exist multiple publications which a more narrow focus:
Different languages, e.g., Java~\cite{2015_a_large_study_on_the_effect_of_code_obfuscation_on_the_quality_of_java_code},
analysis avoidance~\cite{2010_analysis_avoidance_techniques_of_malicious_software,2012_reverse_code_engineering_state_of_the_art_and_countermeasures,2019_obfuscation_where_are_we_in_anti_dse_protections_a_first_attempt},
malware~\cite{2010_malware_obfuscation_techniques_a_brief_survey,2013_binary_code_obfuscations_in_prevalent_packer_tools,2018_anti_emulation_trends_in_modern_packers_a_survey_on_the_evolution_of_anti_emulation_techniques_in_upa_packers},
specific SPs e.g., control flow obfuscations~\cite{2006_a_survey_of_control_flow_obfuscations}, call-flow~\cite{2010_theory_and_practice_of_program_obfuscation}, instruction substitution~\cite{2010_theory_and_practice_of_program_obfuscation}, self modifying code~\cite{2011_a_taxonomy_of_self_modifying_code_for_obfuscation}, and indistinguishability obfuscation~\cite{2016_hopes_fears_and_software_obfuscation,2020_cryptographic_obfuscation_a_survey}.
Three publications survey the opposing side of SP --- analysis and deobfuscation of SP~\cite{2012_a_survey_on_automated_dynamic_malware_analysis_techniques_and_tools,2016_testing_android_malware_detectors_against_code_obfuscation_a_systematization_of_knowledge_and_unified_methodology,2021_sok_automatic_deobfuscation_of_virtualization_protected_applications}.
Several papers assess uses of SPs in practice for Android apps~\cite{2018_a_large_scale_empirical_study_on_the_effects_of_code_obfuscations_on_android_apps_and_anti_malware_products,2018_a_large_scale_investigation_of_obfuscation_use_in_google_play,2018_understanding_android_obfuscation_techniques_a_large_scale_investigation_in_the_wild,2019_obfuscated_android_application_development,2020_a_large_scale_study_on_the_adoption_of_anti_debugging_and_anti_tampering_protections_in_android_apps}, iOS apps~\cite{2018_software_protection_on_the_go_a_large_scale_empirical_study_on_mobile_app_obfuscation}, or malware~\cite{2012_scientific_but_not_academical_overview_of_malware_anti_debugging_anti_disassembly_and_anti_vm_technologies}.
\citeauthor{collberg2016repeatability} published a large-scale study~\cite{collberg2016repeatability} on reproducibility and repeatability of experiments in computer science.
For only 37\% of the surveyed publications code was made available by the authors, which is in line with our findings.

\section{Conclusions}\label{sec:conclusion}
This survey on software obfuscation is based on the largest collection of papers ever studied in the SP domain. While the mix of protection targets is similar to what we observed in other surveys, in terms of types of programming languages targeted, our distinct focus on the measurements performed in the surveyed papers is novel. In the aftermath of the 2019 Dagstuhl Seminar on Software Protection Decision Support and Evaluation Methodologies, where participants expressed subjective worries about evaluation methodologies in SP research, we systematically searched for evidence of these concerns and indeed found a number of issues.

We see a concerning focus on cost measurements while the strength of SPs is way less often measured. Evaluation sample sets suffer from major shortcomings regaring availability, diversi\-fication, and complexity. Furthermore, we identified a troubling number of papers using no or just a single analysis technique for evaluation, and we observed a limited exploration of obfuscation combinations. Next, an identified gap between analysis methods used in analysis papers and evaluations of newly proposed protections raises the concern whether appropriate analysis methods are used for the evaluation of newly proposed protections. We also identified a gap between tools used in published research and the commercial state of the art. This casts a shadow on the real-world relevance of the published research, and confirms the continuing reliance on security-through-obscurity in this domain. Finally, the reproducibility of research results is low as too many papers do not specify versions and configurations of used tools. Interestingly, most of these issues are prevalent even in A and A* publications, although there they tend to occur less frequently.

In summary, our work serves as a robust confirmation of the worries expressed in Dagstuhl. The prevalence of these issues across different tiers of publications underscores the urgent need for a broad reconsideration of evaluation methodologies in the SP research field. \changed{To that end, we formulated concrete recommendations, with a focus on goodware SP research. Beyond those recommendations that individual research teams can implement,} we strongly advocate for the development of a community consensus on shared sample sets, akin to practices in other research domains. \changed{In particular the Trust Hub set of labeled hardware obfuscation benchmarks~\cite{2018_development_and_evaluation_of_hardware_obfuscation_benchmarks} can serve as an example.} Furthermore, the creation of a flexible and easily re-usable protected software analysis toolbox could benefit both the reproducibility and comparability of results across studies, thereby driving the field forward and establishing a solid foundation for future explorations.

\begin{acks}
    This research was funded by the Austrian Science Fund (FWF): I 3646-N31, by The Research Foundation – Flanders (FWO): 3G0E2318, and by the Cybersecurity Research Program Flanders. Further, the financial support by the Austrian Federal Ministry of Labour and Economy, the National Foundation for Research, Technology and Development and the Christian Doppler Research Association is gratefully acknowledged. Likewise, the funding by Ghent University is appreciated. 

For the purpose of open access, the authors have applied a CC BY public copyright licence to any Author Accepted Manuscript version arising from this submission.

The authors would like to acknowledge the help of Armin Huremagic.

\end{acks}

\printbibliography[]

\pagebreak
\section{Supplemental Material}\label{sec:spplemental_material}

The supplemental material in this section lists our definitions of software protection methods (\Cref{sec:protection_definitions}), analysis methods (\Cref{sec:analysis_definitions}), and measurements aspects and categories (\Cref{sec:measurements_definitions}).
Furthermore, \cref{sec:sample-no-sample} provides additional information about the use of samples in papers, and \cref{sub:venue_ranking-sample_set_size-supp} on the correlation of sample sets with publication venues.
\Cref{sec:protection_crosstabs,sec:suppl_analysis_deployment} provide additional data on the papers' deployment of protection methods and analysis methods, respectively. \Cref{sec:suppl_tools} describes the actual tools that were most commonly used in the surveyed papers and provides some recommendations on their use in evaluations.
\Cref{sec:suppl_human} provides some additional information on the protections layered in experiments involving human subjects.
Finally, \cref{sec:related_work_supplemental} presents a more extensive discussion of other surveys in the domain of software protection, complementary to the shorter related work discussion in \cref{sec:related_work}.

\subsection{Protection Definitions}\label{sec:protection_definitions}
For each surveyed paper, we recorded which types of novel protections were presented, which protections were included in the samples of their experimental evaluation, which protections were discussed in their (theoretical) security analysis, and which protections were surveyed (in the case of surveys).
So for each paper in scope, one or more of the protections listed below were marked.

Importantly, concrete protection transformations, as presented in papers, can match multiple classes.
For example, a tool can inject opaque predicates based on aliasing pointer computations as a means to inject spurious data dependencies.
Such a protection is classified as \enquote{opaque predicates}, \enquote{aliasing}, and \enquote{data flow transformation}.
Some classes are conceptually subclasses of others.
For example, \enquote{control flow indirections} such as branch functions are a specific type of \enquote{control flow transformations}.
Whenever some protection discussed in a paper matches a more concrete subclass, we only mark that subclass, not the superclass.
Furthermore, if some transformation is merely a side-effect of another one, we do not mark it.
For example, \enquote{parallelization} inherently implies control flow and data flow transformations. However, we only mark the latter if the parallelization is used as an enabler of a specific flow transformation, for example when implicit data flow is implemented by measuring timing differences between fast and slow threads~\cite{2021_search_based_local_black_box_deobfuscation_understand_improve_and_mitigate}.

Furthermore, it is essential to raise the issue of potential bias.
We collected data by interpreting the information and claims in the papers.
Some papers present explicitly and exhaustively which protections the authors deployed; others do not.
For instance, many malware papers only mention where they got their malware samples but do not discuss in detail how those samples were generated or obfuscated, in many cases because that ground-truth information is not public.
Furthermore, some authors detail how they configured publicly available obfuscation tools; others do not.
Some such tools have minimal or default configurations; others do not.
For some (malware) data sets, descriptions of the deployed protections are available in the literature that we are aware of; for others, this is not the case.
For each paper, we only marked protections if we could determine without reasonable doubt that they had been deployed on samples.
Our discovery process mostly involved the papers themselves, any references in them to other papers discussing the used data sets, and publicly available descriptions of the used tools.
When individual malware samples or families were mentioned without discussing the protections used in them, we did not, however, go as far as looking up external reverse engineering blogs or analysis reports on those individual samples.
As a result, the reader should be aware that for quite some papers, in particular malware analysis papers, protections deployed on their samples might not be recorded as such.
For goodware papers, in particular papers in the obfuscation perspective (for which authors typically generated their protected samples themselves), this is less of an issue.

\paragraph{Data encoding/encryption (DEN)}
Encrypting or changing the type, encoding, and bit representation of data in a program.
This includes replacing static data (e.g., strings) with compressed or encrypted versions.
It also excludes the special case of white-box cryptography in which the keys are replaced by other data or code.

\paragraph{Static data to code conversion (D2C)}
Replacing constant data available in a static program representation by computations that reproduce the data at run time.
If those computations merely involve decryption or decompression, we consider it data encoding/encryption, not static data to code conversion.

\paragraph{White-box cryptography (WBC)}
Implementation of cryptographic primitives such that stored and used keys no longer occur in plain text in the binaries or in the program state during execution.

\paragraph{Mixed Boolean-arithmetic (MBA)}
Code and data obfuscation technique in which computations are replaced by more complex mixes of boolean and arithmetic operations.

\paragraph{Data flow transformation (DFT)}
Introducing fake, covert, or implicit data dependencies with the goal of obfuscating the data dependencies.
Includes anti-taint protections (over-tainting as well as under-tainting), but not MBA which has its own category.

\paragraph{Data transformation (DTR)}
Splitting, merging, and reordering of structured data; changing the shape or layout with which data is stored in memory.

\paragraph{Aliasing (ALI)}
Obfuscations building on or targeting potential aliases among pointers, e.g., to make data analyses imprecise or to yield larger points-to sets.
Includes replacing direct data accesses with indirect ones, e.g., through reflection.

\paragraph{Class based transformations (CBT)}
Class hierarchy transformations, class type hiding, class refactoring, transformations with overriding methods.
This includes adding fake/dead methods to classes to hamper class-based analyses.

\paragraph{Code reordering (CRE)}
Reordering of instructions, functions, basic blocks, expressions, or other code elements; changing the alignment/padding of binary code fragments.

\paragraph{Code diversity (DIV)}
Replacing instructions or sequences thereof with other equivalent ones in the same ISA (e.g., CALL replaced by PUSH \& JMP).
Diversity can be spatial (diversified instances exist at the same point in time), or temporal (code within a software instance is diversified over time).

\paragraph{Control flow flattening (CFF)}
Obscuring structured control flow graphs by replacing them with a flat structure and a data-controlled dispatcher that controls the order in which code fragments are executed.
The goal is to hide the original structure.

\paragraph{Opaque predicates (OPP)}
Using boolean-valued expressions whose values or other invariant properties are known at obfuscation time but difficult for an attacker to figure out, e.g., to steer execution around inserted bogus control flow transfers.
The goal is to inflate the complexity of the code's CFGs.

\paragraph{Control flow indirections (CFI)}
Replacing direct control flow transfers by indirect ones, which can be, e.g., indirect calls or jumps, faulting instructions and exception handlers, uses of reflection.
The goal is to hide the call graph's edges or the control flow graph's edges from static analysis, not to change those graphs.

\paragraph{Control flow transformation (CFT)}
All other ways of obfuscating the control flow, except control flow flattening, opaque predicates, and control flow indirections.
The goal here is to change the control flow graph, not to hide edges from static analyses.

\paragraph{Function transformation (FUT)}
Transformations at the granularity of functions, such as wrapping, cloning, splitting, merging, inlining, and outlining of functions or methods.
This excludes protections based on adding fake/dead methods to classes to hamper class-based analyses.

\paragraph{Identifier renaming (IRE)}
Renaming identifiers such as the names of variables, sections, functions, methods, classes, registers, etc.\ to remove information useful for an attacker or to complicate analyses.

\paragraph{Junk code insertion (JCI)}
Inserting junk bytes, dead code (which can be executed but without impact on the program semantics), or unreachable code (which cannot be executed).

\paragraph{Library hiding (LIH)}
Hiding which library functions are called, e.g., by making calls indirect or by inlining library functions.
If this is achieved by means of specific transformations, such as encrypting the names of the functions, those transformations are also marked.

\paragraph{Loop transformations (LOT)}
Transformation of loops, including fusion, unrolling, splitting, reordering iterations, as well as altering counters, exit conditions, etc.

\paragraph{Overlapping Code (OLC)}
Overlapping instructions or functions.
Native overlapping instructions share bytes in the binaries or in text segments.
Overlapping functions share instructions, i.e., some instructions are executed as part of multiple functions.

\paragraph{Parallelizing (PAR)}
Injecting additional threads or processes or other forms of parallelism into a program in order to complicate analysis, i.e., to make it harder to use analysis tools and to obtain precise models of the software's data flow and control flow.

\paragraph{Repacking (RPA)}
Recompilation, reassembly, or repackaging of code taken from one software component into another.

\paragraph{Anti-debugging (ADB)}
Techniques for preventing the use of debuggers and debugging techniques.

\paragraph{Code mobility (CMO)}
Removing static code in a distributed program by equivalent code that is downloaded dynamically from a (secure) server to hamper static analysis.

\paragraph{Server side execution (SSE)}
Removing code fragments from a distributed program and instead having equivalent code executed on a remote (secure) server to hamper static and dynamic analysis.

\paragraph{Dynamic code modification (DCM)}
Just-in-time compilation, self-modifying code, custom dynamic class loading, and use of functionality such as JavaScript's \texttt{eval()} function that allows to execute code extracted from (dynamically generated) strings.
This excludes packing/encryption techniques, as discussed below.

\paragraph{Packing/encryption (ENC)}
Packing/encryption/compression of software components at different levels of granularity (binaries, text sections, functions, basic blocks, \ldots) to be unpacked/decrypted/decompressed at run time.

\paragraph{Environmental requirements (ERE)}
Checks to identify being emulated, checks for path variables, and for other properties of the host system on which the software runs to detect and block the use of certain tools.
Excludes anti-debugging techniques that check for the presence of a debugger or try to prevent being debugged.
Includes techniques to delay execution to avoid detection in short-running dynamic analysis.

\paragraph{Hardware-assisted protection (HWO)}
Hardware-assisted protections such as USB-dongles, SGX enclaves, hardware-supported software encryption, etc.

\paragraph{Virtualization (VIR)}
Virtualization-based obfuscation, i.e., replacing code in a native, well-known real or virtual ISA by code in a custom virtual ISA, and an interpreter thereof.
This includes lightweight forms in which only opcodes are remapped from their standard values to a custom numbering.

\subsection{Analysis Definitions}\label{sec:analysis_definitions}
Similar to how we analyzed the deployment of protections, we classified the papers' use of code analysis methods and features thereof to evaluate the strength of obfuscations against attackers' toolboxes.

Earlier remarks on our evaluation of deployed protections in the previous section regarding sub-/superclasses of methods and possible biases also apply here.

\paragraph{Abstract interpretation (AI)}
General theory of the sound approximation of a program's semantics.
Parts of the program are abstracted (simplified) and then interpreted step-by-step.

\paragraph{Constraint based analysis (CBA)}
Defining the specifications of a program analysis in a constraint language and using a constraint solver to automate the implementation of the analysis.

\paragraph{Control flow analysis (CFA)}
Analyzing a program's control flow on the basis of reconstructed control flow graphs or call graphs.

\paragraph{Cryptanalysis (CRA)}
Custom analyses techniques of cryptographic algorithms, this includes algebraic attacks.

\paragraph{Disassembly / CFG reconstruction (DIS)}
Techniques to disassemble a program (i.e., identify its instructions) and to (re)construct the functions and their control flow graphs.
This is only marked for publications that specifically aim for hampering or improving the disassembly process and the CFG reconstruction, not for papers that merely consider reconstructed control flow graphs or disassembled code as a starting point for further analysis.

\paragraph{Diffing (DIF)}
Similarity detection; use of diffing tools; looking for similar patterns or corresponding code fragments in one or more software versions.
This excludes the identification of a priori determined patterns, in which case the technique would be classified as pattern matching.

\paragraph{Data flow analysis (DFA)}
Analyzing the data flow and data dependencies in a program.
This includes performing alias analysis. Excludes slicing.

\paragraph{Dynamic event monitoring (DEM)}
Monitoring of events during the execution of a program.
This includes API call and system call monitoring, function hooking (e.g., via interposers or detours), code breakpoints, and data watches in debuggers, etc.

\paragraph{Fuzzing (FUZ)}
Analyzing the operation or behavior of a program when executing it on numerous inputs, which might be invalid, unexpected, or random data.
This includes brute-forcing.

\paragraph{Human analysis (HUA)}
Manual analysis conducted by a human, manual reverse engineering activities such as studying code fragments to comprehend them, or browsing through lists of filtered fragments.

\paragraph{Library dependency analysis (LIB)}
Analyses to identify used/imported/exported libraries and invocations of their APIs.

\paragraph{Lifting (LIF)}
Decompilation of code in a concrete low-level representation to a more abstract (intermediate) representation.
This excludes the mere disassembling of code.

\paragraph{Machine Learning (ML)}
The use of machine learning for optimizing or training an analysis.
Can be combined with other analyses.
For example, when a pattern matcher is optimized using machine learning techniques, the paper is classified as using both pattern matching and machine learning.

\paragraph{Memory dumping (MD)}
Obtaining one or more snapshots of (part of) the memory state of a program under execution for analysis.

\paragraph{Model checking (MCH)}
Using formal methods to prove or disprove the correctness of a certain system.
Excludes the use of model checking techniques to identify path conditions as part of symbolic execution or fuzzing.

\paragraph{Network analysis (NEA)}
Sniffing of network traffic, looking at the network data from outside of the binary.

\paragraph{Normalization (NOR)}
Performing a code normalization/canonicalization step.

\paragraph{Out-of-context execution (OOC)}
Executing parts of a program on chosen inputs without executing the whole program as is.
This can be achieved with a debugger by injecting a custom \enquote{main} function into an existing program, by interposing function calls, by extracting code from a program and repacking it into another program, etc.

\paragraph{Pattern matching (PAT)}
Signature-based analysis, pattern matching on, e.g., data, call sequences, instruction sequences, regular expressions obtained somehow, etc.

\paragraph{Sandboxing (SAN)}
Emulating or simulating a program, or running it in a virtualized environment to observe its execution.

\paragraph{Slicing (SLI)}
Generating a subset of a program that includes all code statements that might affect the value of a variable at a certain statement for all possible inputs.

\paragraph{Statistical analysis (SAN)}
All kinds of program analyses that rely on statistics.
This includes, e.g., entropy measurements and counting the occurrences of certain instructions or values or events, etc.
Importantly, this only includes the use of statistical methods during an attack or analyses, but excludes the a priori use of statistics during the learning phases of analyses based on machine learning.
Furthermore, a paper is not classified as using statistical analyses if statistics are merely used to evaluate a presented analysis technique in the paper.

\paragraph{Symbolic execution (SYM)}
Exploring many possible execution paths by interpreting the execution of code on symbolic rather than concrete inputs.
Analyzing which paths can be executed and under which conditions.

\paragraph{Taint analysis (TNT)}
Marking and tagging variables/input/output or other data of a program and tracking which data and computations depend on the tagged data or on which the tainted data depends.

\paragraph{Tampering (TAM)}
Statically or dynamically altering a program or its state.
In this survey, we only consider tampering with the goal of enabling some analysis (e.g., to circumvent anti-debugging techniques or to revert obfuscations), not tampering to obtain a program with altered functionality.
Furthermore, tampering excludes transforming an obtained program representation without altering the program or its state itself.

\paragraph{Theorem proving (THE)}
Theorem proving includes SAT/SMT analysis.
We do not classify a paper as relying on theorem proving if it is only used to determine path execution conditions during symbolic execution or fuzzing.

\paragraph{Tracing (TRA)}
Generating or analyzing a trace of a program's execution, i.e., a list of the executed operations and possibly of other features, such as the data on which they operate.

\paragraph{Type analysis (TYP)}
This includes type inference and analyses of the class hierarchy.

\subsection{Measurements Definitions}\label{sec:measurements_definitions}

For each surveyed paper, we also recorded which aspects of samples, which effects of protections on samples, and which properties of analyses and analysis results were measured in the experimental evaluation, and why those were measured.
We consider four possible reasons to perform measurements: to quantify stealth, potency, resilience, and costs.
The first three, relating to protection effectiveness, can share concrete metrics, depending on the considered analyses and protections.
In other words, the same or similar concrete measurements can be used to evaluate stealth, potency, or resilience, depending on the perspective and context.
For costs, there is less ambivalence: all measurements of different forms of costs serve only one purpose: measuring cost.

\paragraph{Stealth}
The stealth of a protection is a measure of how difficult it is to detect that the concrete protection, its class, or particular aspects of are present in a program and where.
Protections in malware are stealthy if the presence of the protection is hard to detect, not if the malware is hard to classify as malware.
We mark a paper as measuring stealth if we interpret the paper as evaluating it through some measurement, independent of whether or not the authors explicitly mention stealth.

\paragraph{Potency}
The potency of a protection is the effect that the protection has to make some analysis harder, meaning the analysis will require more time or resources to reach the same result or that the result will be less precise or less useful.
Potency is typically obtained by increasing the (apparent) complexity of the object to be analyzed and/or by lowering its suitability as input for an analysis.
Potency is hence always measured or defined with respect to one or more concrete or conceptual analyses or classes thereof.
Historically, only human, manual analysis are considered for potency, but we extend it to any analysis.
For example, when facing malware detection techniques, a deployed protection is potent if it succeeds in thwarting one or more (state-of-the-art) malware detection and classification techniques.
We mark a paper as measuring potency if we interpret the paper as evaluating potency through some measurement.

\paragraph{Resilience}
The resilience of a protection is a measure of how difficult it is to counter-attack the protection, i.e., to lower its potency with respect to some analysis.
Counter-attacks can come in the form of adaptations to that analysis or in the form of analyses and transformations that can be executed a priori.
Resilience can also be measured with respect to one or more analyses.
For example, a potent protection used to protect malware against detection is also resilient if it is hard to come up with improved or alternative detection and classification techniques to mitigate the protection's potency or if that potency can only be mitigated partially.
We mark a paper as measuring resilience if we interpret the paper as evaluating resilience through some measurement, even in case the paper does not explicitly mentions resilience.

\paragraph{Classification statistics}
These are the traditional metrics derived from false rates and true rates of identification techniques.
We mark papers in this category if the present precisions, recalls, F-scores, etc.
This excludes malware classification, for which we have a separate category.

\paragraph{Malware classification statistics}
When a paper reports classification metrics for the binary classification of samples as malware or goodware, we put it in this category.

\paragraph{Deltas/similarity}
Does the paper measure how similar two or more samples or derivations thereof are, such as original, unprotected programs vs.\ deobfuscated protected programs?

\paragraph{Entropy}
Does the paper measure or discuss entropy (or related statistical properties) of some artifacts of a sample?

\paragraph{Size}
Does the paper measure or discuss the impact of a protection on the size of a sample (e.g., file size)?

\paragraph{Manual analysis}
Does the paper discuss or evaluate the complexity of a manual, human analysis?

\paragraph{Controlled human experiment analysis}
Does the paper present the result of a controlled experiment with human subjects (e.g., with students, reverse engineers, developers, penetration testers, etc.) related to software protection?

\paragraph{Memory usage}
Does the paper report or discuss the impact of a protection on the memory consumption of a sample?
Measurements on memory usage of a performed analysis were put into the \textit{Other costs} category.

\paragraph{Opcode distribution}
Does the paper evaluate the distribution of opcodes/instructions in a sample?

\paragraph{Code complexity}
Does the paper report code complexity measurements, such as branching complexity, cyclomatic complexity, points-to set sizes, etc.

\paragraph{Applicability (code coverage)}
Does the paper report on what fraction of the samples' code a protection can be deployed?

\paragraph{Power consumption}
Does the paper evaluate a protection's effect on power consumption?
Measurements on power consumption of a performed analysis were put into the \textit{Other costs} category.

\paragraph{Attack/analysis overhead time}
Does the paper evaluate how much (more) time certain attacks or analyses require (except for manual effort) on samples?

\paragraph{Compilation time}
Does the paper evaluate how long it takes to add a protection to samples?

\paragraph{Performance overhead time}
Does the paper report the run-time overhead of protection?

\paragraph{Other costs}
Does the paper present measurements of any other costs of a protection/analysis not listed above?

\subsection{To Sample or not to sample}
\label{sec:sample-no-sample}
\begin{figure}
    \centering
    \includegraphics[width=0.99\textwidth]{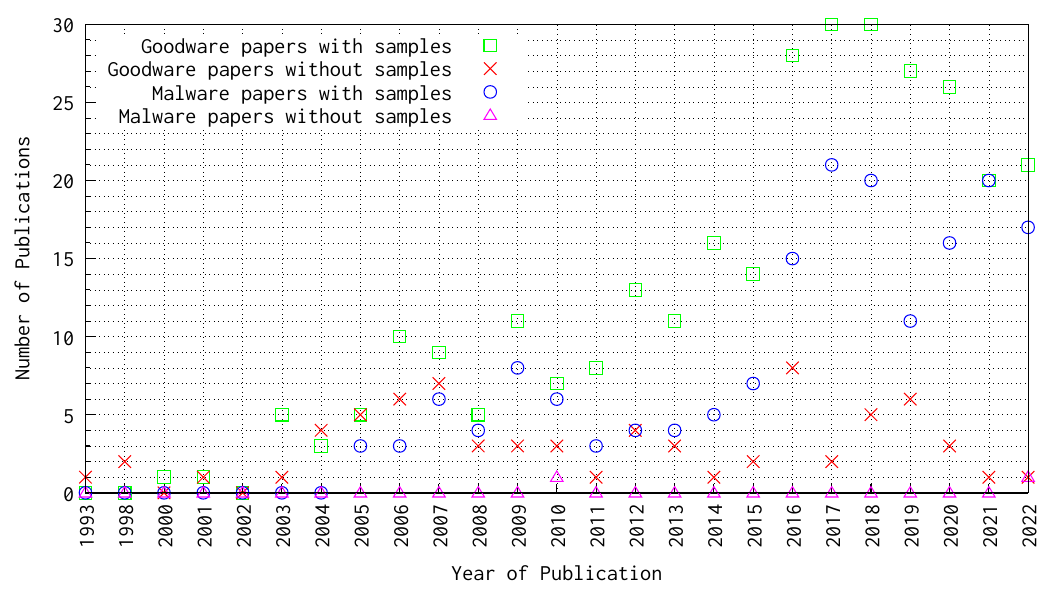}
    \caption{Number of papers per year with and without samples. Although the number of papers increases because of the methodology (see \cref{sec:methodology}), the number of papers without samples in total per year does not increase.}\label{fig:samples_no_sample}
    \Description[Number of papers per year with and without samples.]{TODO add a long Description for visually challenged readers.}
\end{figure}

\Cref{fig:samples_no_sample} shows, per paper category, how many papers perform an evaluation on samples and how many lack such an evaluation.
This information complements that of \cref{fig:sample_types}.
The low number of malware papers without samples is not surprising, given that one of the rules we used to categorize papers as malware papers is specifically that the paper uses malware samples in its evaluation, as discussed in \cref{sec:top_level_categories}.
Over the years, we do not see a general trend in the absolute number of papers without samples. The relative numbers (i.e., comparing papers without samples to total number of papers per year) indicate a downward trend.

\subsection{Correlation between Publication Venue Ranking and Sample Set Sizes}\label{sub:venue_ranking-sample_set_size-supp}

\begin{figure}[t]
    \centering
    \includegraphics[width=0.99\textwidth]{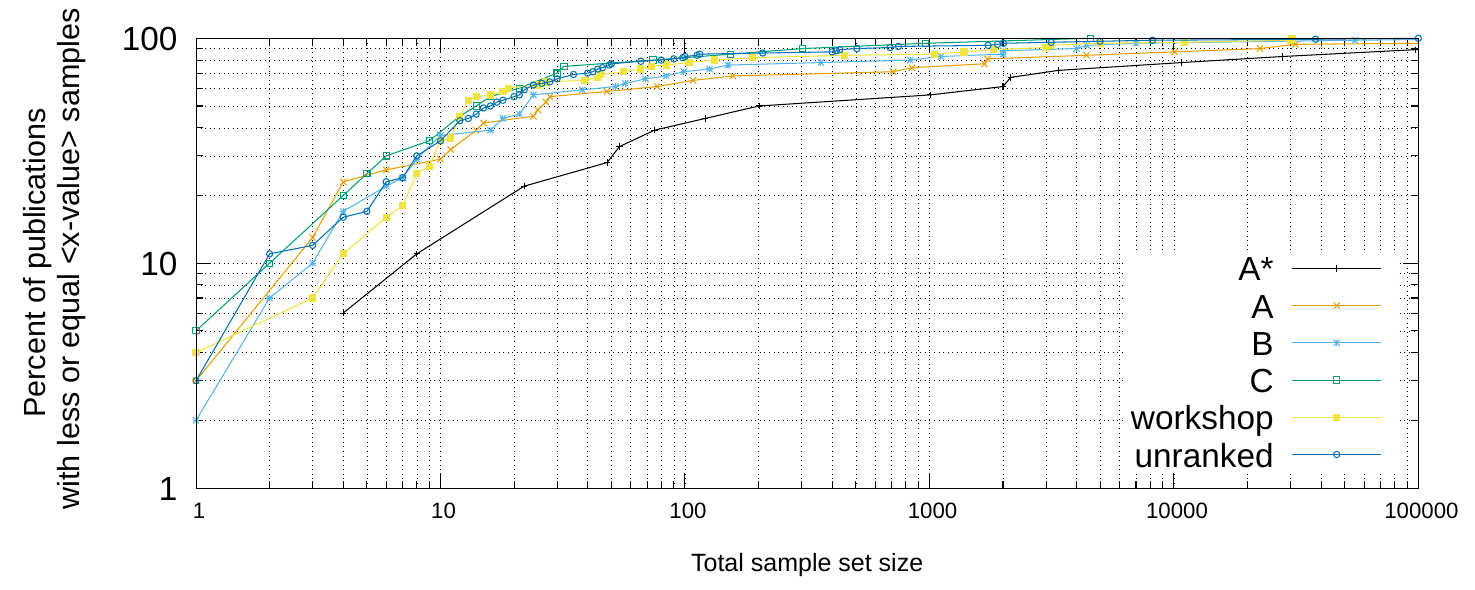}
    \caption{Cumulative graph of the sample size for goodware publications. On the x-axis we have the sample set size. On the y-axis we have the percentage of papers lower or equal to the current x-value.}\label{fig:samplesizes}
    \Description[Number of publications excluded and included per year.]{TODO add a long Description for visually challenged readers.}
\end{figure}

The cumulative graph for total sample set sizes in goodware papers in \cref{fig:samplesizes} further illustrates the observation in \cref{sub:venue_ranking-sample_set_size} on the correlation between sample total set size and publication venue ranking.
In particular, for A* venues, but also for A and B venues, it is clear that their papers consider larger sample sets than lower ranked venues.

\subsection{Deployment of Protections}\label{sec:protection_crosstabs}

\begin{figure}
    \centering
    \includegraphics[trim=0.67cm 12.41cm 0.832cm 12.514cm,clip,width=0.98\textwidth]{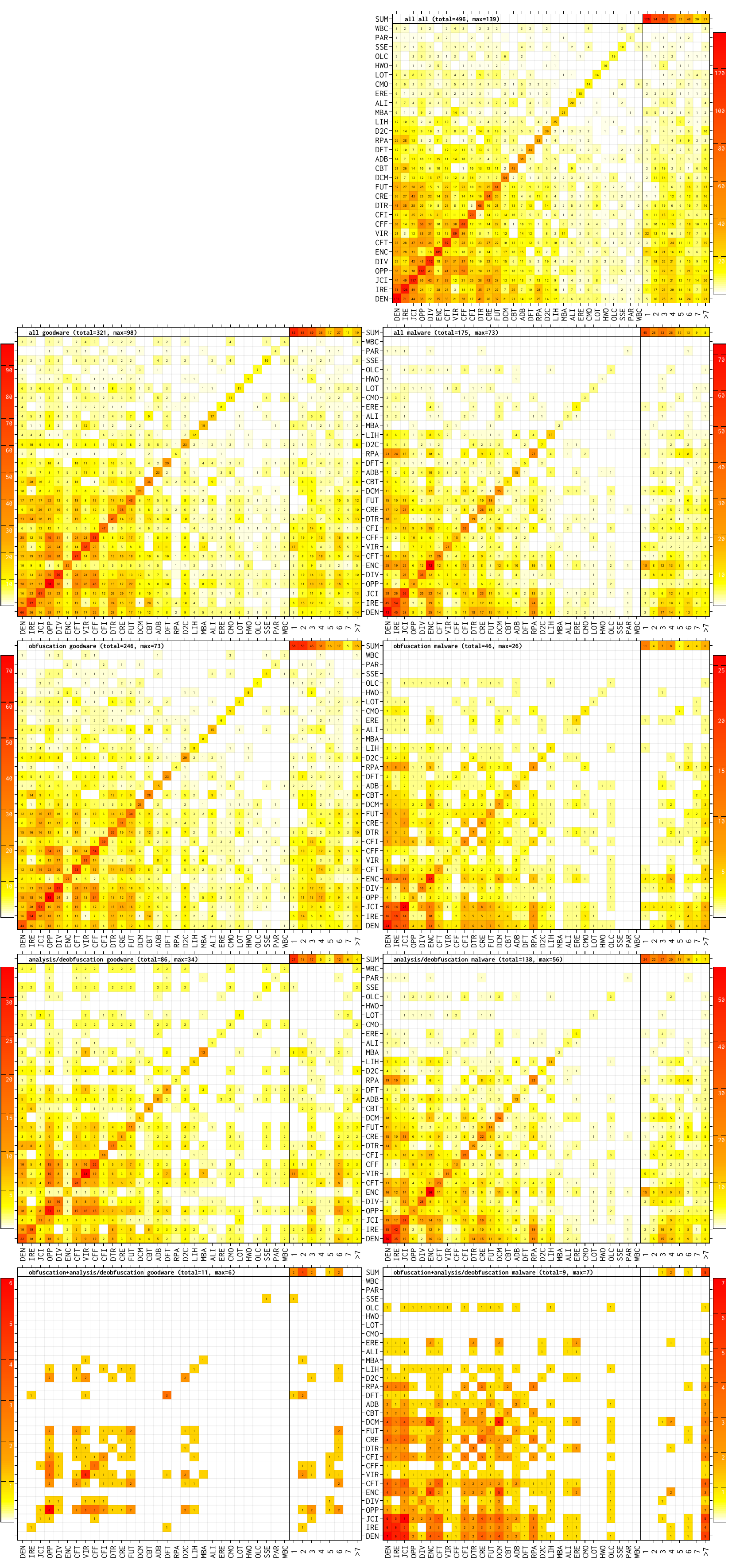}
    \caption{Protections: Six crosstabs visualizing how publications combine or layer protections. The columns on the right indicate for each protection how many protections are in the papers that deploy that protection. The acronyms are defined in \cref{sec:protection_definitions}.}\label{fig:crosstab_prot}
    \Description[TODO add description for visually challenged readers.]{TODO add a long Description for visually challenged readers.}
\end{figure}

\Cref{fig:crosstab_prot} visualizes how many papers feature implementations of protections.
It features six crosstabs for different categories of papers.
Next to each category inside the crosstab, the total number of papers reported in the crosstab is given, i.e., how many papers in that category had at least one protection implemented, as well as the top value of the color scale, i.e., the largest number occurring in the crosstab, which differs for each crosstab.
Note that we included the eleven goodware papers and seven malware papers with both obfuscation and deobfuscation/analysis contributions in both of the corresponding crosstabs. These low numbers of multi-perspective papers with protection implementations confirm our earlier discussion in \cref{sec:top_level_categories} about few papers presenting both cat and mouse moves in the ongoing war between defenders and attackers.

Each cell in the left part of each crosstab on \cref{fig:crosstab_prot} reports the number of papers that deploy at least the protection classes mentioned in its column and row. The diagonal shows how many papers deploy each type of protection; the other cells show in how many papers at least two protection classes were deployed.

For each row, the right part of each crosstab in \cref{fig:crosstab_prot} presents the distribution of the number of protections deployed in the papers that also deploy that row's protection. The top row of the right part sums up the number of papers that deploy 1, 2, \ldots{}, up to \enquote{7 and more} different protections.

\begin{figure}
    \centering
    \includegraphics[trim=0.65cm 12.4cm 0.85cm 12.5cm,clip,width=0.98\textwidth]{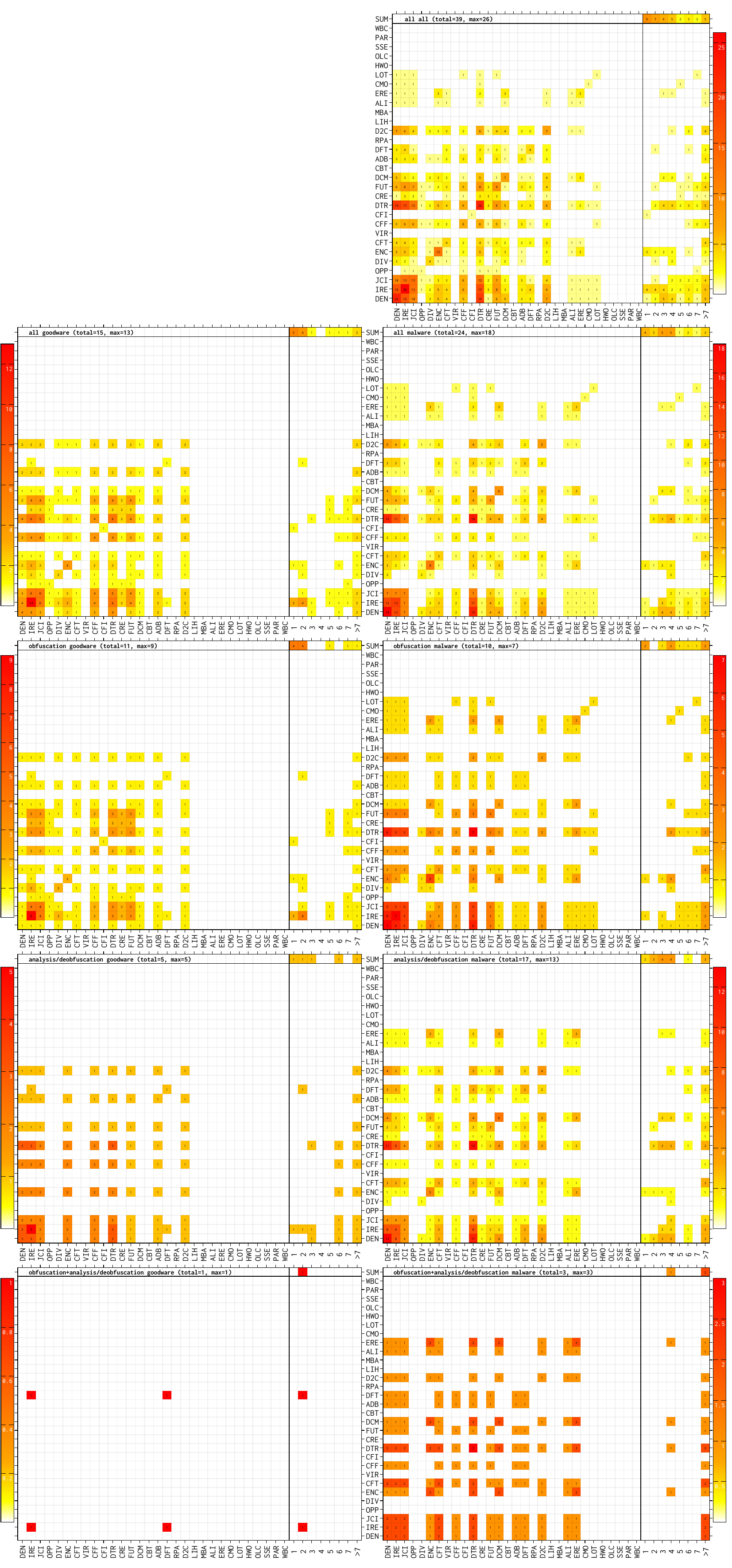}
    \caption{Script language protections: Six crosstabs visualizing how publications targeting script languages combine or layer protections, similar to \cref{fig:crosstab_prot}. The acronyms are defined in \cref{sec:protection_definitions}.}\label{fig:crosstab_prot_script}
    \Description[TODO add description for visually challenged readers.]{TODO add a long Description for visually challenged readers.}
\end{figure}

\begin{figure}
    \centering
    \includegraphics[trim=0.65cm 12.4cm 0.83cm 12.5cm,clip,width=0.98\textwidth]{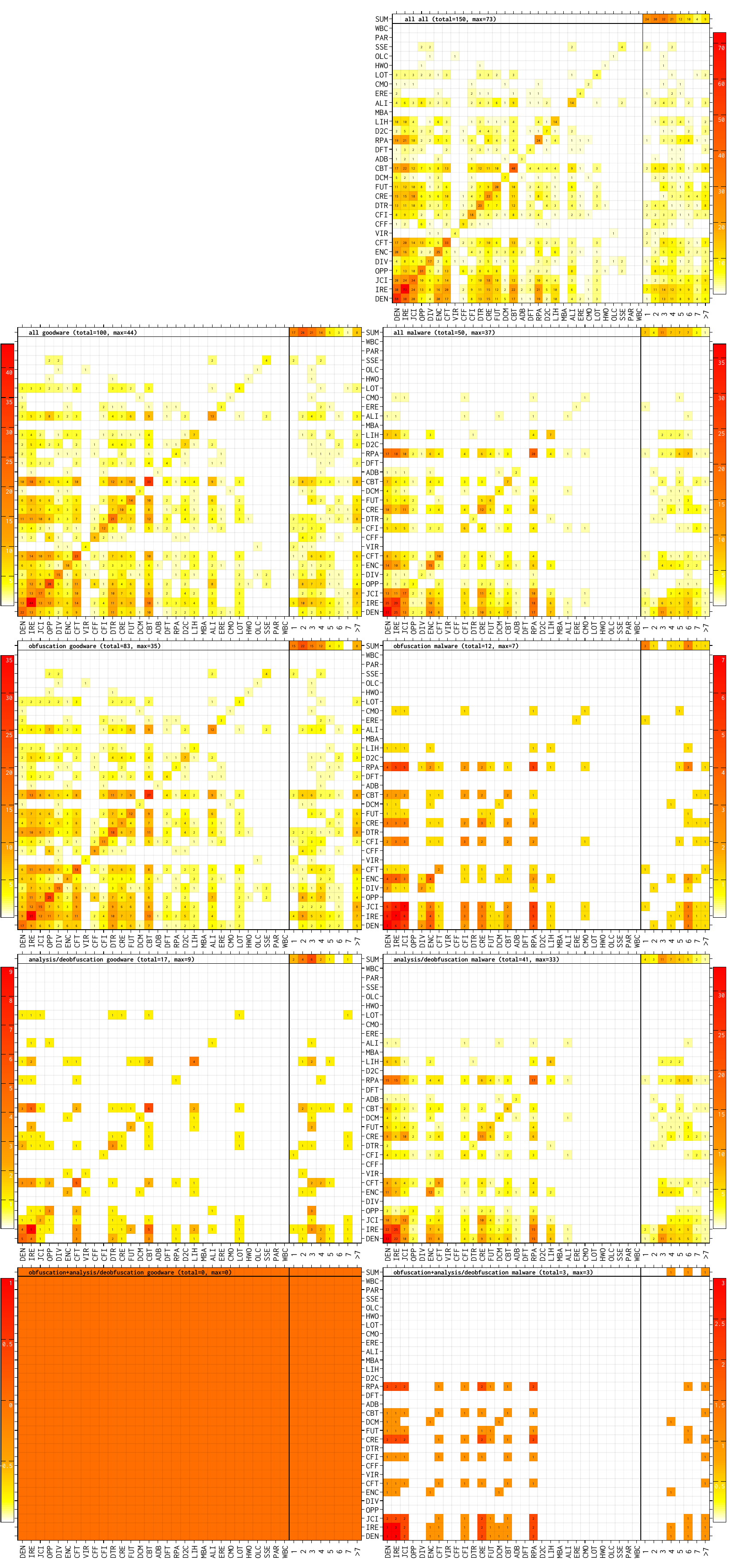}
    \caption{Managed language protections: Six crosstabs visualizing how publications targeting managed languages combine or layer protections, similar to \cref{fig:crosstab_prot}. The acronyms are defined in \cref{sec:protection_definitions}.}\label{fig:crosstab_prot_managed}
    \Description[TODO add description for visually challenged readers.]{TODO add a long Description for visually challenged readers.}
\end{figure}

\begin{figure}
    \centering
    \includegraphics[trim=0.65cm 12.4cm 0.85cm 12.5cm,clip,width=0.98\textwidth]{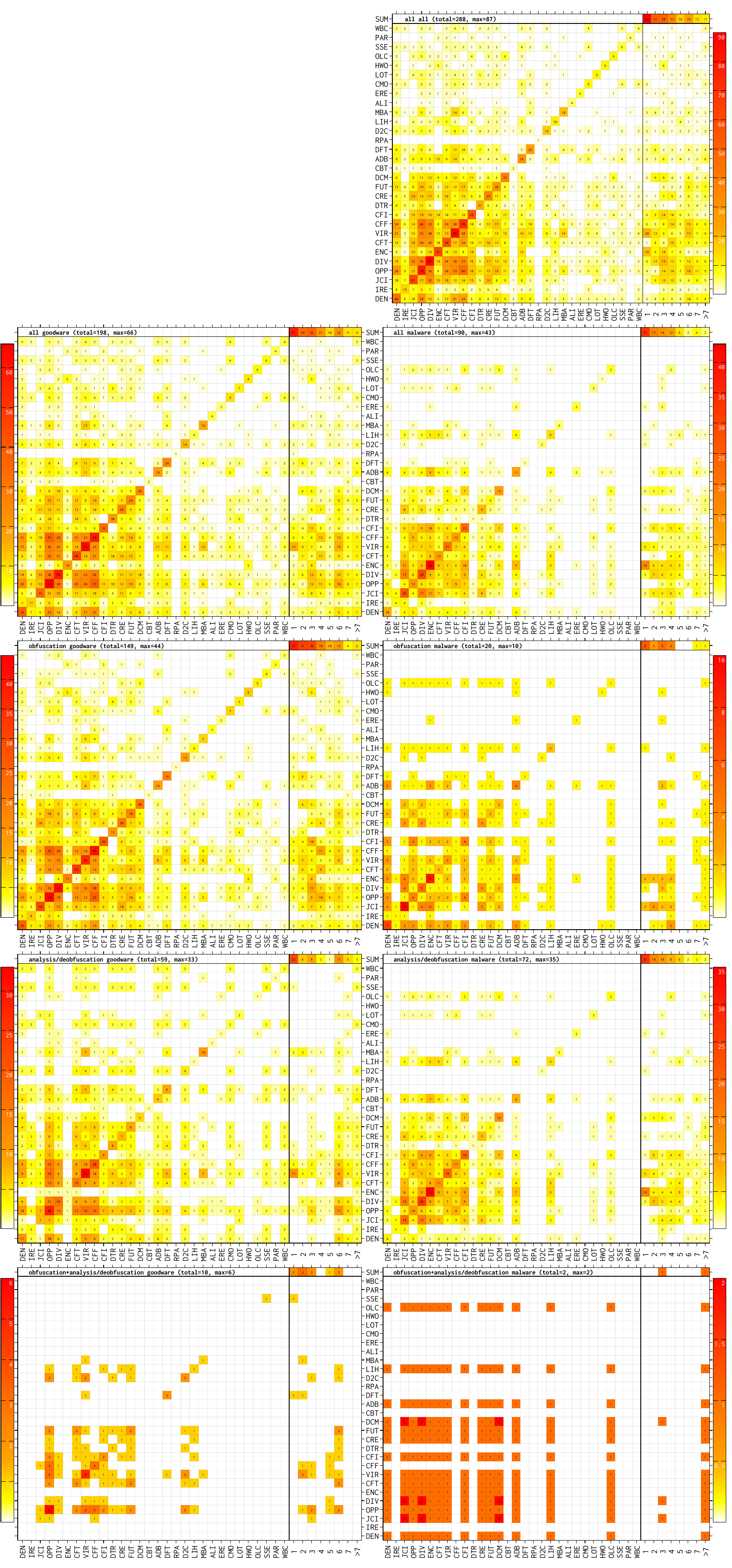}
    \caption{Native language protections: Six crosstabs visualizing how publications targeting native languages combine or layer protections, similar to \cref{fig:crosstab_prot}. The acronyms are defined in \cref{sec:protection_definitions}.}\label{fig:crosstab_prot_native}
    \Description[TODO add description for visually challenged readers.]{TODO add a long Description for visually challenged readers.}
\end{figure}

The studied protections differ from one type of programming language to another. This can be observed by comparing the refinements of \cref{fig:crosstab_prot}, where similar crosstabs are presented in \cref{fig:crosstab_prot_script,fig:crosstab_prot_managed,fig:crosstab_prot_native} per type of programming language: script, managed, and natively compiled.

\subsubsection{Individual Protections}
Complementary to the main observations discussed in \cref{sec:protections}, this section analyzes the popularity of different protections for different types of languages.

Identifier renaming is popular in script and managed languages such as JavaScript and Java, obviously because symbolic information such as class, field, and method names cannot be stripped from such software, they can only be obfuscated.
It is not popular in native languages, although some papers do research the obfuscation of C source code for protecting it when it is distributed as source code.

For script languages, few other protections are frequently studied. Only data encoding/encryption, data-to-code-conversion, and data transformation are deployed frequently on scripts, typically for obfuscating strings by encoding (as data or code) and splitting them. And in fact many protections have not been researched at all for script languages.
For some techniques, this is probably the case because they are not (directly) applicable to script languages.
Hardware-based obfuscation and overlapping code are examples.
For other techniques such as parallelization, which are not frequently researched for any type of language, the reason for those not showing up in script obfuscation papers is probably the relatively low sample size, i.e., the low number of papers.
One clear exception is virtualization, which is frequently researched for natively compiled languages and even a couple of times for managed languages, but not once for script languages.
We consider this a useful opportunity for future research, complementary to existing techniques, which, e.g., compile JavaScript to WebAssembly as a form of obfuscation~\cite{2022_wobfuscator_obfuscating_javascript_malware_via_opportunistic_translation_to_webassembly}.

Almost all papers that consider class-based transformations target managed languages, as do the vast majority of the papers that consider repackaging.
In the case of repackaging, the reason is that it is mostly used in the context of Android malware apps, which consist mostly of Java bytecode.
In the case of class-based transformations, this is due to the object-oriented nature of Java and the fact that papers targeting natively compiled code focus on obfuscations that can be deployed on imperative C code.
This is the case, e.g., for all papers that rely on the Tigress, Obfuscator-LLVM, and Diablo-based obfuscators.
While the latter two can, to some extent, also protect IR and binary code originating from C++ source code, they feature no protections specifically targeting classes.

Regarding control flow obfuscations, we notice that opaque predicates, control flow indirections, and the more general category control flow transformation are less popular for script languages than for managed and native languages.
The one exception is control flow flattening, which is far less popular for managed languages than for script, let alone native languages.
We can only speculate on the reason for this, taking into account that managed language software is obfuscated mainly by transforming bytecode, in particular, Java bytecode (and the DEX-variant thereof) as we observed in \cref{sec:obfuscation_representation}.
While we see no fundamental issue for implementing global (i.e., function-wide) obfuscating transformations such as control flow flattening for Java or DEX bytecode, we do think that it is more complex than implementing control flow flattening in script source code or in native code compiled from C source code. In particular, we think that the omnipresence of exception handling constructs in Java code could be a reason. Exception handling is implemented in Java bytecode by means of exception tables that contain an entry for each exception that is caught by each try block.
Randomly mixing code fragments from a method body, as is done for control flow flattening, requires rewriting and extending the exception tables.
This adds complexity to the obfuscators, but it might also reduce the strength of the protection, as the rewritten exception table allows one to identify which flattened fragments likely belonged together in the original code.

Looking specifically at natively compiled languages, we notice that data transformation is much less popular there than for other types of languages.
We conjecture this is due to pointer aliasing complicating data flow analysis, a prerequisite to deploying data transformation.

Regarding dynamic code modification, we observe that it is quite popular for script and native languages but less so for managed languages.
Importantly, the nature of this protection is very different in the three types of languages.
In native languages, dynamic code modification denotes binary code being updated/rewritten as the program executes.
In managed languages, dynamic code modification denotes custom class loading to alter which code gets loaded and executed.
In script languages, it denotes dynamic generation of source code not present statically, such as with an \texttt{eval()} function that is provided a string decrypted at run time.

Anti-debugging is much less popular in managed language research than in script and natively compiled languages research, despite anti-debugging being used in the vast majority of Android apps~\cite{2020_a_large_scale_study_on_the_adoption_of_anti_debugging_and_anti_tampering_protections_in_android_apps}.
The reason for this discrepancy is not clear.

\subsubsection{Combinations of Protections}

We also observed some interesting relations between protections.
For example, almost all malware papers that deploy repackaging also deploy data encoding/encryption and identifier renaming.
This is not surprising, as data encoding/encryption and identifier renaming are absolutely necessary to avoid trivial detection of repackaging.
In this regard, these papers do show maturity.
Likewise, the vast majority of the papers deploying data transformation or identifier renaming also deploy data encoding/encryption.
This is also not surprising, as all of them are typically needed to mitigate trivial malware detection.

For goodware papers, we notice that of those deploying control flow flattening, \percentGoodwareCfgOpa{}\%~$\hat{=}$~\goodwareCfgOpa{}/\goodwareCfg{} also deploy opaque predicates.
Both are popular control flow obfuscations of which the implementation requires similar kinds of analyses and transformations, so this correlation is not surprising.
Of the \goodwareLoo{}~goodware papers that deploy loop transformations, \percentGoodwareLooDat{}\%~$\hat{=}$~\goodwareLooDat{}/\goodwareLoo{} also deploy data transformation.
This is also no surprise: transforming how data is stored in arrays and transforming the loops iterating over the array's elements go hand in hand.
Perhaps more surprising is that of the \goodwareMba{}~goodware papers in which MBA is deployed, \percentGoodwareMbaVir{}\%~$\hat{=}$~\goodwareMbaVir{}/\goodwareMba{} also deploy virtualization.
This is not because MBA or virtualization have any similarity or dependence in how they are deployed.
Instead, they are related through the analysis with which they are attacked, such as dynamic symbolic execution or code synthesis techniques that can be used to deobfuscate, and are hence evaluated on, both MBA expressions and bytecode instruction handlers~\cite{2022_loki_hardening_code_obfuscation_against_automated_attacks,2021_search_based_local_black_box_deobfuscation_understand_improve_and_mitigate,2020_qsynth_a_program_synthesis_based_approach_for_binary_code_deobfuscation,2019_obfuscation_where_are_we_in_anti_dse_protections_a_first_attempt,2017_syntia_synthesizing_the_semantics_of_obfuscated_code,2017_predicting_the_resilience_of_obfuscated_code_against_symbolic_execution_attacks_via_machine_learning}.

\subsection{Deployment of Analyses}\label{sec:suppl_analysis_deployment}

\begin{figure}[p]
    \centering
    \includegraphics[trim=0.67cm 12.41cm 0.832cm 12.513cm,clip,width=0.98\textwidth]{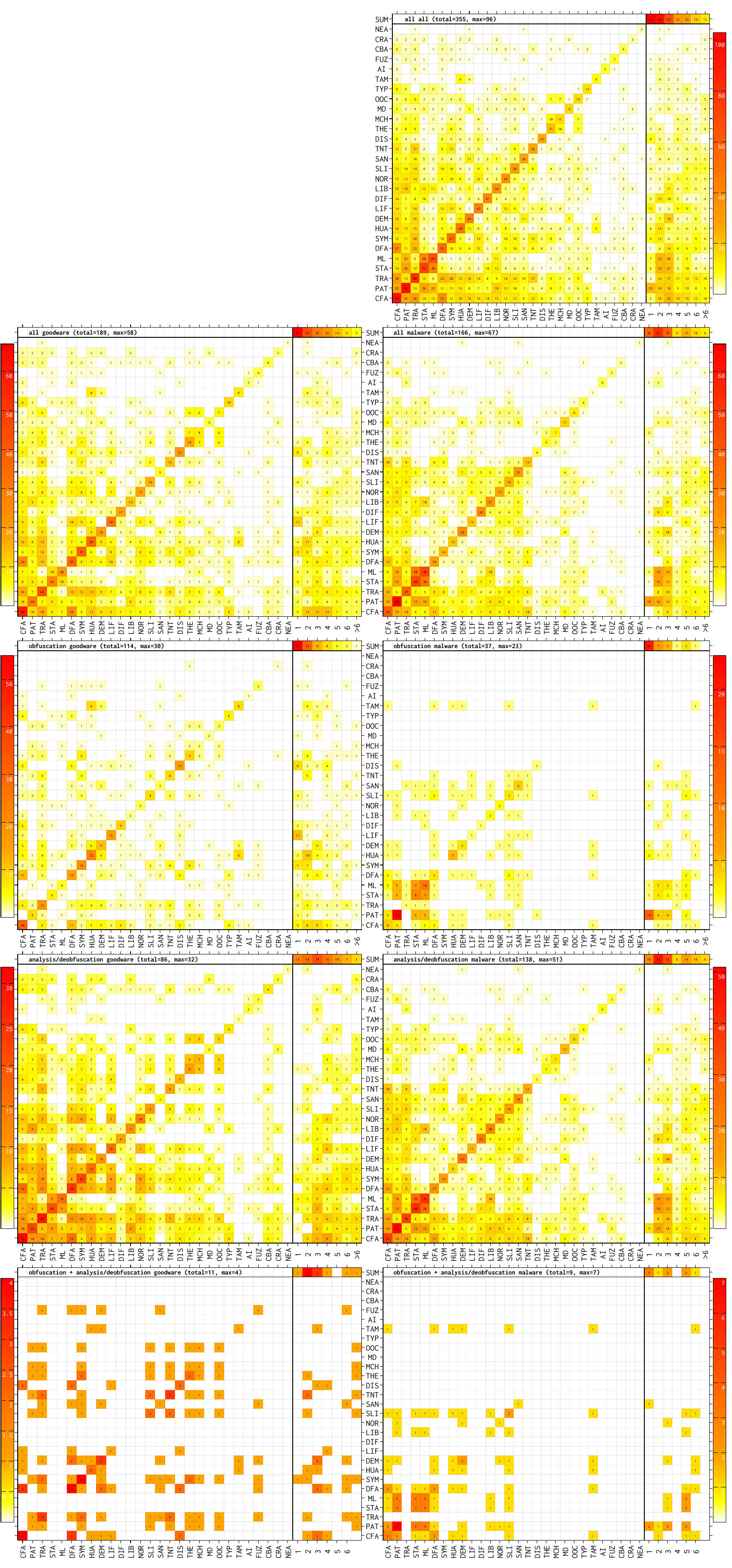}
    \caption{Six crosstabs visualizing how publications combine analysis methods. The columns on the right indicate the total numbers of papers that combine a particular number of analysis methods (1 to 8 and more). The acronyms are defined in \cref{sec:analysis_definitions}.}\label{fig:crosstab_ana}
    \Description[TODO add description for visually challenged readers.]{TODO add a long Description for visually challenged readers.}
\end{figure}

\begin{figure}[p]
    \centering
    \includegraphics[trim=0.67cm 12.41cm 0.832cm 12.514cm,clip,height=0.94\textheight]{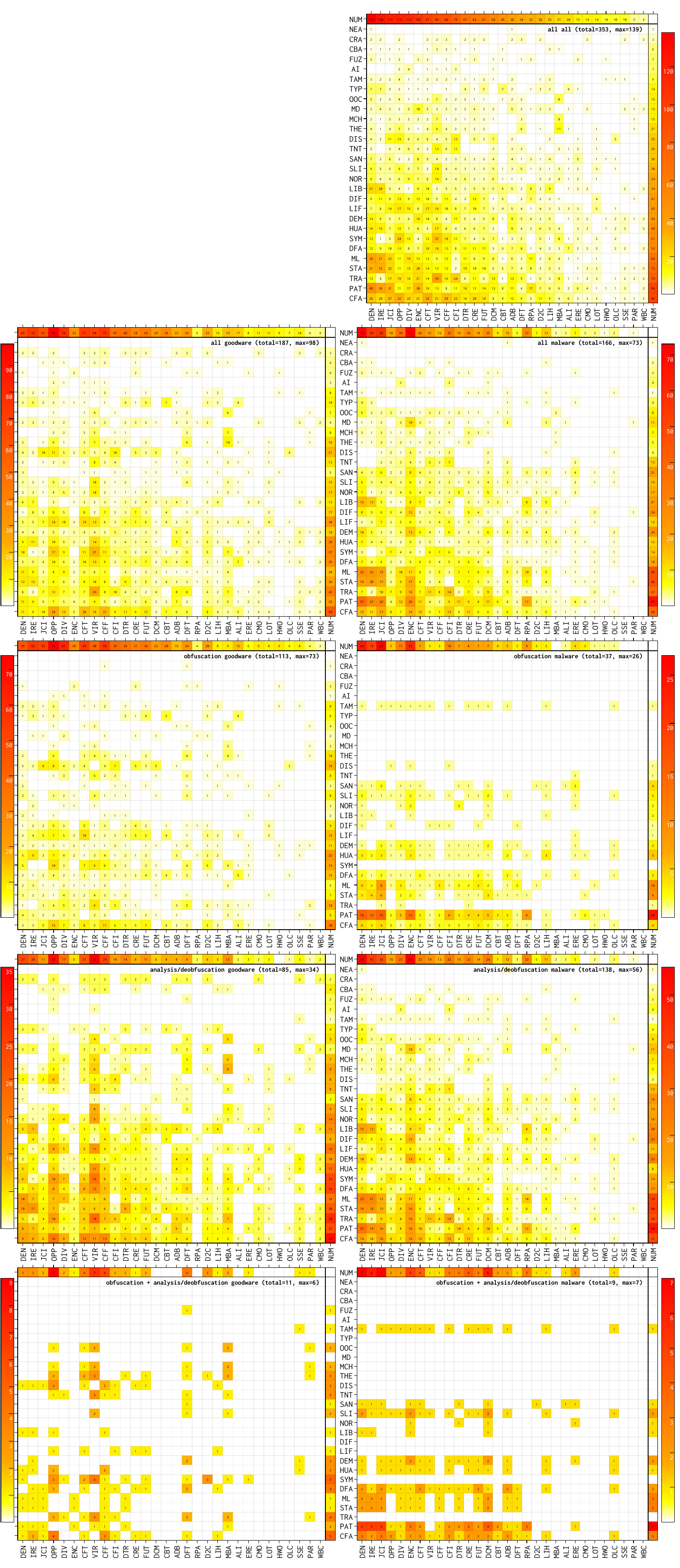}
    \caption{Crosstab analysis vs.\ protections. These crosstabs show how many papers deploy the listed analysis techniques (rows) on samples protected with the listed protections (columns) for evaluating them. The acronyms are defined in \cref{sec:analysis_definitions} and \cref{sec:protection_definitions}.}\label{fig:crosstab_anaprot}
    \Description[TODO add description for visually challenged readers.]{TODO add a long Description for visually challenged readers.}
\end{figure}

\Cref{fig:crosstab_ana} depicts the use of individual analysis methods and combinations of them for each paper category, similar to how previous crosstab figures visualized protections. Furthermore, \Cref{fig:crosstab_anaprot} depicts which protections are being evaluated with which analysis techniques in the surveyed papers.
The analysis in \cref{sec:analysis_methods} is mostly based on the numbers in these crosstabs. The ones in~\Cref{fig:crosstab_anaprot} can help researchers who design new obfuscations to select the most appropriate analysis to evaluate their obfuscations's strengths, assuming that popularity relates to appropriateness. Vice versa, those crosstabs can help researchers of new analysis techniques identify the obfuscations with which they can stress-test their analysis. In this regard, these crosstabs complement the survey by Schrittwieser et al.~\cite{2016_protecting_software_through_obfuscation_can_it_keep_pace_with_progress_in_code_analysis}.

\subsection{Software Protection Research Tools}
\label{sec:suppl_tools}
The tools used in at least five surveyed papers and listed in \Cref{tab:tools} can be categorized into compilation, obfuscation, analysis, and anti-virus tools. When some tool can be used for multiple purposes, we list in the category of its most popular usage. Unless noted otherwise, tools listed here are actively being developed and maintained at the time of writing of this paper.

For many tools, we provide concrete recommendations on their usage for SP research. These recommendations are tool-specific instantiations of the general recommendations from \Cref{sec:recommendations}. Importantly, the tool-specific recommendations need to be understood as applying to classes of tools: While we only formulate them for the more popular tools listed in \Cref{tab:tools} and below, they can also apply to similar, non-listed tools. For example, the recommendations provided below for IDA Pro also hold for Binary Ninja and Ghidra. Similarly, the recommendation for TXL can be extended to any source-to-source rewriter with which one could implement source-level obfuscations.

\subsubsection{Compilation tools}\label{sec:suppl_compilers}
\

\textbf{Recommendation:} Whatever compilers with code optimization capabilities you use to generate natively compiled samples, make sure to specify the compiler version being used and the optimization flags. Compilation of natively compiled code without any optimization (i.e., with \texttt{-O0}, which is the default option for most compilers) is absolutely not acceptable, so make sure you provide a proper optimization level flag.

\pargap
\emph{LLVM}\footnote{https://llvm.org/} is an open-source collection of modular and reusable compiler and toolchain technologies. Its core libraries for optimization and code generation are built around a well-specified code representation known as the LLVM IR. This IR is also suitable for deploying obfuscations (e.g., with OLLVM as will be discussed below), for symbolic execution (with KLEE as will be discussed below), and as a target for code lifting techniques. LLVM is hence used in research as a build tool, as a protection tool, and as an analysis tool.

\pargap
\emph{Clang}\footnote{https://llvm.org/} is the open-source frontend of LLVM for C-like languages, including C, C++, and Objective-C. It can also be used for source-to-source transformations and, hence, for protecting software.

\pargap
\emph{GCC} \footnote{http://gcc.gnu.org/} or the open-source GNU Compiler Collection is Linux' default compilation tool flow. As GCC's design is not as modular as LLVM, it is much less popular for implementing protections, and it is used much less for analysis. As for the latter, GCC is only used to assess the resilience of obfuscations against compiler-like code analysis and optimizations.

\pargap
\emph{Visual Studio}\footnote{https://visualstudio.microsoft.com/} is a proprietary IDE developed by Microsoft. It is used primarily for building applications but also as a debugger for dynamic analysis and, in one case, for implementing a protection scheme.

\subsubsection{Obfuscation tools}
\

\emph{Tigress}\footnote{https://tigress.wtf/} is a state-of-the-art academic SP tool for the C programming language. It is developed by C.\ Collberg from the University of Arizona, and by far the most popular tool for obfuscation research. Its non-commercial use is free, and source code is available to academics upon demand. Tigress is capable of a wide variation of transformations (such as virtualization, control flow flattening, opaque predicates, MBA expressions, self-modifying code, etc.) which it can compose in a layered fashion. While Tigress approaches ten years of age, it is still under active development.

\textbf{Recommendation:} Tigress is a complex tool that requires thoughtful decision-making, and hence a considerable effort, to select which protections to deploy, on which program fragments, and with which configurations. Make sure to describe your choices and provide convincing arguments for them. Importantly, over time, the default configuration options for Tigress have evolved, implying that the defaults that will be mentioned in the future on the Tigress website for its latest version might no longer reflect the default options of the version you used at the time of your research. So, make sure to mention what the default options are if you rely on them in your research.

\pargap
\emph{OLLVM}\footnote{https://github.com/obfuscator-llvm/obfuscator} or Obfuscator-LLVM is an open-source, academic software protection tool built upon the LLVM compilation suite~\cite{2015_obfuscator_llvm_software_protection_for_the_masses}. It supports a number of relatively simple control flow obfuscations and diversifications applied to the compiler's IR. Unlike Tigress, OLLVM has only seen active development for a brief period, namely in the years 2015--2017, and has not been maintained since.

\textbf{Recommendation:} Compared to Tigress, OLLVM is a weak and mostly obsolete protection tool, so using it to generate protected samples is discouraged.

\pargap
\emph{VMProtect}\footnote{https://vmpsoft.com/} is a commercial code virtualization tool that transforms the original binary code into bytecode that is interpreted by a VM embedded in the application. The VM code is obfuscated and hardened against code analysis, and on top, this protection is layered with encryption, anti-debugging, and anti-process-virtualization techniques to mitigate reverse engineering.

\textbf{Recommendation:} Compared to Tigress, VMProtect is more focused on a single protection. It still requires configuration, however, to select the level of protection and the code to be protected. So as with Tigress, make sure to do a proper configuration,and report it in your paper.

\pargap
\emph{Code Virtualizer}\footnote{https://www.oreans.com/codevirtualizer.php} is another commercial code virtualization tool.

\textbf{Recommendation:} Similar to VMProtect, Code Virtualizer is mostly focused on a single protection. It still requires configuration, however, so spend the necessary effort to select configurations, and report them in your paper.

\pargap
\emph{Themida}\footnote{https://www.oreans.com/themida.php} is a commercial SP tool that includes a wide range of techniques, including obfuscation, tamperproofing, and preventive techniques such as dynamic encryption, anti-debugging, metamorphic code, code virtualization, and API-wrapping.

\textbf{Recommendation:} Compared to VMProtect and Codevirtualizer, Themida offers more configuration options, and hence requires more configuration effort, so spend the necessary effort to select configurations, and report them in your paper.

\pargap
\emph{UPX}\footnote{https://upx.github.io/} is a simple, free executable packer.

\pargap
\emph{Zelix Klassmaster}\footnote{https://zelix.com/klassmaster/index.html} is a commercial Java obfuscator. Its features include identifier renaming, string encryption, control flow obfuscation, integer constant encryption, and type obfuscation.

\textbf{Recommendation:} As with other configurable commercial SP tools, you need to spend the necessary effort to select configurations, and report them in your paper.

\pargap
\emph{Allatori}\footnote{https://allatori.com/} is a commercial Java obfuscation tool capable of, among others, identifier renaming, control flow obfuscation, string encryption, and debug-info obfuscation, anti-decompilation techniques, and watermarking.

\textbf{Recommendation:} As with other configurable commercial SP tools, you need to spend the necessary effort to select configurations and report them in your paper.

\pargap
\emph{DashO}\footnote{https://www.preemptive.com/products/dasho/} is another commercial Java and Android software protection tool that offers, among others, identifier renaming, control flow obfuscation, string encryption, watermarking, tamperproofing, anti-debugging, anti-emulation, and anti-hook protections.

\textbf{Recommendation:} As with other configurable commercial SP tools, you need to spend the necessary effort to select configurations, and report them in your paper.

\pargap
\emph{Sandmark}\footnote{http://sandmark.cs.arizona.edu/} was an academic SP framework/tool developed at the University of Arizona to study software watermarking, tamper-proofing, and code obfuscation of Java bytecode, including decision support for deploying the supported protections~\cite{2003_sandmark_a_tool_for_software_protection_research}. It has not been maintained or developed since August 2004.

\textbf{Recommendation:} Being unmaintained for 20 years now, Sandmark should no longer be used for SP research.

\pargap
\emph{ProGuard}\footnote{https://github.com/Guardsquare/proguard} is GuardSquare's proprietary but free tool for Android mobile app developers. ProGuard is mainly aimed at shrinking Java and Kotlin apps and obfuscates them as a side-effect. The resulting protection is much weaker than what is available in tools that specifically target software protection, such as GuardSquare's non-free DexGuard tool. While ProGuard is one of the few popular proprietary obfuscators in SP research, its representativeness of real-world obfuscation can hence be questioned.

\textbf{Recommendation:} Given its lack of advanced SPs, the use of ProGuard for state-of-the-art SP research is questionable.

\pargap
\emph{Diablo}\footnote{https://diablo.elis.ugent.be/} is a proof-of-concept, open-source link-time binary rewriting framework developed at Ghent University. It has been used in research to deploy protections on binary code, including as the binary-level protection tool in the already aforementioned ACTC. Furthermore, it has been used for program analysis, such as for implementing program instrumentation in the style of the PIN tool listed below. Contrary to PIN, Diablo-based instrumentation is static. In 2020, the active maintenance and development of Diablo stopped.

\textbf{Recommendation:} Diablo, only supported long outdated versions of Android, GCC, LLVM, GNU binutils, and the GNU C library. Furthermore, it lacks full support for today's most used instruction set architectures such as 64-bit Intel, AMD, and ARM architectures. We therefore advise against using it for SP research.

\pargap
\emph{ACTC} \footnote{https://github.com/aspire-fp7/actc} or the ASPIRE Compiler Tool Chain was a proof-of-concept, open-source toolchain for protecting native ARM Android libraries and native Linux libraries/binaries.
It comprised several source-to-source rewriting stages as well as several binary-rewriting stages for layering a range of SPs that implement an SP reference architecture~\cite{2016_a_reference_architecture_for_software_protection}, with GCC and LLVM compilers used in between for compiling the software.
The ACTC supported a range of control flow and data flow obfuscations, remote attestation, code mobility~\cite{2015_software_protection_with_code_mobility}, code renewability~\cite{2020_code_renewability_for_native_software_protection}, as well as anti-debugging~\cite{2016_tightly_coupled_self_debugging_software_protection,2020_resilient_self_debugging_software_protection}.

\textbf{Recommendation:} As the ACTC builds on Diablo, we advise against using it for SP research, for the same reason.

\pargap
\emph{TXL}\footnote{https://www.txl.ca/} is a domain-specific language and toolset for implementing source-to-source rewriters. It is leveraged in the aforementioned ACTC to implement its source-level protections.

\textbf{Recommendation:} Using source-to-source rewriters for deploying SPs is fine. However, in case your SPs target natively compiled languages such as C or C++, there is the caveat that evaluating the SPs' strength should not rely solely on source-level measurements. Instead, the evaluation needs to include measurements and experiments on properly built binaries (see our recommendations in \Cref{sec:suppl_compilers} on compilers). Critically, check and report how the transformations applied on the source code are reflected in changed binary code. Optimizing compilers can radically transform code, including by eliminating code that is functionally redundant, so it is paramount to validate that the source-level transformations survived the optimizing compiler as intended.

\begin{figure}[p]
    \centering
    \includegraphics[trim=0.67cm 12.41cm 0.832cm 12.514cm,clip,width=0.98\textwidth]{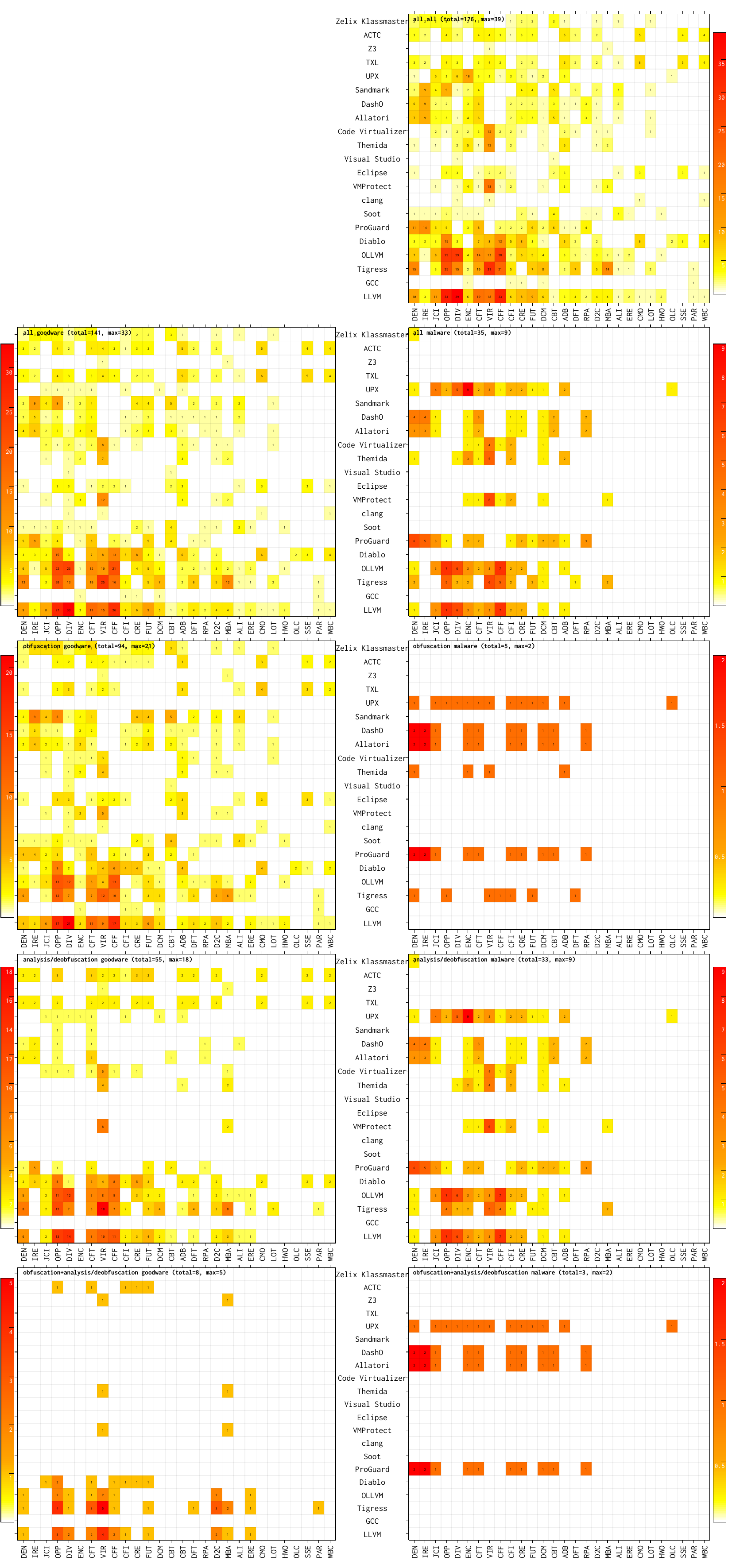}
    \caption{Crosstab Tools and Protection Techniques. For papers that evaluate samples protected with a specific tool, this crosstab shows which protections are deployed in those papers, i.e., how many of those papers deploy the different protections.}\label{fig:crosstab_tools_prot}
    \Description[TODO add description for visually challenged readers.]{TODO add a long Description for visually challenged readers.}
\end{figure}

\pargap
The crosstab in \Cref{fig:crosstab_tools_prot} shows how the deployment of protection tools and the deployment of different types of protections overlaps in the surveyed papers.

\subsubsection{Analysis/Deobfuscation tools}
\

\emph{IDA Pro}\footnote{https://hex-rays.com/IDA-pro/} is the most widely used interactive disassembler for natively compiled code. This proprietary tool from Hex-Rays also serves as a debugger, thus integrating static and dynamic analysis. The user can override IDA Pro's decisions and provide hints to improve the disassembly results. The user can also inspect, query, and edit the disassembled code through a range of views and interfaces. IDA Pro's functionalities can be extended with plug-ins and scripts. Some popular types of plug-ins and tools that build on IDA Pro are differs (e.g., BinDiff), decompilers, and library function identification tools such as F.L.I.R.T.

\textbf{Recommendation:} IDA Pro is extensible and it offers an API to manipulate the data it extracted from a binary, such as the functions it reconstructed, their control flow graphs, and other data. Exploit these features in the way a real reverse engineer would do, such that your evaluation by means of IDA Pro is representative. Hex-Rays' yearly plug-in contest\footnote{http:hex-rays.com/contest} and their plug-in repository\footnote{https://plugins.hex-rays.com/} can provide inspiration for ways to exploit IDA Pro's capabilities, and so do some research papers that included plug-ins in their evaluations~\cite{2008_eureka_a_framework_for_enabling_static_malware_analysis,2009_unpacking_virtualization_obfuscators,2009_visualizing_compiled_executables_for_malware_analysis,2010_experiences_in_malware_binary_deobfuscation,2017_vmattack_deobfuscating_virtualization_based_packed_binaries,2017_backward_bounded_dse_targeting_infeasibility_questions_on_obfuscated_codes,2019_fine_grained_static_detection_of_obfuscation_transforms_using_ensemble_learning_and_semantic_reasoning,2019_defeating_opaque_predicates_statically_through_machine_learning_and_binary_analysis,2019_asm2vec_boosting_static_representation_robustness_for_binary_clone_search_against_code_obfuscation_and_compiler_optimization,2021_obfuscated_integration_of_software_protections,2021_sok_automatic_deobfuscation_of_virtualization_protected_applications}.

In addition, it is critical to be aware that the heuristics that IDA Pro implements to disassemble a program and to model the program in terms of its components are unique to IDA Pro. Radically different approaches exist, however, so relying on only IDA Pro cannot give a complete answer to how disassemblers are impacted by new (anti-disassembly) obfuscations. For example, IDA Pro considers each code byte to be part of, at most, one instruction and each basic block to be part of only one function. Binary Ninja differs in those two aspects, as a result of which it can deal with overlapping instructions much better and handles bogus interprocedural control flow very differently from IDA Pro~\cite{2021_obfuscated_integration_of_software_protections}. In short, be careful when only relying on IDA Pro; instead consider also using Binary Ninja, Ghidra, Radare2, and other comparable tools.

For additional recommendations regarding the evaluation of anti-disassembly obfuscations and disassemblers, we refer to Section 4.1 of the report of the 2019 Dagstuhl Seminar on Software Protection Decision Support and Evaluation Methodologies~\cite{2019_software_protection_decision_support_and_evaluation_methodologies_dagstuhl_seminar_19331}.

\pargap
\emph{PIN}\footnote{https://www.intel.com/content/www/us/en/developer/articles/tool/pin-a-dynamic-binary-instrumentation-tool.html} by Intel is a dynamic binary instrumentation framework for the the IA-32 and x86-64 instruction-set architectures that enables the creation of dynamic program analysis tools. Its main use is the generation of execution profiles and traces. PIN has been used in hundreds of research projects in the domains of computer architecture, programming language design and implementation, and software engineering.

\textbf{Recommendation:} For dynamic analysis techniques, in particular trace-based techniques, it is important to evaluate (experimentally or with an informed argument) their scalability to sufficiently complex dynamic behaviors, such as traces in which the boundaries between relevant and irrelevant trace fragments have been obfuscated, or traces in which dependencies between instructions have been obfuscated with anti-taint techniques.

For additional recommendations regarding the evaluation of trace-based analysis techniques, we refer to Section 4.2 of the report of the 2019 Dagstuhl Seminar on Software Protection Decision Support and Evaluation Methodologies~\cite{2019_software_protection_decision_support_and_evaluation_methodologies_dagstuhl_seminar_19331}.

\pargap
\emph{KLEE}\footnote{https://klee.github.io/} is an open-source dynamic symbolic execution engine that is built on top of the LLVM compiler infrastructure and operates on LLVM bitcode. KLEE has been used in hundreds of research projects on software testing. It is also used for deobfuscation, e.g., to identify unreachable bogus code.

\textbf{Recommendation:} When using KLEE on IR code generated directly from source code with LLVM, e.g., to evaluate how an obfuscation impacts symbolic execution, either provide convincing arguments as to why such experiments are representative for how attackers could use symbolic execution starting from binaries instead of source code, or complement KLEE with native code symbolic execution engines, such as BINSEC/SE~\cite{2016_binsec_se_a_dynamic_symbolic_execution_toolkit_for_binary_level_analysis} or angr~\cite{angr}. Alternatively, deploy KLEE on IR code obtained by decompiling or lifting binary code~\cite{2019_saturn_software_deobfuscation_framework_based_on_llvm}.

\pargap
\emph{Eclipse} \footnote{https://eclipseide.org/} is an open-source IDE for Java development that is extensible with different kinds of plug-ins. Such plug-ins have been used for the analysis as well as the protection of software, and some authors use Eclipse as an interface to their build tools.

\pargap
\emph{Soot}\footnote{https://www.sable.mcgill.ca/soot/} is an open-source static Java analysis and optimization framework. Its functionality to analyze and transform Java bytecode has been used widely in the static software analysis community, in particular to obfuscate and to reverse engineer such bytecode in SP research. Since December 2022, Soot has been officially succeeded by SootUp.

\pargap
\emph{angr}\footnote{https://angr.io/} is an open-source binary analysis framework written in Python~\cite{2016_sok_state_of_the_art_of_war_offensive_techniques_in_binary_analysis}. Its analysis capabilities include static disassembly, lifting, decompilation, and control flow graph recovery, as well as dynamic symbolic execution, a.k.a.\ concolic execution.

\pargap
\emph{OllyDbg}\footnote{http://ollydbg.de/} is a closed-source but free GUI-based Win32 debugger. Its latest release dates from 2013, and offers no functional 64-bit support. OllyDbg is, hence, no longer usable in research. It used to be very popular among hackers, in part because of its open architecture. Many third-party plug-ins were available, including to defeat anti-debugging software protections by interposing those protections' queries to the OS and environment for evidence of ongoing debugging.

\pargap
\emph{Z3}\footnote{https://github.com/Z3Prover/z3} is an open-source theorem prover from Microsoft Research. In the surveyed papers, Z3 is mainly used for solving path conditions during symbolic execution but also for checking the equivalence of obfuscated and deobfuscated code, such as MBA expressions.

\pargap
\emph{objdump}\footnote{https://www.gnu.org/software/binutils/} is part of the GNU Binutils, a collection of tools to inspect and manipulate binaries. It allows for the inspection of object files and executables, their header data as well as their code and data content, and includes a simple (linear-sweep) disassembler.

\pargap
\emph{Syntia}\footnote{https://github.com/rub-sysSec/syntia} is an academic deobfuscation tool~\cite{2017_syntia_synthesizing_the_semantics_of_obfuscated_code}, which is reused as a baseline in several other papers that propose alternative or improved deobfuscation techniques, and as an evaluation of novel or improved obfuscation techniques.

\textbf{Recommendation:} While Syntia is the most commonly evaluated black-box deobfuscation approach in the surveyed papers, better-performing alternatives are now available, such as Xyntia~\cite{2021_search_based_local_black_box_deobfuscation_understand_improve_and_mitigate}, that can hence replace Syntia as a baseline for comparison.

\pargap
\emph{Dex2Jar}\footnote{https://github.com/pxb1988/dex2jar} is a tool to inspect Android Dalvik Executable (DEX) files and to convert them to related bytecode file formats such as Java .class files or Smali, an IR for DEX files. It is used as a basic component in a number of non-trivial analyses/deobfuscation methods approaches, but also as a standalone tool to evaluate the effectiveness of preventive obfuscations that target this type of tool.

\pargap
\emph{BinDiff}\footnote{https://www.zynamics.com/bindiff.html} is a binary differ that builds on IDA Pro. This used to be a proprietary closed-source tool, but as of September 2023, it has been released as open-source code. Obviously, all uses collected in our survey predate that release. BinDiff is mainly used for evaluating obfuscations, but also in the reverse engineering case study of DropBox~\cite{2013_looking_inside_the_drop_box}.

\pargap
\emph{APKTool}\footnote{https://apktool.org/} is a tool for extracting resources from Android apps in the form of APK files, for rebuilding APK files after manipulation of the resources, and for disassembling the code resources into the Smali IR.

\pargap
\emph{Jad}\footnote{http://www.javadecompilers.com/jad} is a now obsolete Java bytecode decompiler that has mainly been used to study the effect of Java obfuscation on decompilation.

\pargap
\emph{AndroGuard}\footnote{https://github.com/androguard/androguard} is a Python tool for analyzing, disassembling, decompiling, and debugging Android DEX files.

\pargap
\emph{QEmu}\footnote{https://www.qemu.org/} is an open-source machine emulator and virtualizer that is popular for dynamic analysis.

\textbf{Recommendation:} Given QEmu's similarity to PIN, the same recommendations apply regarding scalability to complex enough programs.

\pargap
\emph{scikit-learn}\footnote{https://scikit-learn.org/} is a collection of open-source Python tools for machine learning that has some popularity in malware detection research.

\pargap
\emph{Arybo}\footnote{https://github.com/quarkslab/arybo} is a tool for manipulation, canonicalization, and identification of MBA expressions. Given a complex expression, it can return the bit-level symbolic representation thereof. It has been used for reverse engineering MBA expressions and VM handlers.

\pargap
\emph{Triton}\footnote{https://triton-library.github.io/} is a Python library for building dynamic binary analysis tools. It can be used for dynamic symbolic execution, dynamic taint analysis, lifting code to LLVM IR, and more.

\textbf{Recommendation:} Given that Triton can be used, among others, for similar dynamic analyses as PIN and QEmu, the same recommendations apply regarding scalability to complex enough programs.

\subsubsection{Anti-virus tools}
\

\emph{McAfee AV}\footnote{https://www.mcafee.com/} is a range of anti-virus and digital security solutions. In the surveyed papers, it is mainly used to evaluate how (deobfuscating) transformations applied on samples before feeding them to malware detection tools can improve those tools' outcomes.

\pargap
\emph{VirusTotal}\footnote{https://www.virustotal.com/} is a platform for analyzing files to detect malware.

\begin{figure}[p]
    \centering
    \includegraphics[trim=0.67cm 14.32cm 0.832cm 12.514cm,clip,width=0.98\textwidth]{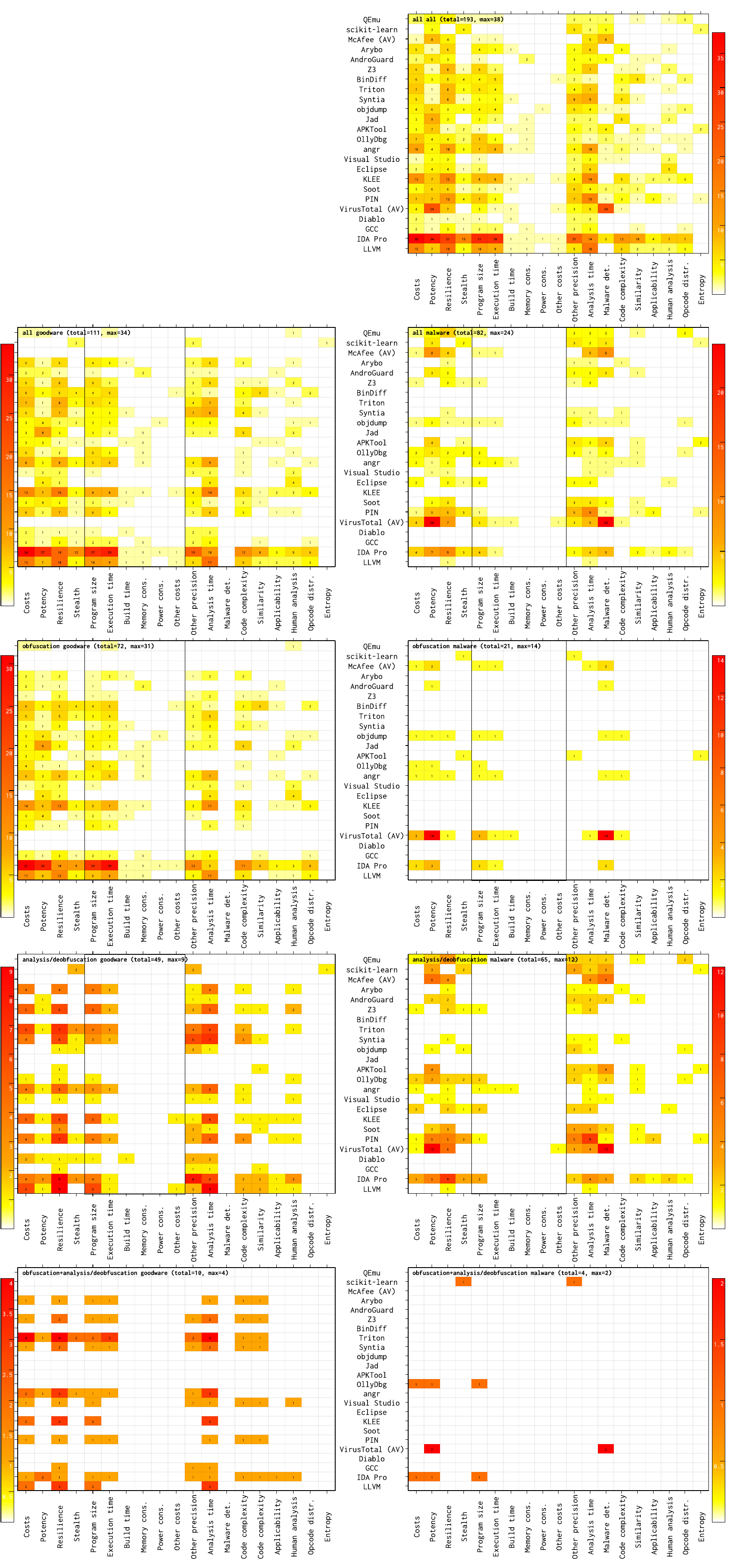}
    \caption{Crosstab Tools and Measurements. For papers that evaluate samples by analysing them with a specific analysis tool, this crosstab shows which measurements are reported in those papers, i.e., how many of those papers report the different measurements.}\label{fig:crosstab_tools_mea}
    \Description[TODO add description for visually challenged readers.]{TODO add a long Description for visually challenged readers.}
\end{figure}

\begin{figure}[p]
    \centering
    \includegraphics[trim=0.67cm 12.41cm 0.832cm 12.514cm,clip,width=0.98\textwidth]{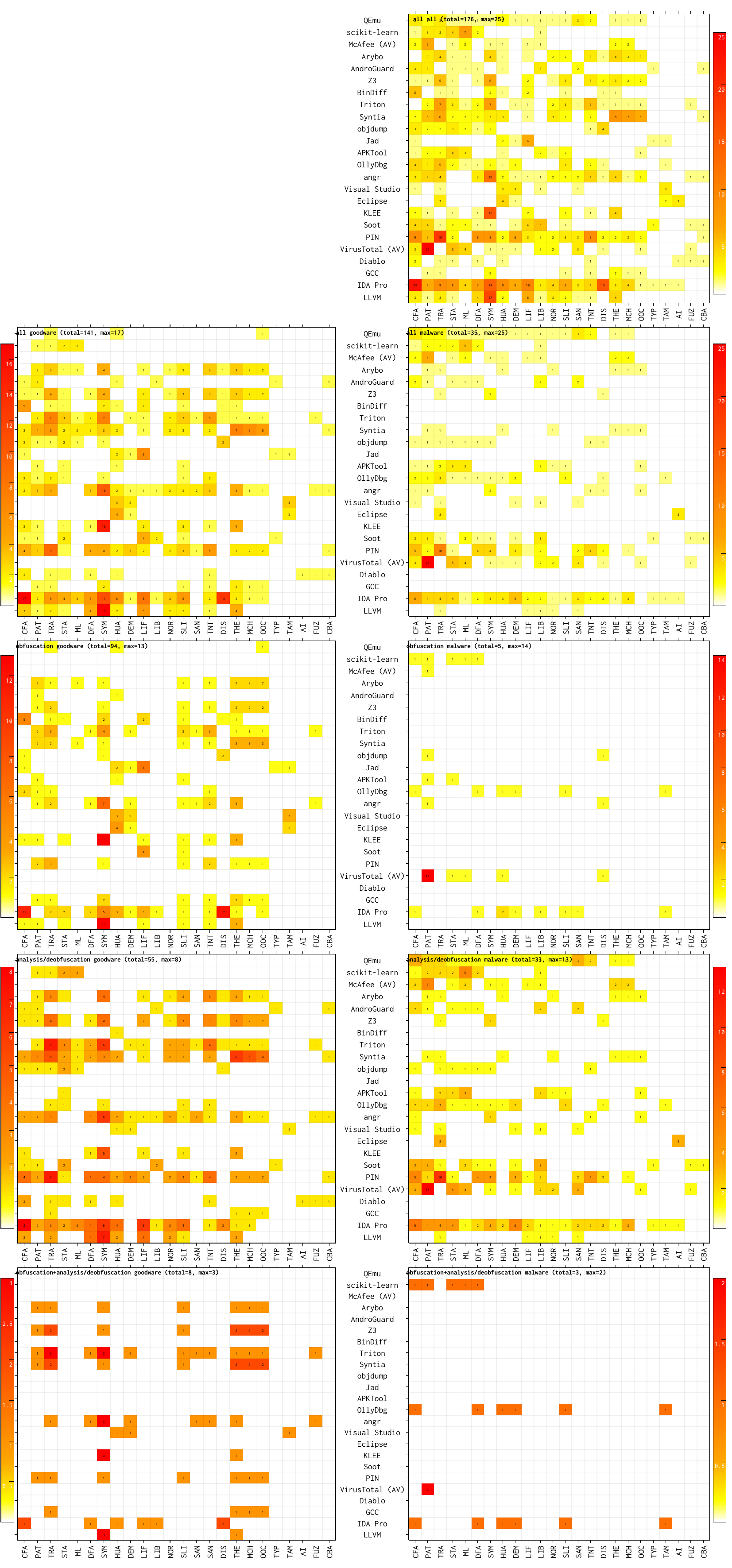}
    \caption{Crosstab Tools and Analysis Methods. For papers that evaluate samples by analysing them with a specific analysis tool, this crosstab shows which analyses are reported in those papers, i.e., how many of those papers report results for the different types of analysis.}\label{fig:crosstab_tools_ana}
    \Description[TODO add description for visually challenged readers.]{TODO add a long Description for visually challenged readers.}
\end{figure}

\pargap
The crosstabs in \Cref{fig:crosstab_tools_mea} and \Cref{fig:crosstab_tools_ana} show how the deployment of analysis tools, deobfuscation tools, and anti-virus tools in the surveyed papers overlaps with the types of measurements and the types of analysis results being reported in those papers.

\subsection{Experiments with Human Subjects}
\label{sec:suppl_human}

\Cref{tab:human_full} extends the information already presented in \Cref{tab:human} with details regarding the specific protections that were composed and layered in the samples handled by humans in the reported experiments. It is clear that few experiments included samples in which many protections were composed and layered.

We want to point out that the three largest combinations and layerings of protections listed for the two papers by Ceccato et al.~\cite{2017_how_professional_hackers_understand_protected_code_while_performing_attack_tasks,2019_understanding_the_behaviour_of_hackers_while_performing_attack_tasks_in_a_professional_setting_and_in_a_public_challenge} were selected by industrial SP experts for protecting the assets in use cases representative for their business. As such, these combinations can serve as inspiration for researchers in search for relevant combinations of protections. Another inspiration can be found on the Tigress website, which lists so-called protection recipes.\footnote{https://tigress.wtf/recipes.html}.

\begin{table}[p]
\caption{Papers that report on (controlled) experiments involving human subjects performing MATE protection and attack tasks. Subjects indicates how many subjects participated of different level of expertise, ranging from \protect\priority{00} bachelor and master students, over \protect\priority{25} PhD students and postgraduate students that are not experts in SP or reverse engineering, \protect\priority{50} students and amateurs with considerable experience in SP or reverse engineering, \protect\priority{75} professional programmers, to \protect\priority{100} professional security experts and pen testers. The samples column indicates what type of samples were handled by the human participants. The asterisks mark samples that are real-world programs rather than being just a \enquote{Complex} program that is somewhere between a toy program and a real-world program. The combinations column indicates which protections are deployed in the different samples, in which combinations, using abbreviations defined in \Cref{sec:protection_definitions}. Commas separate different samples, and + indicates that protections are composed in samples. \enquote{Vanilla} denotes unprotected samples. The \enquote{Lang}~column indicates what type of programming language was targeted: (N)ative, (M)anaged, or (S)cript. The format column indicates the format in which the software was reverse engineered: (S)ource, (N)ative, or (I)ntermediate. The time column indicates how long the experiments lasted. Question marks indicate the information is not available.}
\label{tab:human_full}
\scriptsize
\begin{tabular}{c r p{1cm} p{1.8cm} p{5.7cm} c c r}
\textbf{Paper} & \textbf{Year} & \textbf{Subjects} & \textbf{Samples} & \textbf{Combinations of protections deployed on samples} & \textbf{Lang} & \textbf{Format} & \textbf{Time} \\ \toprule
\cite{2021_input_output_example_guided_data_deobfuscation_on_binary} & 2021 & 5\priorityss{50} & Toy  & MBA, DIV & N & N & 12h \\ \midrule
\cite{2021_experimental_evaluation_of_the_obfuscation_techniques_against_reverse_engineering} & 2021 & 14\priorityss{75} & Toy  & Vanilla, CFF+OPP & M & I & ? \\ \midrule
\cite{2020_empirical_assessment_of_the_effort_needed_to_attack_programs_protected_with_client_server_code_splitting} & 2020 & 87\priorityss{0} & Complex  & Vanilla, SSE & N & S & 2h \\ \midrule
\cite{2020_creative_manual_code_obfuscation_as_a_countermeasure_against_software_reverse_engineering} & 2020 & 22\priorityss{50}\priorityss{100}\priorityss{0} & Complex  & Vanilla, ? & M & S & 1h \\ \midrule
\cite{2019_understanding_the_behaviour_of_hackers_while_performing_attack_tasks_in_a_professional_setting_and_in_a_public_challenge} & 2019 & 6\priorityss{100}\newline\newline\newline 1\priorityss{50} & *Mobile,\newline\newline\newline Toy & D2C+WBC+VIR+ADB+OPP+CFF+SSE+CMO+DIV,\newline D2C+WBC+VIR+ADB+OPP+CFF+CMO+DIV,\newline D2C+ADB+OPP+CFF+SSE+CMO+DIV,\newline OPP+CFF+ADB, DEN+ADB, DEN, WBC, VIR & N & N & 30d \\ \midrule
\cite{2019_the_impact_of_control_flow_obfuscation_technique_on_software_protection_against_human_attacks} & 2019 & 14\priorityss{75} & Toy  & Vanilla, CFF+OPP & M & I & ? \\ \midrule
\cite{2019_resilient_user_side_android_application_repackaging_and_tampering_detection_using_cryptographically_obfuscated_logic_bombs} & 2019 & 4\priorityss{75} & Mobile & DCM & M & I & 40h \\ \midrule
\cite{2018_programming_experience_might_not_help_in_comprehending_obfuscated_source_code_efficiently} & 2018 & 2\priorityss{25} 64\priorityss{0} & Complex  & Vanilla, IRE, OPP & M & I & 1.5h \\ \midrule
\cite{2018_enhance_virtual_machine_based_code_obfuscation_security_through_dynamic_bytecode_scheduling} & 2018 & 2\priorityss{25} 13\priorityss{0} & Complex, Toy & VIR+DIV & N & N & 72h \\ \midrule
\cite{2018_a_large_scale_investigation_of_obfuscation_use_in_google_play} & 2018 & 63\priorityss{75} & Mobile & ? & M & I & ? \\ \midrule
\cite{2017_stochastic_optimization_of_program_obfuscation} & 2017 & 10\priorityss{25} 10\priorityss{0} & Complex  & ? & S & S & ? \\ \midrule
\cite{2017_how_professional_hackers_understand_protected_code_while_performing_attack_tasks} & 2017 & 6\priorityss{100} & *Mobile & D2C+WBC+VIR+ADB+OPP+CFF+SSE+CMO+DIV,\newline D2C+WBC+VIR+ADB+OPP+CFF+CMO+DIV,\newline D2C+ADB+OPP+CFF+SSE+CMO+DIV & N & N & 30d \\ \midrule
\cite{2016_towards_better_program_obfuscation_optimization_via_language_models} & 2016 & 20\priorityss{75} & Complex  & ? & S & S & ? \\ \midrule
\cite{2016_comparing_the_effectiveness_of_commercial_obfuscators_against_mate_attacks} & 2016 & 1\priorityss{75} & Complex, Toy & VIR+ADB+LIH, ENC+ABD, VIR+DEN & N & N & ? \\ \midrule
\cite{2016_assessment_of_source_code_obfuscation_techniques} & 2016 & 1\priorityss{25} 14\priorityss{0} & Complex  & Vanilla, DTR & N & S & 3.5h \\ \midrule
\cite{2014_an_other_exercise_in_measuring_the_strength_of_source_code_obfuscation} & 2014 & 12\priorityss{0} & Complex  & Vanilla, IRE, OPP, IRE+OPP & M & S & 1h \\ \midrule
\cite{2014_a_family_of_experiments_to_assess_the_effectiveness_and_efficiency_of_source_code_obfuscation_techniques} & 2014 & 22\priorityss{25} 52\priorityss{0} & Complex  & Vanilla, IRE, OPP & M & I & 4h \\ \midrule
\cite{2009_visualizing_compiled_executables_for_malware_analysis} & 2009 & 6\priorityss{50} & Malware & ENC & N & N & ? \\ \midrule
\cite{2009_the_effectiveness_of_source_code_obfuscation_an_experimental_assessment} & 2009 & 22\priorityss{25} 10\priorityss{0} & Complex  & Vanilla, IRE & M & I & 4h \\ \midrule
\cite{2008_towards_experimental_evaluation_of_code_obfuscation_techniques} & 2008 & 8\priorityss{0} & Complex  & Vanilla, IRE & M & I & 4h \\ \midrule
\cite{2007_security_strength_measurement_for_dongle_protected_software} & 2007 & 5\priorityss{100} & *Complex  & ?+DEN+CFT+FUT+LIH+HWO & N & N & 80m \\ \bottomrule
\end{tabular}
\end{table}

\subsection{Overview and Description of Related Work}\label{sec:related_work_supplemental}

\Cref{tab:relwork} lists related surveys in specific topics in the SP domain in chronological order by year.

\clearpage

{
\small
\begin{longtable}{
    >{\RaggedRight\arraybackslash}m{0.7cm}
    >{\RaggedRight\arraybackslash}m{0.7cm}
    m{12cm}
}
	\caption{Overview of existing surveys in the SP domain. The second column indicates whether the surveys focus on malware (M) or goodware (G) software protection, as well as if it targets protections (P) and/or analyses including deobfuscation (A).}\label{tab:relwork} \\ \toprule
	\textbf{Ref. Year}&\textbf{M/G P/A}&\textbf{Brief description} \\ \midrule
\endfirsthead%
    \multicolumn{3}{l}{\tiny Continued from the previous page\ldots{}} \\ \toprule
	\textbf{Ref. Year}&\textbf{M/G P/A}&\textbf{Brief description} \\ \midrule
\endhead%
    \multicolumn{3}{l}{\tiny Continues on the next page\ldots{}} \\ \bottomrule
\endfoot%
	\bottomrule
\endlastfoot%
\cite{2002_watermarking_tamper_proofing_and_obfuscation_tools_for_software_protection} 2002&G\newline P&\citeauthor{2002_watermarking_tamper_proofing_and_obfuscation_tools_for_software_protection} gave an overview of SPs and describe obfuscation, watermarking and tamper proofing methods and group argue about their stealth, resilience, obscurity and costs. This work also defines a threat model for each protection.\\ \midrule
\cite{2005_code_obfuscation_literature_survey} 2005&M\newline P&\citeauthor{2005_code_obfuscation_literature_survey} categorized SPs into general, obstructing static analysis, targeting the disassembly phase, and obfuscation related to malware.\\ \midrule
\cite{2006_a_survey_of_control_flow_obfuscations} 2006&P&\citeauthor{2006_a_survey_of_control_flow_obfuscations} shortly gave an overview of the dynamic dispatcher model and opaque predicates as provable control-flow obfuscations.\\ \midrule
\cite{2010_analysis_avoidance_techniques_of_malicious_software} 2010&M\newline P&Brand's PhD thesis and its corresponding publications revised different anti-analysis techniques. While this work heavily focuses on analysis avoidance it also includes obfuscations (control flow and data transformation as well as packing). \citeauthor{2010_analysis_avoidance_techniques_of_malicious_software} implements the different anti-analysis techniques into simple programs and tries to detect or mitigate them.\\ \midrule
\cite{2010_assure_high_quality_code_using_refactoring_and_obfuscation_techniques} 2010&G\newline P&\citeauthor{2010_assure_high_quality_code_using_refactoring_and_obfuscation_techniques} opposed refactoring methods and obfuscation methods regarding their objectives, readability, complexity, uniformity and understandability. They also create a model to link the business or functional goals of the designer and attacker with the refactoring and obfuscation methods.\\ \midrule
\cite{2010_malware_obfuscation_techniques_a_brief_survey} 2010&M\newline P&\citeauthor{2010_malware_obfuscation_techniques_a_brief_survey} described different types of anti-signature techniques for malware (encryption, oligomorphism, polymorphism, and metamorphism) and obfuscation techniques used by polymorphic and metamorphic malware. The following obfuscation are discussed: dead-code insertion, register reassignment, subroutine reordering, instruction substitution, code transposition, and code integration into the target. The focus is on how the techniques look at a binary level.\\ \midrule
\cite{2010_theory_and_practice_of_program_obfuscation} 2010&G\newline P&A book chapter by \citeauthor{2010_theory_and_practice_of_program_obfuscation} describes the theory of obfuscation before the authors present and evaluate a call-flow obfuscation and instruction substitution method.\\ \midrule
\cite{2011_a_taxonomy_of_self_modifying_code_for_obfuscation} 2011&P&\citeauthor{2011_a_taxonomy_of_self_modifying_code_for_obfuscation} presented a taxonomy of self-modifying code. They categorize the methods in concealment, encoding, visibility, and exposure and assess several tools and methods.\\ \midrule
\cite{2012_a_survey_on_automated_dynamic_malware_analysis_techniques_and_tools} 2012&M A&\citeauthor{2012_a_survey_on_automated_dynamic_malware_analysis_techniques_and_tools} covered different automated dynamic analysis tools and techniques for malware. The dynamic analysis techniques entail: function call monitoring, function parameter analysis and information flow tracking.\\ \midrule
\cite{2012_cloud_protection_by_obfuscation_techniques_and_metrics} 2012&G\newline P&\citeauthor{2012_cloud_protection_by_obfuscation_techniques_and_metrics} gauged the strength of obfuscation transformations and argued the usefulness for cloud computing applications. The publication discusses potency, resilience and costs as well as complexity metrics and map them to obfuscation techniques.\\ \midrule
\cite{2012_reverse_code_engineering_state_of_the_art_and_countermeasures} 2012&A\newline P&This special issue article by \citeauthor{2012_reverse_code_engineering_state_of_the_art_and_countermeasures} briefly summarizes static and dynamic program analysis and reverse engineering countermeasures. In regards to obfuscation the authors mention code optimization, opaque predicates, invalid branches, multi threading, import table obfuscation, runtime packers, polymorphism, and virtualization.\\ \midrule
\cite{2012_scientific_but_not_academical_overview_of_malware_anti_debugging_anti_disassembly_and_anti_vm_technologies} 2012&M\newline P&\citeauthor{2012_scientific_but_not_academical_overview_of_malware_anti_debugging_anti_disassembly_and_anti_vm_technologies} gave an overview of anti-debugging and anti-disassembly as well as anti-vm used by malware. The identified techniques include the use of push and pop instructions instead of returns, converting static data to procedures, dead and fake code insertion, instruction substitutions, code reordering, register reassignment and library hiding.\\ \midrule
\cite{2013_binary_code_obfuscations_in_prevalent_packer_tools} 2013&M\newline A\newline P&\citeauthor{2013_binary_code_obfuscations_in_prevalent_packer_tools} analyzed different malware packers. The authors took several code measurements and grouped those into categories, namely dynamic code, instruction obfuscation, code layout, anti-rewriting, anti-tracing. Measurements in the \enquote{instruction obfuscation} class are e.g., percentage of indirect calls, number of indirect jumps or non-standard uses of call and ret instructions.\\ \midrule
\cite{2015_a_large_study_on_the_effect_of_code_obfuscation_on_the_quality_of_java_code} 2015&G\newline P&A study by \citeauthor{2015_a_large_study_on_the_effect_of_code_obfuscation_on_the_quality_of_java_code} quantified 44 obfuscation methods for Java by applying them on 18 applications and measuring 10 code metrics. The metrics consider the modularity, size and complexity of the code. Their work shows, that some obfuscations substantially alter the potency of the code.\\ \midrule
\cite{2016_a_study_review_on_code_obfuscation} 2016&P&A short review by \citeauthor{2016_a_study_review_on_code_obfuscation} described the areas of application, possible obfuscation measurements, available techniques and drawbacks of code obfuscation. They classify different obfuscation techniques into layout transformations, control flow transformations, data abstraction and procedural abstractions.\\ \midrule
\cite{2016_a_survey_on_aims_and_environments_of_diversification_and_obfuscation_in_software_security} 2016&P&\citeauthor{2016_a_survey_on_aims_and_environments_of_diversification_and_obfuscation_in_software_security} conducted a survey to compare the different programming languages, aims and environments of publications in the areas of diversification and obfuscation. The systematic search and selection process yielded 209 publications. One of their results is a categorization of the programming language (e.g., Java, C, Machine code, JavaScript, SQL,\ldots) the publications apply the protection techniques to. \citeauthor{2016_a_survey_on_aims_and_environments_of_diversification_and_obfuscation_in_software_security} mapped the languages into language classes (e.g., native code, domain specific languages, scripts,\ldots). Their identified aims of the publications in the field of software protection cover: prevent reverse engineering, resist tampering, hide data and prevent exploitation. The targeted environments (e.g., server, cloud, desktop, mobile, any) are mostly independent according to the two biggest categories in their results being any and any native environments.\\ \midrule
\cite{2016_hopes_fears_and_software_obfuscation} 2016&P&The review article \citetitle{2016_hopes_fears_and_software_obfuscation} surveys research on indistinguishability obfuscation and its potential impact.\\ \midrule
\cite{2016_protecting_software_through_obfuscation_can_it_keep_pace_with_progress_in_code_analysis} 2016&A\newline P&\citeauthor{2016_protecting_software_through_obfuscation_can_it_keep_pace_with_progress_in_code_analysis} described the arms race between SPs and code analysis. The authors used analysis goals (locating data, locating code, extracting data, extracting code, understanding code, and simplifying code) and methods (pattern matching, automatic static, automatic dynamic, automatic symbolic, and manual analysis) to classify the field of analysis.\\ \midrule
\cite{2016_testing_android_malware_detectors_against_code_obfuscation_a_systematization_of_knowledge_and_unified_methodology} 2016&M\newline P&\citeauthor{2016_testing_android_malware_detectors_against_code_obfuscation_a_systematization_of_knowledge_and_unified_methodology} tested the effectiveness and influence of obfuscation techniques on the detection rate of anti-malware products for Android apps. Their results indicate malware detectors are not able to recognize obfuscated malware versions.\\ \midrule
\cite{collberg2016repeatability} 2016& &\citeauthor{collberg2016repeatability} published a large-scale study on repeatability in the general field of computer science.
For more than 600 publications they tried to get access to the code for the experiments from the authors and to reproduce the experiments.
For only 37\% of the publications the code for the experiments could be obtained, and only just over half of them were executable (20\% of the total number of papers).
Their lack of success of obtaining the results can also be seen in our data from little sample reuse.\\ \midrule
\cite{2017_a_tutorial_on_software_obfuscation} 2016&P&In 2017 \citeauthor{2017_a_tutorial_on_software_obfuscation} classified obfuscations against MATE attacks based on abstraction level (source, intermediate, binary machine code), unit (instruction, basic block, function, program, system level), dynamics (static, dynamic) and target (data, code). MATE attacks based on the attack type (code understanding, data item recovery, location recovery), dynamics (static, dynamic analysis), interpretation (syntactic, semantic attacks) and alteration (passive, active attacks).\\ \midrule
\cite{2017_on_secure_and_usable_program_obfuscation_a_survey} 2017&P&\citeauthor{2017_on_secure_and_usable_program_obfuscation_a_survey} surveyed the field of SP and split it into code (source, binary) vs. model (circuits, Turing machines) oriented obfuscation.\\ \midrule
\cite{2018_a_large_scale_empirical_study_on_the_effects_of_code_obfuscations_on_android_apps_and_anti_malware_products} 2018&M\newline A&In 2018 \citeauthor{2018_a_large_scale_empirical_study_on_the_effects_of_code_obfuscations_on_android_apps_and_anti_malware_products} compared different obfuscation tools for Android. They run the obfuscation tool on benign and malicious apps and afterwards check if apps are installable and runnable. Furthermore the obfuscated Malware is run through anti-malware products to check their detection rate. The authors group the obfuscation techniques into trivial (disassembly/reassembly, modifying the AndroidManifest, Moving the uncompressed data, repackaging) and non-trivial (junk code insertion, class renaming, member reordering, string encryption, identifier renaming, control flow manipulation, reflection).\\ \midrule
\cite{2018_a_large_scale_investigation_of_obfuscation_use_in_google_play} 2018&G\newline A&\citeauthor{2018_a_large_scale_investigation_of_obfuscation_use_in_google_play} measured the usage of obfuscation used by Android apps in the Google Play store and visualized the details about obfuscation depending on last updated, number of downloads, number of apps per account. The second part of their investigation targeted the developer. Of the around 300 participants 78\% had heard of obfuscation, while only 48\% had used protections. 78\% of a subgroup of 70 people were not able to protect an app using ProGuard.\\ \midrule
\cite{2018_a_systematic_study_on_static_control_flow_obfuscation_techniques_in_java} 2018&G\newline P&\citeauthor{2018_a_systematic_study_on_static_control_flow_obfuscation_techniques_in_java} studied control flow obfuscation techniques in Java. They looked at the control flow obfuscation techniques for java used in papers and tools, tested 13 obfuscators on 16 programs and manually evaluated the outcome. They identified 36 unique techniques in literature and seven from tools. The different methods were put into classes based on the transformed component: expression, statement, basic block, method, class. Their studies indicate a gap between theory and practice as only 13 of the 36 are implemented by tools.\\ \midrule
\cite{2018_anti_emulation_trends_in_modern_packers_a_survey_on_the_evolution_of_anti_emulation_techniques_in_upa_packers} 2018&M\newline A&\citeauthor{2018_anti_emulation_trends_in_modern_packers_a_survey_on_the_evolution_of_anti_emulation_techniques_in_upa_packers} shared their knowledge on malware packers resulting from monitoring the development of packers for almost two years. The authors described three packers used by malware families like Upatre, Gamarue, Hedsen and explained the current trends in malware packers.\\ \midrule
\cite{2018_diversification_and_obfuscation_techniques_for_software_security_a_systematic_literature_review} 2018&P&The systematic literature review by \citeauthor{2018_diversification_and_obfuscation_techniques_for_software_security_a_systematic_literature_review} from 2018 analyzed 357 publications from 1993--2017. The authors surveyed metadata, environment and aims of publications in the areas of diversification and obfuscation. Analyzed metadata includes involved countries and universities, publication type (journal article, conference paper,\ldots) and the associated organizations' sector (academic, industrial, both, unknown).\\ \midrule
\cite{2018_software_protection_on_the_go_a_large_scale_empirical_study_on_mobile_app_obfuscation} 2018&G\newline A&\citeauthor{2018_software_protection_on_the_go_a_large_scale_empirical_study_on_mobile_app_obfuscation} analyzed the usage of obfuscation in the iOS ecosystem. They developed a system to compute the likelihood of an app being obfuscated and evaluated it on over a million apps. The system identifies symbol renaming, string encoding, decompilation disruption and control flow flattening by means of a natural language processing model, CFG analysis and the number of IDA Pro failures. Their results suggest an upwards trend for obfuscation and a correlation between the likelihood of an app being obfuscated and the type of application (finance, privacy intellectual properties and monetization are more likely).\\ \midrule
\cite{2018_tutorial_an_overview_of_malware_detection_and_evasion_techniques} 2018&A&The tutorial by \citeauthor{2018_tutorial_an_overview_of_malware_detection_and_evasion_techniques} describes static and dynamic malware detection techniques and tools such as PEiD, DIE, YARA, Cuckoo, and angr. The authors show the basic usage for each tool and describe their limitations in the presence of SPs.\\ \midrule
\cite{2018_understanding_android_obfuscation_techniques_a_large_scale_investigation_in_the_wild} 2018&M\newline G A&In 2018 a huge study by \citeauthor{2018_understanding_android_obfuscation_techniques_a_large_scale_investigation_in_the_wild} investigated the dispersiveness of obfuscation in over 110.000 Android apps. The authors used machine learning on apps from Google Play and six 3rd-party markets as well as VirusShare and VirusTotal to detect identifier renaming, encryption, reflection and packing. The machine learning approach uses the F-Droid Android app repository for the learning phase.\\ \midrule
\cite{2019_a_taxonomy_of_software_integrity_protection_techniques} 2019&P&\citeauthor{2019_a_taxonomy_of_software_integrity_protection_techniques} developed a taxonomy for the field of software protections. The taxonomy describes the view from a view of the defender, the system and the attacker and portray how these views interact with each other. The authors applied the taxonomy to around 50 publications. \\ \midrule
\cite{2019_obfuscated_android_application_development} 2019&M\newline G\newline A&\citeauthor{2019_obfuscated_android_application_development} studied the development of obfuscation techniques over time. They used goodware (GOOD dataset) as well as malware (AndroZoo, AMD and Drebin dataset) and compared the different datasets. The observed techniques are packing (encrypting bytecode), native (usage of native code), DCL (dynamic code loading) and reflection (for hiding method and fields).\\ \midrule
\cite{2019_obfuscated_code_quality_measurement} 2019&A&\citeauthor{2019_obfuscated_code_quality_measurement} described available code complexity metrics for obfuscated code and proposed several improvements. Furthermore, for the different types of metrics the work outlines retrieval algorithms using graphs.\\ \midrule
\cite{2019_obfuscation_where_are_we_in_anti_dse_protections_a_first_attempt} 2019&P&\citeauthor{2019_obfuscation_where_are_we_in_anti_dse_protections_a_first_attempt} explored a subset of SPs against dynamic symbolic execution. For each protection the publication outlines the strength, cost, correctness, stealth, and known mitigations as well as experimental evaluations from previous literature.\\ \midrule
\cite{2020_a_large_scale_study_on_the_adoption_of_anti_debugging_and_anti_tampering_protections_in_android_apps} 2020&G\newline A&\citeauthor{2020_a_large_scale_study_on_the_adoption_of_anti_debugging_and_anti_tampering_protections_in_android_apps} looked at the state of protections against debugging and tampering in Android apps. The work analyzed 23610 apps, out of which 75.62\% employ two to four protections and only 7.49\% implemented no protection. The protection present in most apps were signature checking (88.90\%), installer verification (74.42\%) and emulator detection (49.83\%).\\ \midrule
\cite{2020_a_survey_of_android_malware_detection_with_deep_neural_models} 2020&M\newline A&\citeauthor{2020_a_survey_of_android_malware_detection_with_deep_neural_models} compared the performance and resilience of Android malware detection and classification approaches against code obfuscation.\\ \midrule
\cite{2020_cryptographic_obfuscation_a_survey} 2020&G\newline P&In their book \citeauthor{2020_cryptographic_obfuscation_a_survey} surveyed and explained concepts and methods for cryptographic obfuscation, such as virtual black box obfuscation, indistinguishability obfuscation.\\ \midrule
\cite{2020_layered_obfuscation_a_taxonomy_of_software_obfuscation_techniques_for_layered_security} 2020&P&\citeauthor{2020_layered_obfuscation_a_taxonomy_of_software_obfuscation_techniques_for_layered_security} created a taxonomy for layered obfuscation. It splits SPs into four layers: code element, software component, inter-component and application layer. Each layer contains several sub-layers with the individual targets (layout, controls, data, methods, classes, library calls, resources, DRM systems, neural networks).\\ \midrule
\cite{2021_a_survey_of_android_malware_static_detection_technology_based_on_machine_learning} 2021&M\newline A&\citeauthor{2021_a_survey_of_android_malware_static_detection_technology_based_on_machine_learning} surveyed the features of machine learning approaches to detect malicious Android applications, as well as which malware datasets were used for evaluation and the achieved accuracy. Furthermore, the authors discussed the various obfuscation techniques present in the data sets and how they influence the features.\\ \midrule
\cite{2021_a_survey_of_binary_code_similarity} 2021&M\newline G\newline A&\citeauthor{2021_a_survey_of_binary_code_similarity} compared different approaches for binary code similarity detection. They focused on the platform, targeted architectures, availability, datasets, methodology, and if the approach was tested against obfuscated code.\\ \midrule
\cite{android_app_forensics_survey} 2021&A\newline P&\citeauthor{android_app_forensics_survey} survey Android obfuscation and deobfuscation tools. They categorize, describe each tool and publication and oppose the obfuscation techniques with the deobfuscation tools and analysis methods.\\ \midrule
\cite{2021_measuring_software_obfuscation_quality_a_systematic_literature_review} 2021&M\newline G\newline P&\citeauthor{2021_measuring_software_obfuscation_quality_a_systematic_literature_review} analyzed 67 studies in the field of software obfuscation in a systematic literature review and look at the measurements, evaluation of the obfuscation quality. They reported numbers on the venues with at least two publications (top being Computers \& Security with six), the type of study (51 experiments, one case study, 16 simulations) dataset (top two are: 33 in-the-wild, 10 manually written) as well as which obfuscation quality the examined studies measured (20 potency, nine resilience, 21 cost, eight stealth, 36 similarity).\\ \midrule
\cite{2021_sok_automatic_deobfuscation_of_virtualization_protected_applications} 2021&P&\citeauthor{2021_sok_automatic_deobfuscation_of_virtualization_protected_applications} classified deobfuscation frameworks for virtualization-protected applications based on the degree of automation, extracted artifacts, and analysis methods. The work also presents experiments with existing tools on a uniform sample set for comparable results.\\ \midrule
\end{longtable}
}

\end{document}